\documentclass{article}

\usepackage{arxiv}
\usepackage[square,numbers]{natbib}
\usepackage[hidelinks]{hyperref}
\usepackage{packages_and_macros}
\usepackage{amssymb,mathtools}
\usepackage{titlesec}
\titleformat{\subsubsection}[runin]
  {\normalfont\bfseries} 
  {\thesubsubsection}    
  {1em}                  
  {}                     
  [.]                    

\title{SoK: Harmonizing Attack Graphs and Intrusion Detection Systems}

\date{}

\author{Andrea Agiollo\thanks{Authors contributed equally to this research.}\\
        Delft University of Technology \\
        Van Mourik Broekmanweg 5\\
        Delft, The Netherlands, 2628 XE\\
        \texttt{A.Agiollo-1@tudelft.nl}
        \And
        Enkeleda Bardhi\footnotemark[1]\\
        Delft University of Technology \\
        Van Mourik Broekmanweg 5\\
        Delft, The Netherlands, 2628 XE\\
        \texttt{E.Bardhi-1@tudelft.nl}
        \And
        Alessandro Palma\footnotemark[1]\\
        Sapienza Univerisity of Rome \\
        Via Ariosto 25 \\
        Rome, Italy, 00185 \\
        \texttt{palma@diag.uniroma1.it}
        \And
        Riccardo Lazzeretti\\
        Sapienza Univerisity of Rome \\
        Via Ariosto 25 \\
        Rome, Italy, 00185 \\
        \texttt{lazzeretti@diag.uniroma1.it}
        \And
        Silvia Bonomi\\
        Sapienza Univerisity of Rome \\
        Via Ariosto 25 \\
        Rome, Italy, 00185 \\
        \texttt{bonomi@diag.uniroma1.it}
        \And
        Fernando Kuipers\\
        Delft University of Technology \\
        Van Mourik Broekmanweg 5\\
        Delft, The Netherlands, 2628 XE\\
        \texttt{F.A.Kuipers@tudelft.nl}
}

\renewcommand{\shorttitle}{SoK: Harmonizing Attack Graphs and Intrusion Detection Systems}
\renewcommand{\undertitle}{This manuscript is currently under review at ACM Transactions on Privacy and Security.}
\renewcommand{\headeright}{Under Review}

\begin{document}
\maketitle

\begin{abstract}
    Detecting and responding to cyber attacks is increasingly difficult as high-volume, complex network traffic allows threats to remain concealed. 
While Intrusion Detection Systems (IDSs) identify anomalous behavior, Attack Graphs (AGs) serve as the primary threat model for analyzing attacker strategies and informing any response. 
Despite the conceptual connection being recognized in early research, the field of AG and IDS integration lacks a common structure. 

This paper presents the first systematic analysis of AG-IDS integration, reviewing a \totalcount{} comprehensive works in literature.
We introduce a novel taxonomy revealing that current research is dominated by specialized, single-purpose integrations—such as using AGs to filter IDS false positives or using IDS alerts to prune AGs.
Our analysis highlights a critical gap: the absence of a unifying framework that treats IDSs and AGs as a cohesive, integrated system.
 
To address this gap, we propose a formal AG-IDS lifecycle. 
This framework establishes a continuous feedback loop where IDSs refine the accuracy of AG models, and those updated models, in turn, enhance IDS detection capabilities.
We provide a proof-of-concept implementation demonstrating the practical advantages of this lifecycle for threat detection and incident response. 
Finally, we conclude by elaborating on significant opportunities for future development within the AG-IDS domain.

\end{abstract}

\newacronym{nist}{NIST}{National Institute of Standards and Technology}

\newacronym{ag}{AG}{Attack Graph}
\newacronym{ids}{IDS}{Intrusion Detection System}
\newacronym{nids}{NIDS}{Network-level Intrusion Detection System}
\newacronym{hids}{HIDS}{Host-level Intrusion Detection System}
\newacronym{sids}{SIDS}{Signature-based Intrusion Detection System}
\newacronym{aids}{AIDS}{Anomaly-based Intrusion Detection System}
\newacronym{cps}{CPS}{Cyber Physical System}

\newacronym{ddos}{DDoS}{Distributed Denial of Service}
\newacronym{cve}{CVE}{Common Vulnerabilities and Exposures}

\newacronym{ml}{ML}{Machine Learning}
\newacronym{dl}{DL}{Deep Learning}
\newacronym{nn}{NN}{Neural Network}
\newacronym{dnn}{DNN}{Deep Neural Network}
\newacronym{cnn}{CNN}{Convolutional Neural Network}
\newacronym{gnn}{GNN}{Graph Neural Network}
\newacronym{lstm}{LSTM}{Long Short-Term Memory}
\newacronym{ais}{AIS}{Artificial Immune System}
\newacronym{mc}{MC}{Markov Chain}
\newacronym{dt}{DT}{Decision Tree}

\newacronym{iot}{IoT}{Internet of Things}
\newacronym{sdn}{SDN}{Software-Defined Networks}
\newacronym{ami}{AMI}{Advanced Metering Infrastructure}

\newacronym{fpr}{FPR}{False Positive Rate}
\newacronym{cti}{CTI}{Cyber Threat Intelligence}

\section{Introduction}
\label{sec:intro}

%
Attack detection and response are two essential phases for securing computer networks~\cite{opdenbusch2025we,vermeer2023,althunayyan2026survey}. 
Detection involves monitoring system states, identifying anomalous activities, and assessing their impact on the network. 
In this context, \gls{ids} are the state-of-the-art tools for identifying and alerting anomalous behavior~\cite{KhraisatCybersecurity2019,LambertsArxiv2023,JeuneSpace2021,teuwen2024ruling}.
On the other hand, responding to potential cyber risks in the network involves reacting to detected incidents, defining mitigation strategies, and updating response plans. 
Effective response relies on threat modeling to anticipate potential attacks~\cite{xiong2019threat}.
\gls{ag}~\cite{kaynar_taxonomy_2016,palma_behind_2025} supports response by modeling possible attackers' strategies in a graph-based structure that enables the analysis of attack vectors and the planning of countermeasures in distributed computer networks~\cite{palma2023workflow,KonstaCompsec2024,almazrouei2023review}.
The potential to use these techniques in combination is widely recognized in the literature~\cite{CapobiancoNspw2019,NavarroCompsec2018,10190540,nadeem2021sage,rose2022ideres}.
\glspl{ids} can identify individual anomalies but lack the ability to correlate them across nodes, whereas \gls{ag} captures the relationships among vulnerabilities and attack steps, enabling the reconstruction and early detection of complex, multi-stage intrusions~\cite{10.1145/3154273.3154311,palma_shield_2025}.
With this motivation, different research efforts combined these techniques to correlate alerts~\cite{shittu2015intrusion}, learn attacker strategies~\cite{ning2003learning}, or plan optimal countermeasure in complex networks~\cite{shameli2016dynamic,palma_sparq_2025}.
For instance, \glspl{ag} often suffer from scalability issues due to the combinatorial number of strategies an attacker may exploit in a network~\cite{SaburCompnet2022,palma_behind_2025,yang2026self}.
In this case, research focused on addressing such issues by leveraging intrusion alerts~\cite{ZHANG2008188,jia2023artificial,haque2020integrating,mjihil_improving_2017}.
Similarly, \glspl{ids} must process increasing volumes of traffic, resulting in high false positive rates~\cite{NeupaneAccess2022}.
Hence, research focused on addressing the problem of false positive reduction by leveraging \glspl{ag}~\cite{JADIDI2022103741,RAMAKI2015206,10.1145/3154273.3154311,CapobiancoNspw2019}.
Despite clear benefits, efforts to combine \glspl{ag} and \glspl{ids} remain highly fragmented. Existing approaches often tackle isolated problems -- such as reducing \gls{ids} false positives or improving \gls{ag} scalability -- without a unified vision for deployment in dynamic, real-world networks. Integration is further complicated by heterogeneous \gls{ids} technologies, diverse \gls{ag} models, and inconsistent alert semantics. To address these gaps, this paper provides the first comprehensive Systematization of Knowledge (SoK) on \gls{ag}–\gls{ids} integration, categorizing existing methods, and identifying opportunities for advancing adaptive and scalable solutions.

Beyond consolidating a fragmented field, our SoK critically evaluates the strengths and limitations of combining \gls{ag} and \gls{ids} and reveals how far their integration has truly been explored.
To guide our systematization, we pose the following research questions that enable identifying a novel taxonomy:
\begin{enumerate}[label=\textbf{RQ\arabic{*}}]
    \item\label{item:rq1} -- \emph{``How are \glspl{ag} and \glspl{ids} coupled together in existing approaches?''}
    \item\label{item:rq2} -- \emph{``To what end are the current approaches combining \glspl{ag} and \glspl{ids}?''}
\end{enumerate}
Based on these questions, we systematized \totalcount{} papers, whose analysis resulted in the identification of four categories of \gls{ag}-\gls{ids} combination approaches, namely: 
\begin{inlinelist}
    \item \emph{\gls{ids}-based \gls{ag} generation} approaches ($AG|IDS$): constructing  meaningful \gls{ag}(s) from \gls{ids} output;
    \item \emph{\gls{ag}-integrated \gls{ids}} approaches ($IDS[AG]$): embedding \gls{ag} to optimize the \gls{ids} detection pipeline; 
    \item \emph{\gls{ag}-based \gls{ids} refinement} approaches ($IDS \rightarrow AG$): leveraging \gls{ag} to validate or enhance the \gls{ids} outputs; and
    \item \emph{hybrid approaches}: combining any two of the above-mentioned categories.
\end{inlinelist}

Our systematization reveals that hybrid \gls{ag}-\gls{ids} integration approaches offer promising avenues for enabling continuous, adaptive threat detection and response.
However, most of the analyzed works typically assume detection environments where integration occurs once and remains unchanged during the detection process.
This assumption fails to resemble real-world computer networks, where security threats continuously evolve, and network characteristics change.
To address this gap, we envision a novel \gls{ag}-\gls{ids} \emph{lifecycle} that supports continuous updates and iterative refinement to support prompt responses and accurate detection.
The term \emph{lifecycle} aptly captures the ongoing, structured process through which these techniques evolve in response to emerging cybersecurity threats.
We illustrate the proposed lifecycle through a proof-of-concept, designed to elucidate the potential of its key stages and core functionalities.
In summary, we contribute:
\begin{itemize}
    \item a comprehensive systematization of \gls{ag}-\gls{ids} integration approaches, covering \totalcount{} papers;
    \item the definition of a new taxonomy based on integration methods and objectives; 
    \item a novel \gls{ag}-\gls{ids} lifecycle, showcasing its envisioned effectiveness experimentally; and
    \item the systematization of the current research gaps and opportunities.
\end{itemize}
%

The remainder of the paper is structured as follows.
\Cref{sec:background_related} introduces the preliminary concepts and notations on \glspl{ids} and \glspl{ag}, as well as the related work.
The systematization methodology and obtained taxonomy are presented in \Cref{ssec:methodology,ssec:taxonomy}.
\Cref{ssec:ag_generation_survey,ssec:ag_integrated_ids_survey,ssec:ag_based_ids_refine_survey,ssec:hybrid_survey} delve into the core categorization of our systematization, presenting each category of approaches.
\Cref{sec:lifecycle,sec:case_study} expand on the categorization, by envisioning an \gls{ag}-\gls{ids} lifecycle, which is implemented as a proof-of-concept.
\Cref{sec:opportunities} lists the opportunities for future research directions that emerge from our systematization.
Finally, \Cref{sec:conclusions} provides the conclusions and outlooks.
\section{Background and Related Work}
\label{sec:background_related}

\subsection{Background}
\label{ssec:background}
In this section, we provide a background on the two main security components considered in this SoK, namely \acrlong{ids} (\Cref{sssec:intrusion_detection_systems}) and \acrlong{ag} (\Cref{sssec:attack_graphs}).

\subsubsection{Intrusion Detection Systems}
\label{sssec:intrusion_detection_systems}
An \acrfull{ids} monitors computer networks, devices, and data traffic to detect known malicious activities or policy violations~\cite{KhraisatCybersecurity2019}.
Traditionally, \glspl{ids} are categorized by their detection technique and monitoring scope.

\paragraph{\emph{Detection perspective}}
\glspl{ids} are either \emph{\gls{sids}} or \emph{\gls{aids}}~\cite{liao2013intrusion}.
\glspl{sids} rely on pattern-matching to identify known threats using signature databases~\cite{hubballi2014false,kumar2012signature}, such as specific malware code sequences~\cite{kaur2013automatic}.
\glspl{aids} detect anomalies modeling normal system behavior and flagging significant deviations~\cite{aldweesh2020deep,liao2013intrusion}, using statistical, knowledge-based, or data-driven methods.
Given the rise of \gls{ml}-based approaches, we further classify \glspl{ids} into four types: 
\begin{inlinelist}
    \item \gls{ml}-based \glspl{aids},
    \item \gls{ml}-based \glspl{sids},
    \item non-\gls{ml} \glspl{aids}, and 
    \item non-\gls{ml} \glspl{sids}.
\end{inlinelist}
\paragraph{\emph{Monitoring process}}
\glspl{ids} also differ by their placement and scope.
\emph{\glspl{nids}} monitor network-wide traffic from strategic points -- typically near firewalls -- to detect external threats~\cite{ahmad2021network,kumar2021research}.
\emph{\glspl{hids}}, installed on individual devices -- e.g., laptops, routers --, monitor local activity such as file access and device-specific traffic~\cite{liu2018host}.
\glspl{nids} offer broad visibility, while \glspl{hids} provide fine-grained monitoring; combining both yields comprehensive protection.
Despite advances in targeted \gls{ids}---e.g., for \gls{ddos}~\cite{sahu2021structural,melo2019cloud,wang2022end,bardhi2025mine}, they still suffer from false positives/negatives due to limited host-level correlation~\cite{NavarroCompsec2018}, hindering holistic threat analysis.
\subsubsection{Attack Graphs}
\label{sssec:attack_graphs}
An \acrfull{ag} is a graph-based model representing potential attack paths an adversary might follow by exploiting sequences of vulnerabilities to compromise a network. 
It serves as a comprehensive threat model supporting tasks such as cyber risk assessment~\cite{zenitani2022attack,palma2023time}, network hardening~\cite{bonomi_improving_2025,lagraa2024review}, and threat detection~\cite{NavarroCompsec2018,palma_progressive_2025}.
The core of \glspl{ag} lies in the interconnection of vulnerability exploits, which define multi-step attacks~\cite{mubasshirgotta}.
%
Based on the representation of the attacker strategies and their relationships, four main \gls{ag} models are identified in the state of the art taxonomy~\cite{kaynar_taxonomy_2016}: 
\begin{itemize} 
    \item \emph{State-based \glspl{ag}} represents network state nodes indicating networked software configuration, attacker privileges, network performance measures, etc.
    \item \emph{Vulnerability-based \glspl{ag}} focus on vulnerability exploits.
    \item \emph{Host-based \glspl{ag}} represent hosts and topological information of the network.
    \item \emph{Attack Scenario-based \glspl{ag}} represent attack plans and strategies composed of coordinated attacker actions.
\end{itemize}
Research primarily focuses on \gls{ag} \emph{generation}---i.e., computing possible attack paths~\cite{zenitani2022attack} since it
%
remains a key challenge due to the combinatorial number of attack paths in the network size and vulnerabilities~\cite{LallieCompscierev2020}.
As a result, applying \glspl{ag} in dynamic, real-time, and heterogeneous environments remains an open research problem~\cite{palma_behind_2025}. 
%
\subsection{Related Work} 
\label{ssec:related}
With the growing interest in supporting detection and analysis for network security, different studies have reviewed the integration of \gls{ag} and \gls{ids}.
From a modeling standpoint, recent works survey graph-based approaches, emphasizing the potential of \glspl{ag} for alert correlation~\cite{lagraa2024review,albasheer2022cyber}.
In particular, Almazrouei et al.~\cite{almazrouei2023review} provide a comprehensive review of \glspl{ag} for vulnerability assessment in \gls{iot} networks, highlighting their role in intrusion detection.
Focusing on detection, Kovavcevic et al.~\cite{kovavcevic2020systematic} review techniques for attack scenario detection but lack coverage of diverse \gls{ag} models.
Capobianco et al.~\cite{CapobiancoNspw2019} explore the use of \glspl{ag} for intrusion detection, particularly in computing attack surfaces, states, and actions.
Overall, existing surveys remain fragmented, each addressing specific aspects of the \gls{ag}-\gls{ids} interplay without offering a unified systematization.
Additionally, some works address multi-step attack detection and its challenges -- such as modeling, automation, dataset design, and reproducibility~\cite{NavarroCompsec2018,ponnumani2026multi} -- but do not fully explore the \gls{ag}-\gls{ids} combination in this context, as this SoK aims to do. 
\section{Systematization}
\label{sec:systematization}
In this section, we collect and analyze the relevant literature concerning the \gls{ag}-\gls{ids} integration, thus answering \ref{item:rq1} and \ref{item:rq2}.
To this end, we first describe the methodology used to collect the relevant literature in \Cref{ssec:methodology}.
Then, we introduce a novel taxonomy for the \gls{ag}-\gls{ids} integration (see~\Cref{ssec:taxonomy}) and categorize the collected relevant papers accordingly (see~\Cref{ssec:ag_generation_survey,ssec:ag_integrated_ids_survey,ssec:ag_based_ids_refine_survey,ssec:hybrid_survey}).
%
\subsection{Methodology}
\label{ssec:methodology}
To systematize the literature, we defined a set of queries to be executed on widely used bibliographic search engines.
The queries incorporate the following keyword combinations ($\wedge$ is AND, $\vee$ is OR):
\begin{itemize}
    \item `intrusion detection' $\wedge$ (`attacks graph' $\vee$ `attack path')
    \item (`anomaly detection' $\vee$ `alert') $\wedge$ (`attack graph' $\vee$ `attack path')
    \item `attack graph generation' $\wedge$ (`anomalies' $\vee$  `alerts')
    \item (`attack mitigation' $\vee$ `secure network') $\wedge$ (`attack graph' $\vee$  `attack path')
\end{itemize}
As far as bibliographic search engines are concerned, we leverage ACM Digital Library, IEEE Xplore, Springer Link, Science Direct/Scopus, and aggregators like Google Scholar and DBLP.
For each search engine and query pair, we reviewed the first three pages of results.
Our focus on top-ranked results prioritizes high-impact work.
The research reached saturation after the first three pages of all digital libraries, where no additional relevant papers appeared in any case. 
This first search process led us to $102$ initial articles.
For each result, the three equally contributing authors inspect the title, abstract, and introduction to determine whether the paper addresses the topic of \gls{ag}-\gls{ids} integration.
We classify each paper according to three disjoint circumstances:
\begin{inlinelist}
    \item\label{step:primary} \emph{Primary work}, which proposes a methodology for integrating \glspl{ag} and \glspl{ids},
    \item\label{step:secondary} \emph{Secondary work}, which surveys literature related to either \gls{ag}, \gls{ids}, or both,
    \item \emph{Unrelated work} which does not pertain to either \gls{ag} or \gls{ids}, and is thus excluded from our analysis.
\end{inlinelist}
Notably, secondary works identified in step \ref{step:secondary} served as valuable sources for discovering additional primary works.
We recursively explored their bibliographies to identify further relevant contributions.
In this phase, we particularly relied on comprehensive surveys such as \cite{almazrouei2023review,lagraa2024review,albasheer2022cyber,kovavcevic2020systematic,NavarroCompsec2018}, which we recognize as significant references in the \gls{ag} and \gls{ids} domains (see~\Cref{ssec:related}).
While informative, these secondary works typically focus on only one of the two topics -- either \gls{ag} or \gls{ids} -- and often lack detailed schematics outlining integration aspects.
To refine our selection of primary works, we assessed each paper based on the following questions:
\begin{itemize}
    \item ``Are the \gls{ids} specifics -- e.g., detection engine, scale, monitoring process -- clearly described?"
    \item ``Are the \gls{ag} specifics -- e.g., category, scale -- clearly described?" 
    \item ``Is the integration purpose explicitly stated?" 
\end{itemize}
%
Papers that failed to provide clear answers to all three questions were excluded from further analysis due to insufficient specificity.
This selection process ultimately yielded \totalcount{} articles, which are examined in this SoK.
Each of the equally contributing authors deeply analyzed one of the categories (see~\Cref{ssec:taxonomy}). 
Iterative sessions were organized for further discussions and presentations of the findings of each category.
Finally, any disagreements regarding the classification or inclusion of a paper were resolved through majority voting among the authors.


\noindent \textbf{Disclaimer:} The research methodology presented herein applies exclusively to the \textit{integration} of \gls{ag} and \gls{ids}.
Consequently, literature proposing novel advancements in \textit{standalone IDS}---such as provenance-graph-based systems~\cite{cheng2024kairos,jiang2025orthrus,wang2022threatrace}---is considered outside the scope of this work.
Furthermore, research focusing on generic \textit{graph-based} solutions (e.g., provenance graphs or \gls{gnn} architectures~\cite{wang2024k,ding2023airtag}) that do not align with the specific \gls{ag} definition (see \Cref{sssec:attack_graphs}) is similarly excluded from this study.

\subsection{Taxonomy}
\label{ssec:taxonomy}

\begin{figure}
    \centering
    \resizebox{0.6\linewidth}{!}{
        \begin{tikzpicture}[
        title/.style={font=\fontsize{10}{10}\color{black},align=center,on grid},
        citer/.style={font=\fontsize{18}{18}\color{black},align=center,on grid},
        allround/.style={fill opacity=0.3,
            append after command={%
                \pgfextra
                    \fill[fill=#1!40] (\tikzlastnode.west) [rounded corners] |- (\tikzlastnode.north) -| (\tikzlastnode.east) |- (\tikzlastnode.south) -| cycle;
                    \draw[color=#1,rounded corners] (\tikzlastnode.west) |- (\tikzlastnode.north) -| (\tikzlastnode.east) |- (\tikzlastnode.south) -| cycle;
                \endpgfextra}},
        allroundborder/.style={fill opacity=0.3,
            append after command={%
                \pgfextra
                    \draw[color=#1,rounded corners] (\tikzlastnode.west) |- (\tikzlastnode.north) -| (\tikzlastnode.east) |- (\tikzlastnode.south) -| cycle;
                \endpgfextra}},
        bottomflat/.style={fill opacity=0.3,
            append after command={%
                \pgfextra
                    \fill[fill=#1!40] (\tikzlastnode.south west) [rounded corners] |- (\tikzlastnode.north) -| (\tikzlastnode.east) [sharp corners] |- cycle;
                    \draw[color=#1,rounded corners] (\tikzlastnode.south west) |- (\tikzlastnode.north) -| (\tikzlastnode.south east);
                \endpgfextra}},
        bottomflatbutline/.style={fill opacity=0.3,
            append after command={%
                \pgfextra
                    \fill[fill=#1!40] (\tikzlastnode.south west) [rounded corners] |- (\tikzlastnode.north) -| (\tikzlastnode.east) [sharp corners] |- cycle;
                    \draw[color=#1,rounded corners] (\tikzlastnode.south west) |- (\tikzlastnode.north) -| (\tikzlastnode.south east) [sharp corners] (\tikzlastnode.south east) -| (\tikzlastnode.south west);
                \endpgfextra}},
        topflat/.style={fill opacity=0.3, 
            append after command={%
                \pgfextra
                    \fill[fill=#1!40] (\tikzlastnode.north east) [rounded corners] |- (\tikzlastnode.south) -| (\tikzlastnode.west) [sharp corners] |- cycle;
                    \draw[color=#1,rounded corners] (\tikzlastnode.north east) |- (\tikzlastnode.south) -| (\tikzlastnode.north west);
                \endpgfextra}},
        leftflat/.style={fill opacity=0.3, 
            append after command={%
                \pgfextra
                    \fill[fill=#1!40] (\tikzlastnode.north west) [rounded corners] -| (\tikzlastnode.east) |- (\tikzlastnode.south) [sharp corners] -| cycle;
                    \draw[color=#1,rounded corners] (\tikzlastnode.north west) -| (\tikzlastnode.east) |- (\tikzlastnode.south west);
                \endpgfextra}},
        rightflat/.style={fill opacity=0.3, 
            append after command={%
                \pgfextra
                    \fill[fill=#1!40] (\tikzlastnode.south east) [rounded corners] -| (\tikzlastnode.west) |- (\tikzlastnode.north) [sharp corners] -| cycle;
                    \draw[color=#1,rounded corners] (\tikzlastnode.south east) -| (\tikzlastnode.west) |- (\tikzlastnode.north east);
                \endpgfextra}},
        leftandbottomflat/.style={fill opacity=0.3, 
            append after command={%
                \pgfextra
                    \fill[fill=#1!40] (\tikzlastnode.north west) [rounded corners] -| (\tikzlastnode.east) [sharp corners] |- (\tikzlastnode.south) -| cycle;
                    \draw[color=#1,rounded corners] (\tikzlastnode.north west) -| (\tikzlastnode.south east);
                \endpgfextra}},
        leftandtopflat/.style={fill opacity=0.3, 
            append after command={%
                \pgfextra
                    \fill[fill=#1!40] (\tikzlastnode.north east) [rounded corners] |- (\tikzlastnode.south) [sharp corners] -| (\tikzlastnode.west) |- cycle;
                    \draw[color=#1,rounded corners] (\tikzlastnode.north east) |- (\tikzlastnode.south west);
                \endpgfextra}},
        rightandbottomflat/.style={fill opacity=0.3, 
            append after command={%
                \pgfextra
                    \fill[fill=#1!40] (\tikzlastnode.south west) [rounded corners] |- (\tikzlastnode.north) [sharp corners] -| (\tikzlastnode.east) |- cycle;
                    \draw[color=#1,rounded corners] (\tikzlastnode.south west) |- (\tikzlastnode.north east);
                \endpgfextra}},
        topandrightflat/.style={fill opacity=0.3, 
            append after command={%
                \pgfextra
                    \fill[fill=#1!40] (\tikzlastnode.south east) [rounded corners] -| (\tikzlastnode.west) [sharp corners] |- (\tikzlastnode.north) -| cycle;
                    \draw[color=#1,rounded corners] (\tikzlastnode.south east) -| (\tikzlastnode.north west);
                \endpgfextra}},
        onlytop/.style={fill opacity=0.3, 
            append after command={%
                \pgfextra
                    \fill[fill=#1!40] (\tikzlastnode.north west) [sharp corners] -| (\tikzlastnode.east) |- (\tikzlastnode.south) -| (\tikzlastnode.west) |- cycle;
                    \draw[color=#1] (\tikzlastnode.north west) -| (\tikzlastnode.north east);
                \endpgfextra}},
        onlyleft/.style={fill opacity=0.3, 
            append after command={%
                \pgfextra
                    \fill[fill=#1!40] (\tikzlastnode.north west) [sharp corners] -| (\tikzlastnode.east) |- (\tikzlastnode.south) -| (\tikzlastnode.west) |- cycle;
                    \draw[color=#1] (\tikzlastnode.south west) |- (\tikzlastnode.north west);
                \endpgfextra}},
        onlybottom/.style={fill opacity=0.3, 
            append after command={%
                \pgfextra
                    \fill[fill=#1!40] (\tikzlastnode.north west) [sharp corners] -| (\tikzlastnode.east) |- (\tikzlastnode.south) -| (\tikzlastnode.west) |- cycle;
                    \draw[color=#1] (\tikzlastnode.south east) -| (\tikzlastnode.south west);
                \endpgfextra}}
        ]
            \node[citer] (genonly) at (5,3) {
            \cite{nadeem2021sage,buabualuau2024forecasting,fredj2015realistic,sharadqh2023hybrid,mao2021mif,zhao2024graph}\\
            \cite{tayouri2025coral,haque2020integrating,li2019complex,wang2022end,sahu2021structural}\\
            \cite{alrehaili2023attack,8245631,zhang2019multi,bajtovs2020multi,pivarnikova2020early}\\
            \cite{liao2010building,DBLP:journals/compsec/ZhangLCF08,ning2003learning,4406402,saad2013extracting}\\
            \cite{wang2016alert,hoque2016alert,alhaj2023effective,10.1145/3325061.3325062,zhang2023correlating}\\
            \cite{xie2010using,noel2004correlating,mohammadzad2023magd,mouwen2022robust,nadeem2021alert}
            };
            \node[citer] (postonly) at (0,0) {
            \cite{rose2022ideres,hu2020attack,wang2006using,zali2012realarticle,fayyad2013attack}\\
            \cite{shittu2015intrusion,melo2019cloud,JADIDI2022103741,dang2023research,shawly2023detection}\\
            \cite{melo2019novel,chung2013nice,shameli2015orcef,shameli2016dynamic}
            };
            \node[citer] (basedonly) at (-5,4) {
            \cite{zali2012realconf,RAMAKI2015206,franccois2016bayesian,10.1145/3154273.3154311}\\
            \cite{8613951,sen2022holistic,wu2023network,kazeminajafabadi2024optimal}
            };
            \node[citer] (genandpost) at (5, 0.25) {
             \cite{alserhani2010mars,wang2010alert,ghasemigol2016comprehensive,sen2021towards,DBLP:journals/access/AminSNTK21}\\
             \cite{sen2022using,melo2022ism,jia2023artificial,PresekalSRP23,presekal2025anomaly}
            };
            \node[citer] (genandbased) at (0, 4) {
            \cite{4511565,ahmadian2016causal,holgado2017real}\\
            \cite{yang2022multi,bhardwaj2024attack}
            };
            \node[citer] (basedandpost) at (-5, 1) {
            \cite{tabia2010bayesian,roschke2011new,ahmadinejad2011hybrid}\\
            \cite{bendiab2020advanced,eom2020framework,LANIGAN2025200606}
            };

            \node[citer] (lifecycle) at (0, 2.2) {
            \small{\textit{AG-IDS Integration}}\\\small{\textit{Lifecycle}} (\Cref{sec:lifecycle})
            };
            
            \begin{pgfonlayer}{background}
                \begin{scope}[opacity=.5, transparency group]
                    \path node[topflat=alizarin,minimum width=5cm, minimum height=2.5cm] at (-5, 0.75) {};
                    \path node[rightandbottomflat=alizarin,minimum width=5cm, minimum height=2.5cm] at (-5, 4.25) {};
                    \path node[leftandtopflat=alizarin,minimum width=5cm, minimum height=1.875cm] at (0, 2.125) {};
                    \path node[onlyleft=alizarin,minimum width=5cm, minimum height=1.875cm] at (-5, 2.125) {};
                    \path node[leftandbottomflat=alizarin,minimum width=5cm, minimum height=2.5cm] at (0, 4.25) {};
                    \path node[allround=white,minimum width=2.5cm, minimum height=0.5cm] at (-4.75, 5.5) {};
                    \path node[allroundborder=alizarin,minimum width=2.5cm, minimum height=0.5cm] at (-4.75, 5.5) {};
                \end{scope}
    
                \begin{scope}[opacity=.5, transparency group]
                    \path node[topandrightflat=bleudefrance,minimum width=5cm, minimum height=2cm] at (-4.75, -0.125) {};
                    \path node[leftflat=bleudefrance,minimum width=5cm, minimum height=2cm] at (4.75, -0.125) {};
                    \path node[leftandbottomflat=bleudefrance,minimum width=4.5cm, minimum height=2cm] at (0, 1.875) {};
                    \path node[onlybottom=bleudefrance,minimum width=4.5cm, minimum height=2cm] at (0, -0.125) {};
                    \path node[rightandbottomflat=bleudefrance,minimum width=5cm, minimum height=2cm] at (-4.75, 1.875) {};
                    \path node[allround=white,minimum width=2.5cm, minimum height=0.5cm] at (-4.75, -1.125) {};
                    \path node[allroundborder=bleudefrance,minimum width=2.5cm, minimum height=0.5cm] at (-4.75, -1.125) {};
                \end{scope}
                
                \begin{scope}[opacity=.5, transparency group]
                    \path node[topflat=darkpastelgreen,minimum width=5cm, minimum height=2.25cm] at (5.1, 0.375) {};
                    \path node[leftandbottomflat=darkpastelgreen,minimum width=5cm, minimum height=3.5cm] at (5.1, 3.25) {};
                    \path node[topandrightflat=darkpastelgreen,minimum width=5.2cm, minimum height=1.5cm] at (0, 2.25) {};
                    \path node[rightandbottomflat=darkpastelgreen,minimum width=5.2cm, minimum height=2cm] at (0, 4) {};
                    \path node[allround=white,minimum width=2.5cm, minimum height=0.5cm] at (5.25, 5) {};
                    \path node[allroundborder=darkpastelgreen,minimum width=2.5cm, minimum height=0.5cm] at (5.25, 5) {};
                \end{scope}
            \end{pgfonlayer}
    
            \node[title] at (5.25, 5){$AG|IDS$};
            \node[title] at (-4.75, -1.125){$IDS \rightarrow AG$};
            \node[title] at (-4.75, 5.5){$IDS[AG]$};
            
        \end{tikzpicture}

    }
    \caption{Distribution of papers across the three taxonomy categories of \gls{ag}-\gls{ids} integration.}
    \label{fig:class_categorization}
\end{figure}
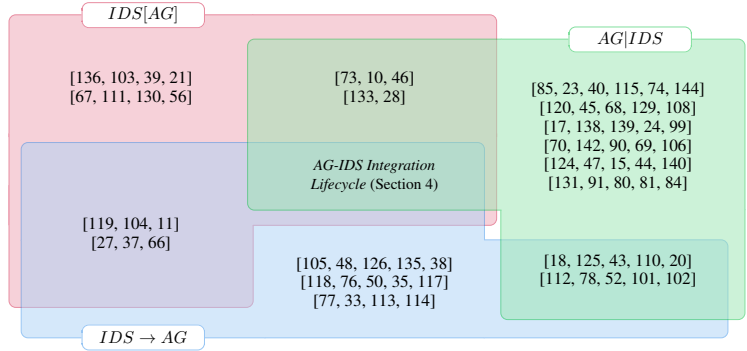

Analyzing the set of papers identified using our search methodology, a clear categorization of \gls{ag}-\gls{ids} combination approaches naturally emerged, based on the scope of integration.
More specifically, we classify these approaches according to how \glspl{ag} and \glspl{ids} are combined (\ref{item:rq1}) to achieve a particular objective (\ref{item:rq2}).
Accordingly, we identify four main categories:
\begin{inlinelist}
    \item[a)] \emph{IDS-based AG generation}, denoted as $AG|IDS$; and,
    \item[b)] \emph{AG-integrated IDS}, denoted as $IDS[AG]$; 
    \item[c)] \emph{AG-based IDS refinement}, denoted as $IDS \rightarrow AG$; 
    \item[d)] \emph{Hybrid approaches}, representing works that span two categories.
\end{inlinelist}
These categories reflect different strategies for combining \glspl{ag} and \glspl{ids}, aiming to enhance attack detection capabilities, address scalability challenges, or improve system reliability.
\Cref{fig:class_categorization} illustrates the classification of the surveyed papers according to this taxonomy.
Subsequently, we provide a definition of each category.

\paragraph{$\mathbf{AG}|\mathbf{IDS}$}
%
Approaches in the $AG|IDS$ category use \gls{ids} outputs to generate \glspl{ag} and compute potential attack paths inferred from intrusion alerts.
These methods support challenging tasks like alert correlation and vulnerability analysis, with the generated \gls{ag} providing useful support for both.
Techniques employed to generate \gls{ag} from \gls{ids} alerts include data mining~\cite{8245631,alhaj2023effective}, causal knowledge~\cite{ahmadian2016causal,ghasemigol2016comprehensive}, \gls{cti} techniques~\cite{bhardwaj2024attack,LANIGAN2025200606}, and probability computation~\cite{sahu2021structural}.
%
%

\paragraph{$\mathbf{IDS}[\mathbf{AG}]$}
The $IDS[AG]$ category encompasses approaches that incorporate \glspl{ag} or their underlying knowledge -- expressed in any form -- into the \gls{ids} inference pipeline.
These methods aim to enhance the detection capabilities or reliability of \gls{ids} systems.
Techniques in this category integrate the \gls{ag}'s network exposure knowledge -- or a relevant subportion of the \gls{ag} -- primarily for correlating \gls{ids} alerts with attack paths \cite{4511565,yang2022multi,bhardwaj2024attack}, refining \gls{ids} detection parameters based on \gls{ag} data \cite{roschke2011new,ahmadinejad2011hybrid,sen2022holistic}, or leveraging the \gls{ag} as a runtime detection module directly \cite{franccois2016bayesian,wu2023network,kazeminajafabadi2024optimal}.
%
%
The \gls{ag} may be either predefined or dynamically generated during system execution or optimization.

\paragraph{$\mathbf{IDS} \rightarrow \mathbf{AG}$}
Approaches in the $IDS \rightarrow AG$ category utilize \glspl{ag} to refine, analyze, or validate the predictions made by \gls{ids} components.
Common objectives include improving detection accuracy~\cite{melo2022ism,JADIDI2022103741}, correlating intrusion alerts for comprehensive analysis~\cite{sen2022using,hu2020attack,melo2019cloud}, and enabling adequate response based on alerts~\cite{dang2023research,shawly2023detection,bendiab2020advanced}.
Similarly to $IDS[AG]$, here the \gls{ag} may be either known a-priori or dynamically constructed during system execution.
Generally, the \gls{ag} is assumed to be complete and reliable, serving as a basis for verifying or analyzing the output of the \gls{ids}.

\paragraph{\textbf{Hybrid approaches}}
%
As illustrated in \Cref{fig:class_categorization}, the three categories are not entirely distinct.
Different works propose hybrid approaches that address multiple aspects of \gls{ag}-\gls{ids} integration.
Some methods generate \glspl{ag} to represent system states and use them to optimize the \gls{ids} detection pipeline or refine its output to catch the potential of $AG|IDS$, $IDS[AG]$, and $IDS \rightarrow AG$.
However, no existing work comprehensively integrates \emph{all} three categories (see \Cref{fig:class_categorization}). 
Such a gap highlights both the diversity of challenges tackled by each category -- such as \gls{ids} detection accuracy and \gls{ag} scalability -- and the complexity of designing multi-layered systems that integrate \glspl{ag} and \glspl{ids}.
It also underscores the need for further investigation into the strengths and limitations of current approaches, paving the way for future innovations in \gls{ag}-\gls{ids} integration.
Accordingly, in this paper, we propose to fill this gap by envisioning a continuous \gls{ag}-\gls{ids} integration lifecycle in \Cref{sec:lifecycle} and implement it in \Cref{sec:case_study} as a proof-of-concept.

\mypara{\textbf{Research subquestions}}
While our systematization of knowledge primarily aims to address \ref{item:rq1} and \ref{item:rq2}, throughout our investigation, we also consider a set of research subquestions (\textbf{RSQ}) that support the analysis of the state of the art. These questions help identify the strengths, limitations, and open challenges of existing techniques.
Specifically, we consider the following research subquestions:
\begin{enumerate}[leftmargin=1.1cm,label=\textbf{RSQ\arabic{*}}]
    \item\label{item:rsq1} {\bf Integration} -- \emph{``For what purpose are \gls{ag} and \gls{ids} combined?''}
    %
    \item\label{item:rsq2} {\bf IDS Type} -- \emph{``Which types of \gls{ids} approaches are used?''}
    \item\label{item:rsq3} {\bf AG Type} -- \emph{``Which type of \glspl{ag} are used?''}
    \item\label{item:rsq4} {\bf Threats} -- \emph{``What kinds of threats are \glspl{ag} and \glspl{ids} combined to address?''}
    \item\label{item:rsq5} {\bf Application} -- \emph{``In which application scenarios are the approaches applied?''}
    \item\label{item:rsq6} {\bf ML} -- \emph{``Among the techniques leveraging \gls{ml} for integrating \glspl{ag} and \glspl{ids}, which \gls{ml} models are used?''}
\end{enumerate}

Subsequently, we elaborate an in-depth analysis for each category based on the above research subquestions (\Cref{ssec:ag_based_ids_refine_survey,ssec:ag_integrated_ids_survey,ssec:ag_generation_survey,ssec:hybrid_survey}). 
Additionally, the reader is invited to refer to~\Cref{tab:systematization} for a detailed categorization of the systematized papers.

\subsection{IDS-based AG Generation ($\textbf{AG}|\textbf{IDS}$)}\label{ssec:ag_generation_survey} 
\begin{figure}[t]
    \centering
    \includegraphics[width=0.6\linewidth]{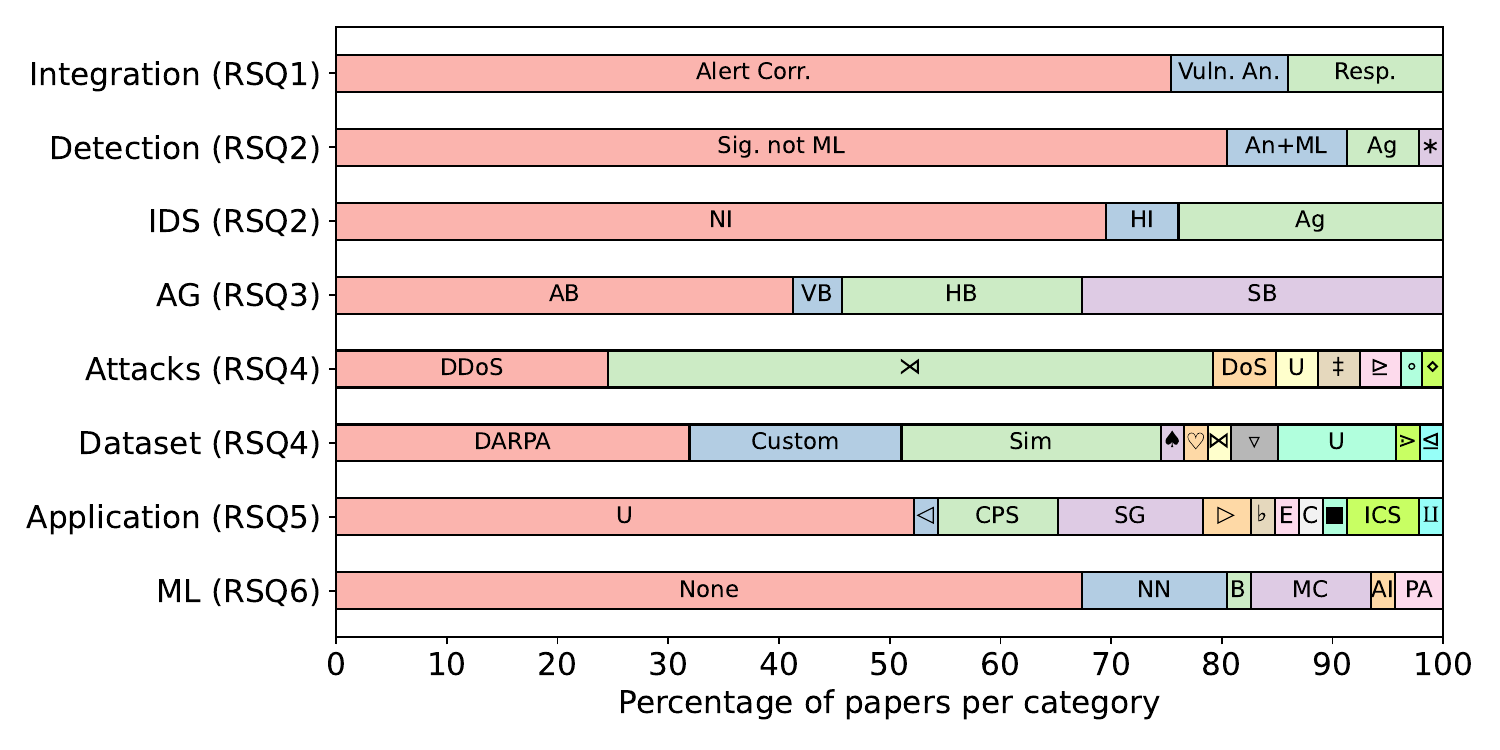}
    \caption{Statistical analysis of the \gls{ids}-based \gls{ag} generation literature. 
    Alert Corr. = Alert Correlation,
Vuln. An. = Vulnerability Analysis,
Resp. = Response,
Det. Ref. = Detection Refinement,
IDS Optim. = IDS Optimization,
Run. Det. = Runtime Detection,
Sig. not ML = Signature not ML,
An+ML = Anomaly and ML,
Ag = Agnostic,
$\ast$ = Signature and ML,
An = Anomaly not ML,
NI = Network,
HI = Host,
HB = Host-based,
SB = State-based,
VB = Vulnerability-based,
AB = Attack scenario-based,
DDoS = DDoS,
$\rtimes$ = Multi-step attacks,
$\dagger$ = Remote Code Execution,
DoS = DoS,
$\ddagger$ = U2R,
$\unrhd$ = R2L,
K = Key Loggers,
$\bullet$ = OS scan,
$\circ$ = Probing,
$\diamond$ = Port scan,
$\clubsuit$ = SSH Brute Force,
DARPA = DARPA2000,
Sim = Simulation,
$\spadesuit$ = Defcon CTF'17,
$\heartsuit$ = CSE-CIC-IDS-2018,
$\bowtie$ = ISCXIDS2012,
$\bigstar$ = NLS-KDD,
$\triangle$ = CTU-13,
$\blacktriangle$ = CICIoT2023,
$\triangledown$ = CPTC-2018,
Custom = Custom,
$\pitchfork$ = DARPA-CT-2019,
$\Vdash$ = StreamSpot,
$\gtrdot$ = 4SICS-2015,
$\unlhd$ = CCDC-2018,
U = Unspecified,
$\triangleleft$ = Cloud computing,
CPS = Cyber-Physical Systems,
SG = Smart Grids,
$\triangleright$ = Internet of Things,
$\natural$ = AMI System,
$\flat$ = Software Defined Networking,
C = Smart Cities,
$\sharp$ = Enterprise network system,
$\blacksquare$ = Smart home system,
ICS = Industrial Control Systems,
$\amalg$ = SOCs,
E = Edge computing,
None = None,
NN = Neural Network,
B = Bayesian Network,
MC = Markov Chain,
AI = Artificial Immune System,
DT = Decision Tree,
SV = Support Vector Machine,
PA = Probabilistic Automaton.
    }
    \label{fig:ag_generation_stats}
\end{figure}

We classify as $AG|IDS$ those approaches that leverage \gls{ids} outputs to construct \glspl{ag} and represent potential attack paths inferred from intrusion alerts.
Our systematization identified \greencount{} papers in this category.
\Cref{fig:ag_generation_stats} summarizes the key statistical findings, while \Cref{tab:systematization} depicts the detailed systematization for each paper.

\subsubsection*{\ref{item:rsq1} \textbf{Integration}}
Most approaches that generate \glspl{ag} leverage integration with \glspl{ids} primarily for alert correlation. \glspl{ids} typically produce an extensive number of alerts, many of which are inaccurate or redundant, making them difficult to utilize for improving detection. Alert correlation mitigates this issue by converting noisy, low-level alerts into structured, causally linked, high-level events. Once correlated, these alerts become suitable for constructing \glspl{ag} that represent complex, multi-step attacks.
Common approaches for alert correlation employ data mining~\cite{4406402,8245631,alhaj2023effective}, modeling of causal knowledge~\cite{ahmadian2016causal,DBLP:journals/compsec/ZhangLCF08,ghasemigol2016comprehensive}, \gls{cti} techniques~\cite{sen2022using,bhardwaj2024attack}, and probability computation~\cite{sahu2021structural,xie2010using}.
Among these works, many integrate \gls{ids} into \gls{ag} generation to support response.
Such works identify active attack locations in the network~\cite{PresekalSRP23} and provide assessment for suitable mitigation actions~\cite{presekal2025anomaly,sharadqh2023hybrid,bhardwaj2024attack}.
A third relevant purpose in this category deals with vulnerability analysis.
In fact, different works generate \gls{ag} from \gls{ids} to handle missed detections through the analysis of network vulnerability dependencies~\cite{noel2004correlating,holgado2017real} and reveal the network security situation~\cite{alrehaili2023attack,tayouri2025coral}.

Finally, only a limited number of studies address real-time detection, probably due to \gls{ag} generation suffering from scalability limitations, making it hard to adjust for real-time tasks.
Among them, ~\citeap{holgado2017real} design algorithmic approaches to generate \glspl{ag} efficiently and leverage them for prompt detection, while ~\citeap{ahmadian2016causal} design a three-phase alert correlation framework, which processes the generated alerts in real time.

\subsubsection*{\ref{item:rsq2} \textbf{IDS Type}}
Most approaches in this category ($69.5\%$) rely on \gls{nids}, as their operational model aligns closely with the structural characteristics of \glspl{ag}.
For example, many works leverage the IP addresses from \gls{nids} alerts to represent nodes of the \gls{ag} and edges to represent possible attack steps between them~\cite{zhang2019multi,nadeem_alert-driven_2022,wang2022end,ghasemigol2016comprehensive}.
Surprisingly, a large portion of the works in this category ($24\%$) do not rely on strict assumptions about the monitoring process -- i.e., \gls{nids} or \gls{hids} -- of the \gls{ids} used~\cite{tayouri2025coral,mohammadzad2023magd,jia2023artificial,zhang2023correlating}. 
This flexibility stems from the ability to extract key information -- such as affected nodes, services, and alert timestamps -- from virtually any \gls{ids}.
Finally, only a small number of studies employ \gls{hids} for \gls{ag} generation~\cite{xie2010using,bhardwaj2024attack,melo2022ism}.
These works posit that attack structures function as abstract knowledge, independently of network-specific details. 
Consequently, they can be modeled and managed considering alerts from specific hosts and independently of any particular network environment~\cite{xie2010using}.

Concerning the detection perspective, the majority of works consider \gls{sids}~\cite{wang2010alert,alhaj2023effective,haque2020integrating,bajtovs2020multi} as signatures align well with the formal structure of \glspl{ag}, facilitating their usage for \gls{ag} generation.
For example, different works incorporate probabilistic methods such as Markov Models~\cite{nadeem_alert-driven_2022,DBLP:journals/access/AminSNTK21} or Bayesian Networks~\cite{mouwen2022robust,yang2022multi}, and may also use semantic analysis via ontologies~\cite{saad2013extracting}.
In contrast, \gls{aids}~\cite{PresekalSRP23,mao2021mif,bhardwaj2024attack,presekal2025anomaly} are less frequently used, probably due to the semantic gap between the \gls{aids} output and the input requirements of an \gls{ag}.
For example, in \gls{aids}, an anomaly may be an abstract statistical deviation (e.g., unusual traffic) that hardly map to a specific \gls{ag} node.
Nonetheless, these methods emphasize the topological structure of \glspl{ag}, capturing asset and vulnerability dependencies. 
Here, alert correlation supports topology-aware clustering and downstream \gls{ml} tasks~\cite{mao2021mif,presekal2025anomaly}.
%
%

Lastly, detection-agnostic techniques are rarer and apply alert aggregation to correlate events and identify relationships indicative of a single attack~\cite{8245631,sahu2021structural,tayouri2025coral}, which then inform \gls{ag} construction.

\subsubsection*{\ref{item:rsq3} \textbf{\gls{ag} Type}}
The various models defined in the \gls{ag} taxonomy~\cite{kaynar_taxonomy_2016} are considered with comparable frequency for \gls{ag} generation using \gls{ids}. This observation is significant as it suggests that existing approaches address multiple perspectives of \gls{ag} modeling. Nevertheless, a slightly greater proportion of works focus on generating attack scenario-based~\cite{ahmadian2016causal,alserhani2010mars,alhaj2023effective,wang2022end,4406402,ning2003learning,10.1145/3325061.3325062} and state-based~\cite{tayouri2025coral,mao2021mif,xie2010using,nadeem2021alert,holgado2017real,ghasemigol2016comprehensive} \glspl{ag}.
These models are natural choices: the former emerge from alert correlation, while the latter facilitate mapping between \gls{ids} signatures and \gls{ag} states.
While promising, these approaches overlook the complexity of the correlation algorithms they rely upon, thus hindering their effectiveness in improving \gls{ag} scalability.

Other works generate host-based \glspl{ag}~\cite{li2019complex,zhang2019multi,liao2010building,mohammadzad2023magd,PresekalSRP23} as they focus on modeling the attacker’s behavior at the host level.
Differently, only a handful of works consider vulnerability-based \glspl{ag}~\cite{ghasemigol2016comprehensive,noel2004correlating} due to the difficulty of extracting vulnerability information from \gls{ids} alerts.
These approaches mainly address vulnerability analysis, foreseeing the vulnerability information (e.g., CVSS metrics of the different CVEs) as an important integration of \gls{ids} alerts.
This disparity suggests that while alert correlation is well-explored~\cite{alrehaili2023attack,nadeem_alert-driven_2022}, vulnerability modeling and prediction for scalable \gls{ag} generation remain underexplored.
Finally, recent efforts in attack path prediction~\cite{buabualuau2024forecasting,zhao2024graph} show promise and warrant further investigation, although inaccurate scalability assumptions may hinder real-world applicability.

\subsubsection*{\ref{item:rsq4} \textbf{Threats}}
A notable observation is that \gls{ag} generation is commonly used to assess cyber risks associated with multi-step attacks.
However, a significant portion of the literature relies on the DARPA2000 dataset~\cite{wang2022end,ning2003learning,alserhani2010mars,mao2021mif,wang2010alert,alhaj2023effective,holgado2017real}, which contains only \gls{ddos} attack instances.
This limits the applicability of the corresponding approaches to non-\gls{ddos} scenarios.
Moreover, DARPA2000 lacks network topology information, which hinders the construction of host-based \glspl{ag}.
As a result, most models in this category define either attack scenario-based or state-based \glspl{ag}, further highlighting the scarcity of suitable datasets.
To compensate, some studies rely on simulated or custom datasets~\cite{PresekalSRP23,li2019complex,tayouri2025coral,DBLP:journals/compsec/ZhangLCF08}, despite \gls{ag} generation typically requiring inputs such as reachability graphs and vulnerability inventories.
This reliance on limited or synthetic data impairs the fair validation and comparison of \gls{ag} generation methods, ultimately affecting their reproducibility and applicability in real-world environments.

\subsubsection*{\ref{item:rsq5} \textbf{Application}}
Our analysis reveals that only half of the approaches ($52\%$) generate \glspl{ag} without specifying a concrete application scenario~\cite{nadeem2021alert,DBLP:journals/compsec/ZhangLCF08,ahmadian2016causal,ning2003learning,mao2021mif,holgado2017real,noel2004correlating}.
These works are in line with \gls{ag} modeling, which is intended to be a general model by definition.
Therefore, here generality is a strength of these approaches.

Surprisingly, the other half of the works consider scenario-specific information to generate \glspl{ag}.
For example, some solutions focus on \gls{iot}~\cite{bhardwaj2024attack,alrehaili2023attack} and smart grids~\cite{yang2022multi,sen2021towards,sen2022using}, mainly due to the availability of DDoS-related datasets in these domains.
The reasoning behind these application-specific works is that customizing \glspl{ag} for a specific application may produce better results due to the consideration of contextual implications of real-world deployments.
This specificity may be relevant for vulnerabilities and attack vectors that can vary significantly across domains.
Therefore, investigating context-specific characteristics of \glspl{ids} and vulnerabilities in greater depth, to build scenario-specific \glspl{ag}, represents a promising research direction.

\subsubsection*{\ref{item:rsq6} \textbf{\gls{ml}}}
Most $AG|IDS$ approaches do not employ \gls{ml} for \gls{ag} generation.
Among those that do, two main \gls{ml} models are employed.
\gls{mc} are used to model probabilistic transitions between alerts~\cite{nadeem_alert-driven_2022,holgado2017real,fredj2015realistic}. \glspl{mc} are particularly well-suited for generating state-based \glspl{ag} since the states of a \gls{mc} can be easily mapped to the states of an \gls{ag}.
On the other hand, \glspl{nn} are particularly used for \gls{ag} generation.
Framing \gls{ag} generation as an attack path prediction problem has shown promise in improving scalability, where popular models include learning from intrusion alerts to predict sequences of compromised devices, with \glspl{cnn} being frequently used in this context~\cite{mao2021mif,sharadqh2023hybrid,PresekalSRP23}.
This approach is flexible and agnostic to the underlying \gls{ids} type.
Few works leverage different \gls{ml} models, like Artificial Immune System~\cite{melo2022ism} and Bayesian networks~\cite{ahmadian2016causal}.
Surprisingly, only a few studies explore \glspl{gnn}~\cite{zhao2024graph,presekal2025anomaly}, despite their suitability for processing and generating graph-structured data, possibly due to their relative novelty.

Overall, \gls{ml}-based approaches for \gls{ag} generation based on \gls{ids} remain underexplored and represent a promising direction for future research.

\begin{systematizationtakeaways}
\phantomsection
{{\bf \faSearch\ \textit{Systematization Takeaways:}\label{synopsis_ag_gen}}
    \begin{itemize}
        \item $AG|IDS$ generation is driven primarily by alert correlation to translate noisy \gls{ids} data into structured attack paths. However, AG scalability limits currently hinder real-time detection.
        \item \gls{nids} and \gls{sids} systems are heavily preferred ($\sim$70\%) because their specific outputs easily map to \gls{ag} nodes, unlike \gls{hids} and \gls{aids} which suffer from a semantic gap.
        \item $AG|IDS$ relies heavily on outdated or narrow datasets (like DARPA2000). This restricts evaluation mostly to DDoS scenarios and severely impairs real-world validation and reproducibility.
        \item The application of \gls{ml} for $AG|IDS$ remains largely underexplored. Specifically, \glspl{gnn} present a highly promising avenue given their natural fit for processing graph-structured data.
        \item While many models prioritize generality, developing context-aware AGs tailored to distinct environments -- e.g., IoT, smart grids -- may capture unique, scenario-specific vulnerabilities.
    \end{itemize}
}
\end{systematizationtakeaways}

\subsection{AG-integrated IDS ($\textbf{IDS}[\textbf{AG}]$)}\label{ssec:ag_integrated_ids_survey}
\begin{figure}[t]
    \centering
    \includegraphics[width=0.6\linewidth]{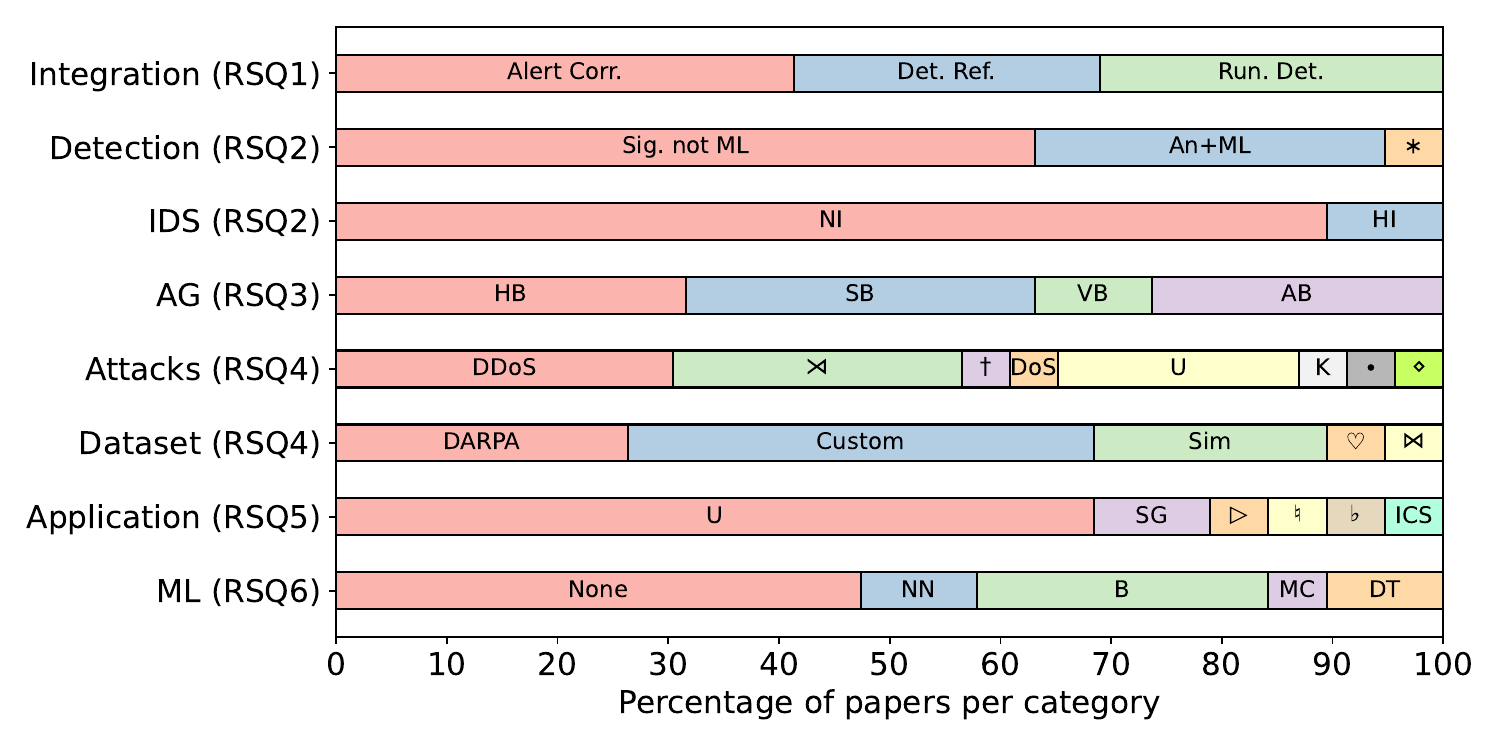}
    \caption{Statistical analysis of the \gls{ag}-integrated \gls{ids} literature. Refer to \Cref{fig:ag_generation_stats} for the legend. 
    }
    \label{fig:ag_integrated_stats}
\end{figure}

We classify as $IDS[AG]$ those approaches that leverage \gls{ag} as a model to support the detection process of \gls{ids}.
Based on our systematization, we identified \orangecount{} papers that fall into this category.
\Cref{fig:ag_integrated_stats} summarizes the key statistical findings, while \Cref{tab:systematization} depicts the detailed systematization for each paper.

\subsubsection*{\ref{item:rsq1} \textbf{Integration}}
The foundational idea behind approaches in this category is to harness \gls{ag} as a means to modify how the \gls{ids} brings about its decision process, aiming at reducing false alarms and improving precision by leveraging \gls{ag}'s ability to comprehensively represent network exposure.
This integration enhances the detection capabilities, reliability, and transparency of the \gls{ids}.
Our analysis reveals that the \glspl{ag} can be incorporated into the \gls{ids} for: 
\begin{itemize}
    \item correlating \gls{ids} alerts with the \gls{ag} to ensure the existence of an attack path and define an \gls{ids} that flags an alarm only if an attack path is exploited (we refer to this sub-class with \emph{alert correlation})~\cite{yang2022multi,bhardwaj2024attack,4511565};
    \item modifying the detection rules or learning process that build the \gls{ids}, for instance, generating a new detection rule from a specific attack path of the \gls{ag} (we refer to this subclass as \emph{detection refinement})~\cite{sen2022holistic,10.1145/3154273.3154311,8613951}; or
    \item leveraging the \gls{ag} in real-time as a detection module -- modeling multi-step intrusions -- where \gls{ag} states are verified based on network/hosts information and alerts are raised only whenever an attack path is complete (we refer to this sub-class as \emph{runtime detection})~\cite{ahmadian2016causal,wu2023network,franccois2016bayesian,kazeminajafabadi2024optimal}.
\end{itemize}
Alert correlation approaches are the most popular solutions since establishing a mapping between \gls{ids} alerts and/or signatures with the \gls{ag} is relatively feasible in simple scenarios, and the mapping can be made efficiently.
On the other hand, detection refinement is underanalyzed, as it requires constructing a methodology to design detection techniques from attack path instances, which may be cumbersome over a complex network.
However, detection refinement may also represent the most effective strategy if used properly; thus, we stress the need to conduct a more thorough investigation.
Finally, runtime detection systems are quite popular as they enable leveraging the \gls{ag} as a detection framework, but suffer scalability issues as they require large graph traversal and state updates.

%

\subsubsection*{\ref{item:rsq2} \textbf{\gls{ids} Type}}
\gls{nids} represents the most commonly adopted detection model, appearing in more than $90\%$ of the reviewed works \cite{bendiab2020advanced,sen2022holistic,eom2020framework,wu2023network,yang2022multi,holgado2017real}.
These approaches differ depending on the \gls{ag} used and on how they combine its knowledge into the \gls{ids}.
For example, \gls{nids}-based runtime detection approaches leverage the \gls{ag} as a standalone detection tool and filter out unwanted \gls{ids} network states to raise alerts only when an attack path is complete \cite{wu2023network,kazeminajafabadi2024optimal,bendiab2020advanced,eom2020framework}.
Meanwhile, \gls{nids}-based detection refinement systems modify the \gls{ids} inner principles -- e.g., detection rules --, matching one or more \gls{ag}'s attack paths to a detection principle \cite{8613951,RAMAKI2015206}.
The overly reliance on \gls{nids} can be attributed to several factors:
\begin{inlinelist}
    \item while \glspl{ag} can correlate with both network-level (e.g., attacker lateral movements) and host-level (e.g., privilege escalation or account-to-account movements) steps, the most common \gls{ag} models are primarily built upon network connectivity and vulnerability chains. This network-centricity makes correlation with \gls{nids} more direct and straightforward for initial integration efforts compared to mapping the intricate, host-specific events tracked by \gls{hids}.
    \item \gls{ag}-integrated \gls{ids} was initially proposed to enhance detection at the network level, which naturally encouraged a focus on \gls{nids}. And, 
    \item\label{item:snort_usage} most \gls{ag}-integrated \gls{ids} approaches rely on available off-the-shelf \glspl{ids} like Snort\footnote{\url{https://www.snort.org}} and Suricata\footnote{\url{https://suricata.io}} due to their ease of use and flexibility. 
\end{inlinelist}

As a consequence of \ref{item:snort_usage}, the majority of \gls{ag}-integrated \gls{ids} approaches rely on \glspl{sids} to easily map alerts and their signatures to the corresponding nodes or states of an \gls{ag}.
Several of these approaches correlate \gls{sids} alerts to the \gls{ag} nodes to verify the state of attack progress and construct a runtime detection scheme, differing by the alert correlation procedure and detection scheme used \cite{zali2012realconf,ahmadian2016causal,yang2022multi}.
Meanwhile, \glspl{aids} are mostly used in systems that leverage \gls{ml} to aggregate alerts and refine the detection or define it at runtime \cite{sen2022holistic,tabia2010bayesian}, since \gls{ml} is well-suited for both correlating alerts -- e.g., via clustering or classification -- and constructing anomaly detection schemes over the classes of aggregated alerts.

\subsubsection*{\ref{item:rsq3} \textbf{\gls{ag} Type}}
Several different \gls{ag} types~\cite{kaynar_taxonomy_2016} are considered for \gls{ag}-integrated \gls{ids}.
State-based \cite{zali2012realconf,ahmadian2016causal,bendiab2020advanced,sen2022holistic} and host-based \glspl{ag} \cite{roschke2011new,franccois2016bayesian,yang2022multi,bhardwaj2024attack} represent popular solutions since they naturally enable mapping with \gls{ids} signatures and/or alerts.
Similarly, attack scenario-based \glspl{ag} are largely used in approaches relying on alert correlation as they assume that similar alerts map to similar attack scenarios \cite{RAMAKI2015206,ahmadian2016causal,8613951}.
These approaches mostly rely on runtime detection and refinement to quickly identify the activation of practical attack scenarios and block them.
Lastly, given the difficulty of mapping vulnerability information to \gls{ids} alerts and vice-versa, only a handful of works falling in this category build on vulnerability-based \glspl{ag} \cite{eom2020framework,ahmadinejad2011hybrid}.
A critical aspect of $IDS[AG]$ models is their usage of \glspl{ag} at runtime---for instance, to detect intrusions by verifying the completion of attack paths or to map alerts over the states of the \gls{ag}.
This implies that current approaches assume the feasibility of attack path computation at runtime.
However, most are validated on simplified network topologies~\cite{bhardwaj2024attack,chung2013nice,ahmadinejad2011hybrid}.
Among \orangecount{} analyzed papers, only three explicitly support \glspl{ag} with over $100$ nodes~\cite{sen2022holistic,RAMAKI2015206,zali2012realconf}.
Given that real-world detection must operate over high-volume data streams, scalability remains a key challenge for deploying $IDS[AG]$ systems.
%
%

\subsubsection*{\ref{item:rsq4} \textbf{Threats}}
%
%
Many approaches are agnostic to specific threats~\cite{kazeminajafabadi2024optimal,wu2023network,yang2022multi}, focusing instead on aggregating alerts and evaluating detection performance at a network-wide level.
This lack of specificity limits the applicability of these approaches since, in practice, \glspl{ids} are typically tailored to specific threat models. 
Among the works that do specify attack types, the most common are \gls{ddos}~\cite{holgado2017real,RAMAKI2015206}, DoS~\cite{bhardwaj2024attack}, and multi-step attacks in general~\cite{8613951,10.1145/3154273.3154311,roschke2011new}.
These multi-step attacks benefit significantly from \gls{ag}-based modeling, as early detection enables timely mitigation.

The datasets used for evaluation support these findings. Most studies rely on:
\begin{inlinelist}
    \item well-known \gls{ddos} and DoS datasets such as DARPA2000~\cite{RAMAKI2015206,zali2012realconf,ahmadinejad2011hybrid,4511565,holgado2017real},
    \item custom-built datasets~\cite{kazeminajafabadi2024optimal,bendiab2020advanced,sen2022holistic,bendiab2020advanced,yang2022multi,roschke2011new}, or 
    \item simulation-based environments~\cite{bhardwaj2024attack,10.1145/3154273.3154311,8613951,franccois2016bayesian}.
\end{inlinelist}
This reliance on narrow or synthetic datasets raises concerns about the generalizability of current frameworks, as their effectiveness may not extend beyond the specific conditions under which they were evaluated.

\subsubsection*{\ref{item:rsq5} \textbf{Application}}
Most of the works falling in this category do not specify the application environment they consider~\cite{kazeminajafabadi2024optimal,wu2023network,mao2021mif,holgado2017real}. 
The root cause for such a trend is to be found on the \gls{ag} notion being fundamentally designed as a general threat model, where generality is a key theoretical strength. 
However, in the context of operational integration with \glspl{ids}, this generality can become a practical constraint: \gls{ag} that lack domain-specific, real-time context often suffer from scalability issues and yield low actionability due to the sheer volume of theoretically possible, but practically improbable, attack paths.
This intuition seems to be validated by the existence of a few works focusing on highly specific, resource-constrained environments like Smart Grids and \gls{iot}, where resource prioritization and highly contextualized response are crucial~\cite{yang2022multi,sen2022holistic,bhardwaj2024attack}.
\subsubsection*{\ref{item:rsq6} \textbf{ML}}
While the majority of analyzed approaches in this category do not rely on any \gls{ml} solution \cite{roschke2011new,zali2012realconf,RAMAKI2015206,eom2020framework,yang2022multi,4511565}, a few recent works consider incorporating \gls{ml} to address scalability concerns of \gls{ag}-\gls{ids} integration.
Most rely on relatively simple models such as Bayesian networks~\cite{franccois2016bayesian,sen2022holistic,tabia2010bayesian,kazeminajafabadi2024optimal} or Markov Chains~\cite{holgado2017real} to define a runtime detection scheme where the network or chain states are used to track attack progress and raise corresponding alerts.
Bayesian networks are particularly suitable for such task due to their compatibility with graph structures like \glspl{ag} with the additional benefit of enabling state transition probabilities and this optimizable detection.
Few recent approaches employ \gls{nn}~\cite{bendiab2020advanced,bhardwaj2024attack}, using \glspl{ag} to enrich training data~\cite{bhardwaj2024attack} or define detection-time topologies~\cite{bendiab2020advanced}.
Despite their potential, \gls{nn}-based solutions remain underexplored, likely due to the difficulty of integrating symbolic knowledge into sub-symbolic learning processes~\cite{CiattoCsur2024,SarkerZEH21,bardhi2024ai}.
As a result, current \gls{ml}-based solutions are relatively simplistic and future research should explore more advanced techniques, such as \glspl{gnn}~\cite{ZhouAiopen20,AgiolloEurosp2023,AgiolloUai2022}, federated learning~\cite{agrawal2022federated,campos2022evaluating,AgiolloEnea2024,AgiolloAnonymous2024}, and neuro-symbolic methods~\cite{CiattoCsur2024,SarkerZEH21}, which offer promising avenues for integrating structured knowledge into learning-based detection systems.

\begin{systematizationtakeaways}
\phantomsection
{{\bf \faSearch\ \textit{Systematization Takeaways:}\label{synopsis_ag_integrated_ids}}
    \begin{itemize}
        \item $IDS[AG]$ integration primarily aims to reduce false alarms via alert correlation and runtime detection. However, severe scalability limits drastically hinder real-world, high-volume deployment.
        \item There is a massive over-reliance on \gls{nids} and \gls{sids} ($>$90\%). This is driven by the convenience of mapping deterministic outputs from off-the-shelf tools (like Snort or Suricata) to network-centric AG structures.
        \item The reliance on outdated and/or synthetic datasets raises concerns regarding the real-world generalizability of current frameworks.
        \item While runtime detection is popular, detection refinement remains a highly effective but notably underanalyzed strategy.
        \item Current \gls{ml} usage relies on simple probabilistic models. Future research must pivot toward cutting-edge solutions (e.g., \glspl{gnn}) and neuro-symbolic AI to merge \glspl{ag} with advanced learning-based detection.
    \end{itemize}
}
\end{systematizationtakeaways}


%

\subsection{AG-based IDS Refinement ($\textbf{IDS} \rightarrow \textbf{AG}$)}\label{ssec:ag_based_ids_refine_survey}
\begin{figure}[t]
    \centering
    \includegraphics[width=0.6\linewidth]{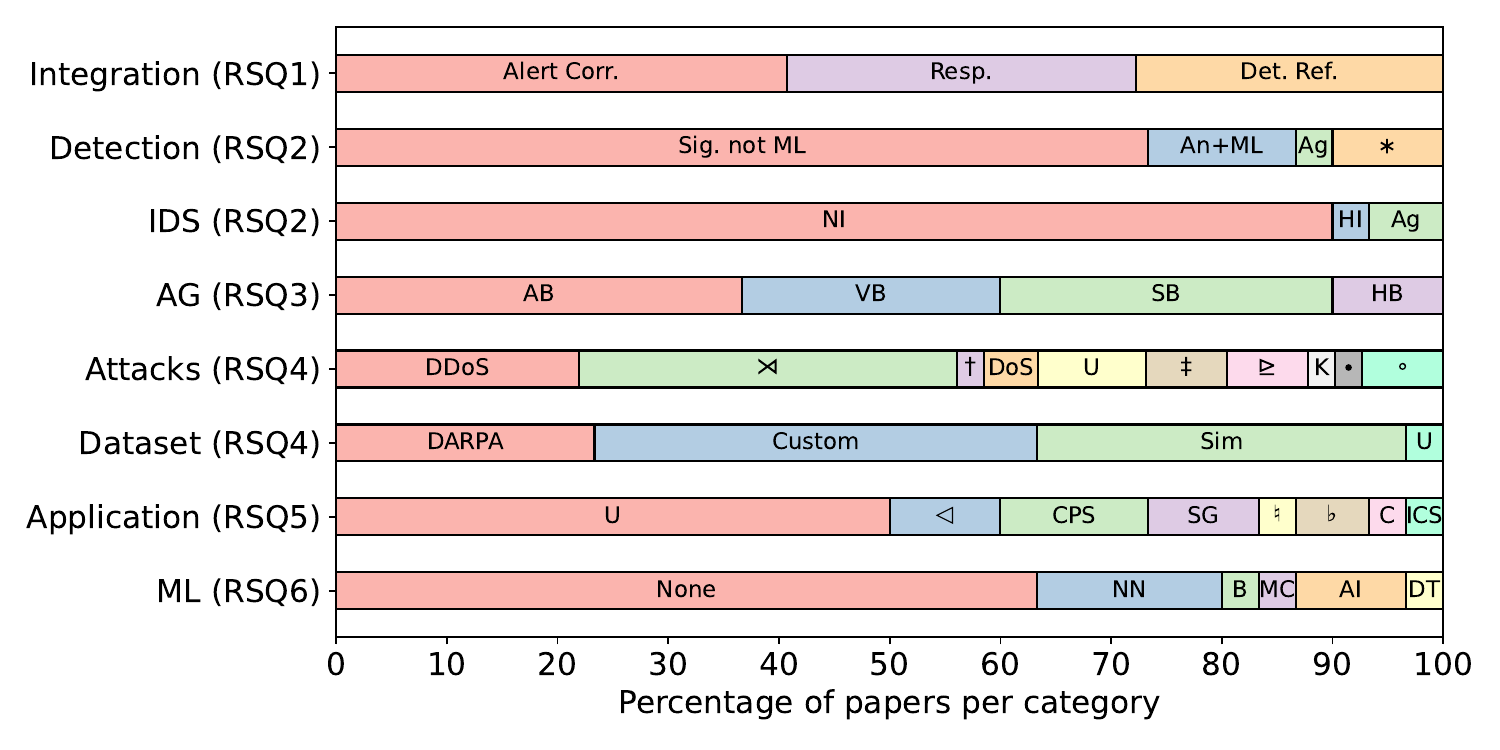}
    \caption{Statistical analysis of the \gls{ag}-based \gls{ids} Refinement literature. Refer to \Cref{fig:ag_generation_stats} for the legend.
    }
    \label{fig:postag_stats}
\end{figure}

We classify as $IDS \rightarrow AG$ those approaches that leverage \gls{ag} as a model to support or enhance the outcomes of \gls{ids}.
We identified \bluecount{} papers that fall into this category.
\Cref{fig:postag_stats} summarizes the key statistical findings, while \Cref{tab:systematization} depicts the detailed systematization for each paper.

\subsubsection*{\ref{item:rsq1} \textbf{Integration}}
%
Integrating post-hoc checks of \gls{ids} alerts onto an \gls{ag} is vital for transforming raw alerts into actionable intelligence. This process facilitates detection refinement by validating and contextualizing suspicious events against known attack paths, enables the creation of adequate responses by revealing the attacker's stage and objective in the attack chain, and ultimately enhances runtime detection by confirming that a series of isolated alerts represents a true, ongoing multi-stage campaign.
The systematization in this category reveals that one of the most straightforward applications of \glspl{ag} in this context is the interpretation of the large volume of alerts produced by \glspl{ids}.
In particular, alert correlation plays a crucial role in accurately identifying multi-step attacks~\cite{sen2022using,melo2019cloud,hu2020attack,roschke2011new, alserhani2010mars,ghasemigol2016comprehensive,DBLP:journals/access/AminSNTK21,sen2021towards,rose2022ideres,jia2023artificial,presekal2025anomaly}, which often require the integration of information from different sources. 
Moreover, alert correlation enables network administrators to reconstruct attack paths or understand the broader attack behavior~\cite{hu2020attack,roschke2011new,ahmadinejad2011hybrid,wang2010alert,PresekalSRP23,wang2006using}, thereby facilitating the deployment of appropriate mitigation and response strategies.
Importantly, alerts themselves serve as a valuable source of information that can be leveraged for the detection refinement, thus improving detection accuracy and reducing false positives~\cite{melo2019cloud,shittu2015intrusion,alserhani2010mars,wang2006using,zali2012realarticle}.
Another prominent use of \gls{ag} in the context of \gls{ids} refinement is the enhancement of intrusion detection reliability.
Some approaches introduce techniques that leverage \glspl{ag} based on causal relations, not only improving detection accuracy but also enabling the prediction of future attacker behaviors~\cite{JADIDI2022103741}. 
%
%

\subsubsection*{\ref{item:rsq2} \textbf{\gls{ids} Type}}
%
The majority of papers in this category -- i.e., ~73\% -- adopt \gls{sids}, mainly leveraging Snort~\cite{rose2022ideres,sen2022using,hu2020attack}, RealSecure~\cite{zali2012realarticle,ahmadinejad2011hybrid} and Amilyzer~\cite{shawly2023detection}.
The large adoption of \gls{sids} can be attributed to their ease of implementation, fast detection, and transparent actionable alerts. 
Therefore, it is relatively simple to interface \gls{ids} alerts to \gls{ag} post-hoc to determine if the alert fits into any pre-calculated attack path or attack scenario.
These post-hoc checks on alerts enable actionable intelligence, including detection refinement, adequate response for the raised alerts, and runtime detection, as elaborated earlier.  
On the other hand, only a limited number of approaches adopt \glspl{aids}~\cite{bendiab2020advanced,JADIDI2022103741,PresekalSRP23,presekal2025anomaly}.
This mainly happens due to the challenges of integrating anomaly-based detection with the formal model of \glspl{ag}.
Meanwhile, all of them leverage \gls{ml} for the alert correlation, detection refinement, or even runtime detection.

Lastly, from the monitoring perspective, the vast majority of contributions -- i.e., 90\% -- of this category, target \glspl{nids}.
The reasons behind such an imbalance are similar to the ones expressed for the $IDS[AG]$ category in \Cref{ssec:ag_integrated_ids_survey} (\ref{item:rsq2}). 

\subsubsection*{\ref{item:rsq3} \textbf{\gls{ag} Type}}
The majority of the works in this category adopt attack scenario-based \glspl{ag}~\cite{zali2012realarticle,melo2019cloud,JADIDI2022103741,tabia2010bayesian,alserhani2010mars,sen2021towards,sen2022using,melo2022ism,jia2023artificial,melo2019novel}. 
Such a finding is mainly attributed to the alert correlation methodologies that these works propose, for which the \gls{ag} maps raised \gls{ids} alerts to known attack scenarios.
Additionally, such a mapping enables fast and efficient detection refinement~\cite{PresekalSRP23,DBLP:journals/access/AminSNTK21} and response~\cite{shameli2016dynamic,shameli2015orcef,rose2022ideres}.
For a similar reasoning, state-based \glspl{ag} are highly adopted~\cite{fayyad2013attack,shittu2015intrusion,hu2020attack,dang2023research,bendiab2020advanced,wang2010alert,DBLP:journals/access/AminSNTK21,presekal2025anomaly,shameli2016dynamic}.
Analyzing \gls{ids} alerts with vulnerability-based \glspl{ag} has also been quite well-adopted. 
The preparation of an adequate response and detection refinement benefits from an extensive vulnerability-based \gls{ag}.
Lastly, host-based \glspl{ag}~\cite{roschke2011new,rose2022ideres,PresekalSRP23} are less adopted than other types of \gls{ag} for this category primarily because of their construction, collection, and maintenance complexity.

%
%
%
%
%

\subsubsection*{\ref{item:rsq4} \textbf{Threats}}

\glspl{ag} are particularly effective for modeling \gls{ddos}~\cite{PresekalSRP23,zali2012realarticle,ahmadinejad2011hybrid,wang2010alert,wang2006using,melo2019cloud} and multi-step attacks~\cite{JADIDI2022103741,hu2020attack,melo2019cloud,sen2022using,alserhani2010mars}, as they capture the sequential and distributed nature of such threats.
Consequently, many studies focus on these attack types.
Furthermore, a significant number of studies are agnostic to the specific attack type~\cite{dang2023research,eom2020framework,shittu2015intrusion,fayyad2013attack}, building systems on top of general-purpose \glspl{ids} without addressing potential compatibility issues. 
Commonly used datasets include DARPA2000~\cite{zali2012realarticle,ahmadinejad2011hybrid,wang2010alert,wang2006using,melo2019cloud,alserhani2010mars,melo2019novel}, which provides communication traces but lacks detailed network topology and vulnerability data.
This raises concerns about the representativeness of such a dataset for evaluating \gls{ag}-based \gls{ids} refinement.
Many also rely on simulations or custom datasets, further limiting the reliability and reproducibility of their findings.
Notably, the literature lacks a systematic analysis of how \gls{ids} outputs can be semantically prepared for \gls{ag}-based processing, which is a critical gap for practical deployment.

\subsubsection*{\ref{item:rsq5} \textbf{Application}}
A considerable amount of $IDS \rightarrow AG$ approaches are applied in critical infrastructure domains, such as \gls{cps}~\cite{shawly2023detection,ghasemigol2016comprehensive,rose2022ideres,presekal2025anomaly}, industrial control systems~\cite{JADIDI2022103741}, smart grids~\cite{PresekalSRP23,sen2021towards,sen2022using}, and \gls{ami}~\cite{bendiab2020advanced}, where the volume of alerts is particularly high.
Other applications include cloud computing~\cite{melo2019cloud,chung2013nice,shameli2016dynamic} and \gls{sdn}~\cite{eom2020framework,melo2022ism}.
However, many studies do not specify the application domain~\cite{dang2023research,hu2020attack,shittu2015intrusion,fayyad2013attack,zali2012realarticle,wang2006using,tabia2010bayesian,roschke2011new,ahmadinejad2011hybrid,alserhani2010mars,wang2010alert,DBLP:journals/access/AminSNTK21,melo2019novel,shameli2015orcef}, limiting the contextual relevance of their findings.

\subsubsection*{\ref{item:rsq6} \textbf{ML}}
%

Only a small subset of the \bluecount{} reviewed papers incorporate \gls{ml} into $IDS \rightarrow AG$ approaches. 
Among these, \glspl{nn} and \gls{ais} are the most common.
Bendiab et al.~\cite{bendiab2020advanced} classify network traffic using ResNet50, Mobilenet, and SOINN, while other works apply \gls{lstm} on multi-device time series for attack detection~\cite{JADIDI2022103741}, propose a hybrid GCN-\gls{lstm} model for anomaly detection~\cite{PresekalSRP23,presekal2025anomaly}, or use \gls{ais}-based techniques like negative and clonal selection~\cite{melo2022ism,melo2019novel,melo2019cloud}.
These approaches primarily use \gls{ml} to enhance \gls{ids} performance, rather than to support the integration with \glspl{ag}.
This highlights a promising research direction: developing advanced \gls{ml}-driven methods for \gls{ag}-based \gls{ids} refinement. 
In particular, the integration of \glspl{gnn} and neuro-symbolic learning~\cite{CiattoCsur2024,SarkerZEH21} could significantly improve the expressiveness and scalability of future solutions.

\begin{systematizationtakeaways}
\phantomsection
{{\bf \faSearch\ \textit{Systematization Takeaways:}\label{synopsis_ag_based_ids_refine}}
    \begin{itemize}
        \item The primary value of $IDS \rightarrow AG$ integration is transforming massive volumes of raw IDS alerts into actionable intelligence, allowing to reconstruct attack paths and the formulation of targeted responses.
        \item Signature-based ($\sim$73\%) and Network-based (90\%) systems heavily dominate. Their alerts are much easier to map onto \glspl{ag} post-hoc compared to \gls{aids}, which struggle with formal model integration.
        \item Because $IDS \rightarrow AG$ approaches excel at managing high alert volumes and tracking multi-step campaigns, they see strong real-world application in domains like cyber-physical systems and smart grids.
        \item $IDS \rightarrow AG$ approaches rely on limited datasets and ignore the systematic issue of how to semantically prepare raw \gls{ids} outputs for seamless \gls{ag} processing.
    \end{itemize}
}
\end{systematizationtakeaways}



\subsection{Hybrid Approaches}\label{ssec:hybrid_survey}
A key finding of our systematization is that the majority of the studied approaches fall into one of the categories proposed in the taxonomy.
Nonetheless, a smaller portion of existing works propose \emph{hybrid} approaches that span more than one of the defined categories.
%
%
These approaches aim to either design effective detection or develop comprehensive frameworks by combining the strengths of different integration paradigms (e.g., to provide real-time detection and response~\cite{bendiab2020advanced}).

In the following, we refer to $X + Y$ approaches as those that simultaneously adopt principles from both category $X$ and $Y$---e.g., $AG|IDS + IDS[AG]$ considers both \gls{ag} generation ($AG|IDS$) and the \gls{ag}-based \gls{ids} ($IDS[AG]$). 
We analyze each hybrid category, focusing on the specific advantages and limitations of hybridization.
Let us note that specific answers to research subquestions (\ref{item:rsq1}-\ref{item:rsq6}) are included in the previous analyses (\Cref{ssec:ag_generation_survey,ssec:ag_integrated_ids_survey,ssec:ag_based_ids_refine_survey}) and are thus are ignored here to avoid redundancy.

\subsubsection{AG$|$IDS + IDS[AG]}
The collection of hybrid approaches combining $AG|IDS$ with $IDS[AG]$ primarily focuses on the dynamic evaluation of known vulnerabilities and system entry points~\cite{ahmadian2016causal,4511565}.
In these approaches, \glspl{ag} are constructed using  $AG|IDS$ principles -- i.e., from \gls{ids} alerts -- and are updated whenever an alarm is triggered and linked to a previously unknown vulnerability~\cite{holgado2017real,yang2022multi}.
This enables dynamic verification and extension of the system's vulnerability set, reducing the efforts of costly expert-driven analysis.
Furthermore, $AG|IDS + IDS[AG]$ approaches also update the detection scheme in response to changes in the \gls{ag}~\cite{bhardwaj2024attack}.
This ensures that newly discovered vulnerabilities -- identified through \gls{ids} alerts -- are incorporated into future iterations of the intrusion detection process, preventing them from being overlooked or forgotten.

It is worth noticing that approaches in this category rely mostly on static \gls{sids}, suggesting opportunities for \gls{ml}-enhanced and adaptive correlation models, such as \gls{gnn} or Reinforcement Learning. 
Existing temporal models like \gls{mc} and Bayesian Networks are outdated, opening the door to modern probabilistic and sequence-learning methods for multi-step attack detection.
Another gap is that most evaluations depend on old or simulated datasets due to the difficulty of retrieving datasets for both detection and \gls{ag} generation.
This highlights the need for realistic benchmarking, especially related to resource-constrained environments.

\subsubsection{AG$|$IDS + IDS$\rightarrow$AG}
Similar to the previous category, the collection of hybrid approaches combining $AG|IDS$ with $IDS\rightarrow AG$ primarily focuses on the dynamic evaluation of known vulnerabilities and system entry points~\cite{ghasemigol2016comprehensive,melo2022ism,PresekalSRP23}.
Reasonably, they are more than the ones in $AG|IDS + IDS[AG]$ (almost twice), as \glspl{ag} are more wired towards improving the detection accuracy by design, rather than optimizing the detection mechanism itself.
Most of AG$|$IDS + IDS$\rightarrow$AG approaches focus on updating the alert correlation schemes used to verify the alerts generated by the \gls{ids}~\cite{presekal2025anomaly,jia2023artificial,sen2022using,DBLP:journals/access/AminSNTK21}.
As such, these systems function as static or dynamic verification tools, ensuring that the alert validation process remains aligned with known vulnerabilities in the networks.

A potential drawback of these approaches is that they are agnostic to the \gls{ids} optimization process, possibly causing susceptibility to \gls{ids} obsolescence~\cite{zhang2024caravan,apruzzese2024adversarial}. 
In particular, the \gls{ids} used to update the \gls{ag} may fail to detect exploits that arise from the combination of multiple, previously unknown vulnerabilities. 
To mitigate the obsolescence risks, it is essential to update not only the alert verification and correlation mechanisms but also the \gls{ids} itself whenever a new vulnerability is discovered.

\subsubsection{IDS[AG] + IDS$\rightarrow$AG}
The collection of hybrid approaches combining $IDS[AG]$ and $IDS \rightarrow AG$ aims to provide comprehensive detection and \gls{ag}-informed responses to detected anomalies.
Many of these works verify whether the detection outcomes produced by the \gls{ids}, when informed by \gls{ag}, satisfy the constraints defined by the \gls{ag} itself~\cite{roschke2011new,eom2020framework,LANIGAN2025200606}.
When the \gls{ag} model is introduced during the \gls{ids} optimization phase, it is essential to ensure that information provided by \gls{ag} is not lost or overwritten during the learning process.
This challenge becomes even more pronounced when the underlying detection mechanism is \gls{ml}-based, as safely incorporating structured knowledge into learning models is non-trivial~\cite{tabia2010bayesian,bendiab2020advanced,LANIGAN2025200606}.
An interesting aspect is that this issue resembles the challenges of neuro-symbolic systems, where injected knowledge may or may not be retained effectively by the model~\cite{GanAaai2021,MuralidharBigdata2018,AgiolloAamas20O23,AgiolloFedcsis2O23,AgiolloWoa2O22}.
This indicates that hybrid \gls{ag}-\gls{ids} pipelines would benefit from methodologies explicitly designed to preserve, reason over, and update symbolic knowledge as part of the detection process.
Moreover, the lack of guarantees regarding the persistence of injected knowledge suggests that current hybrid architectures may only partially exploit the \gls{ag} model, limiting their potential for coherent multi-step attack reasoning.
Addressing this gap calls for research into robust knowledge integration strategies, where the \gls{ag} functions not merely as a post-processing layer or auxiliary input, but as a structural prior guiding the entire decision pipeline.\\

In summary, the current body of research demonstrates several hybrid approaches, each aiming either at dynamic \gls{ag} refinement or at improving \gls{ids} reliability. 
These works address hybridization from only one of these perspectives. Therefore, they do not exploit the full potential of coordinated \gls{ag}-\gls{ids} interaction. 
None of the surveyed approaches implements a continuous \gls{ag}-\gls{ids} integration loop combining all three integration classes proposed in this work (a gap clearly reflected by the empty central cluster in \Cref{fig:class_categorization}).
Nevertheless, existing approaches form a valuable \emph{foundation} for the development of a complete \gls{ag}–\gls{ids} lifecycle. 
Within such a lifecycle, hybrid strategies may enable dynamic optimization and continuous updating of both the \glspl{ag} and the \glspl{ids} used to detect the exploitation of known vulnerabilities. 
To mitigate obsolescence when new vulnerabilities arise, it is crucial to update not only alert verification and correlation mechanisms but also the \gls{ids} itself.

\begin{systematizationtakeaways}
\phantomsection
{{\bf \faSearch\ \textit{Systematization Takeaways:}\label{synopsis_hybrid}}
    \begin{itemize}
        \item While various two-way hybridizations exist, current research lacks a unified framework that merges all three integration classes ($AG|IDS$, $IDS[AG]$, $IDS\rightarrow AG$) into a continuous, self-updating lifecycle.
        \item Hybrid models that focus solely on updating alert correlation schemes remain vulnerable to new exploits. To truly mitigate obsolescence, the \gls{ids} must be dynamically optimized alongside the \gls{ag}.
        \item In \gls{ml}-based \glspl{ids}, retaining injected \gls{ag} knowledge is non-trivial. Future pipelines must treat \gls{ag} not merely as auxiliary inputs, but as structural priors to guarantee coherent multi-step attack reasoning.
        \item The field relies heavily on outdated static/temporal models and simulated datasets. Advancing hybrid integration requires modern techniques and realistic benchmarking.
    \end{itemize}
}
\end{systematizationtakeaways}

\vspace{0.5cm}


\tablecaption{
    Categorization of the systematized papers across the \textbf{RQs} and \textbf{RSQs}. \textbf{Legend}:
    AC = Alert correlation,
    VA = Vulnerability Analysis,
    DR = Detection Refinement,
    RD = Runtime Detection,
    Res = Response,
    Ag = Agnostic, 
    S+ML = Signature and ML, 
    S-ML = Signature not ML, 
    A+ML = Anomaly and ML, 
    A-ML = Anomaly not ML,
    NIDS = Network Intrusion Detection System,
    HIDS = Host Intrusion Detection System,
    ASB = Attack Scenario-based ag,
    HB = Host-based ag,
    SB = State-based ag,
    VB = Vulnerability-based ag,
    ND = Undefined,
    Mul = Multi-Step Attack,
    RCE = Remote Code Execution,
    U2R = User to Root,
    R2L = Remote to Local,
    KL = Key Loggers,
    OS = OS Scan,
    Pr = Probing,
    PS = Port Scanning,
    SSH = SSH Brute Force,
    Sim = Simulation, 
    $\divideontimes$ = DARPA2000,
    $\bowtie$ = ISCXIDS2012,
    $\bigstar$ = NLS-KDD,
    $\heartsuit$ = CSE-CIC-IDS-2018,
    $\spadesuit$ = Defcon CTF'17,
    $\triangle$ = CTU-13,
    $\triangledown$ = CPTC-2018,
    $\blacktriangle$ = CICIoT2023,
    $\gtrdot$ = 4SICS-2015,
    $\unlhd$ = CCDC-2018,
    SDN = Software Defined Networking,
    SG = Smart Grids,
    CC = Cloud Computing,
    IoT = Internet of Things,
    CPS = Cyber-Physical Systems,
    SC = Smart Cities,
    AMI = Advanced Metering Infrastructures,
    ENS = Enterprise Network System,
    SHS = Smart Home System,
    ICS = Industrial Control Systems,
    SOCs = Security Operation Centers,
    EC = Edge Computing,
    \o = None,
    BN = Bayesian Network,
    MC = Markov Chain,
    AIS = Artificial Immune System,
    CNN = Convolutional Neural Network,
    DT = Decision Tree, 
    SVM = Support Vector Machine,
    LSTM = Long Short Term Memory Neural Network,
    GNN = Graph Neural Network,
    MLP = Multi-Layer Perceptron,
    PA = Probabilistic Automaton.
}
\label{tab:systematization}

{\tiny

\tablefirsthead{%
  \toprule
  \multirow{2}{*}{\textbf{Method}} &
  \multirow{2}{*}{\textbf{Year}} &
  \multicolumn{3}{c|}{\ref{item:rq1}} &
  \multirow{2}{*}{\textbf{Aim (\ref{item:rsq1})}} &
  \multicolumn{2}{c|}{\ref{item:rsq2}} &
  \multirow{2}{*}{\textbf{AG (\ref{item:rsq3})}} &
  \multicolumn{2}{c|}{\ref{item:rsq4}} &
  \multirow{2}{*}{\textbf{App (\ref{item:rsq5})}} &
  \multirow{2}{*}{\textbf{ML (\ref{item:rsq6})}}\\
  \cmidrule(lr){3-5}\cmidrule(lr){7-8}\cmidrule(lr){10-11}
  & & \textbf{AG$|$IDS} & \textbf{IDS[AG]} & \textbf{IDS$\rightarrow$AG} &
  & \textbf{Detection} & \textbf{IDS} &
  & \textbf{Attacks} & \textbf{Dataset} &
  & \\
  \midrule\midrule
}

\tablehead{%
  \toprule
  \multicolumn{13}{l}{\small\itshape (continued)}\\
  \midrule
  \multirow{2}{*}{\textbf{Method}} &
  \multirow{2}{*}{\textbf{Year}} &
  \multicolumn{3}{c|}{\ref{item:rq1}} &
  \multirow{2}{*}{\textbf{Aim (\ref{item:rsq1})}} &
  \multicolumn{2}{c|}{\ref{item:rsq2}} &
  \multirow{2}{*}{\textbf{AG (\ref{item:rsq3})}} &
  \multicolumn{2}{c|}{\ref{item:rsq4}} &
  \multirow{2}{*}{\textbf{App (\ref{item:rsq5})}} &
  \multirow{2}{*}{\textbf{ML (\ref{item:rsq6})}}\\
  \cmidrule(lr){3-5}\cmidrule(lr){7-8}\cmidrule(lr){10-11}
  & & \textbf{AG$|$IDS} & \textbf{IDS[AG]} & \textbf{IDS$\rightarrow$AG} &
  & \textbf{Detection} & \textbf{IDS} &
  & \textbf{Attacks} & \textbf{Dataset} &
  & \\
  \midrule
}

\tabletail{%
  \midrule
  \multicolumn{13}{r}{\small\itshape (continues on next page)}\\
}
\tablelasttail{\bottomrule}

\begin{supertabular}{ wc{3.5cm} | wc{0.5cm} | wc{0.75cm} wc{0.75cm} wc{0.75cm} | wc{1.1cm} | wc{0.8cm} wc{0.5cm} | wc{1.1cm} | wc{0.8cm} wc{0.7cm} | wc{1.1cm} | wc{1.1cm}}
    \citeap{ning2003learning} & \citeyear{ning2003learning} & \cmark & \xmark & \xmark & AC & S-ML & NIDS & ASB & DDoS & $\divideontimes$ & ND & \o \\
    \rowcolor{gray!15} 
    \citeap{noel2004correlating} & \citeyear{noel2004correlating} & \cmark & \xmark & \xmark & AC, VA & S-ML & NIDS & VB & Mul & Sim & ND & \o \\
    \citeap{wang2006using} & \citeyear{wang2006using} & \xmark & \xmark & \cmark & DR, AC & S-ML & NIDS & VB & DDoS & $\divideontimes$ & ND & \o \\
    \rowcolor{gray!15} 
    \citeap{4406402} & \citeyear{4406402} & \cmark & \xmark & \xmark & AC & S-ML & NIDS & ASB & DDoS & $\divideontimes$ & ND & \o \\
    \citeap{4511565} & \citeyear{4511565} & \cmark & \cmark & \xmark & AC & S-ML & NIDS & HB & DDoS & $\divideontimes$ & ND & \o \\
    \rowcolor{gray!15} 
    \citeap{ZHANG2008188} & \citeyear{ZHANG2008188} & \cmark & \xmark & \xmark & AC & S-ML & NIDS & SB & Mul & Cus & ND & \o \\
    \citeap{tabia2010bayesian} & \citeyear{tabia2010bayesian} & \xmark & \cmark & \cmark & DR, AC & S+ML & NIDS & ASB & RCE & Cus & ND & BN \\
    \rowcolor{gray!15} 
    \citeap{alserhani2010mars} & \citeyear{alserhani2010mars} & \cmark & \xmark & \cmark & AC, DR, Res & S-ML & NIDS & ASB & Mul & $\divideontimes$ & ND & \o \\
    \citeap{wang2010alert} & \citeyear{wang2010alert} & \cmark & \xmark & \cmark & AC, Res & S-ML & NIDS & SB & DDoS & $\divideontimes$ & ND & \o \\
    \rowcolor{gray!15} 
    \citeap{xie2010using} & \citeyear{xie2010using} & \cmark & \xmark & \xmark & AC & S-ML & HIDS & SB & ND & Sim & ND & \o \\
    \citeap{liao2010building} & \citeyear{liao2010building} & \cmark & \xmark & \xmark & AC & S-ML & NIDS & HB & Mul & Cus & ND & \o \\
    \rowcolor{gray!15} 
    \citeap{roschke2011new} & \citeyear{roschke2011new} & \xmark & \cmark & \cmark & DR, AC & S-ML & NIDS & HB & Mul & Cus & ND & \o \\
    \citeap{ahmadinejad2011hybrid} & \citeyear{ahmadinejad2011hybrid} & \xmark & \cmark & \cmark & DR, AC & S-ML & NIDS & VB & DDoS & $\divideontimes$ & ND & \o \\
    \rowcolor{gray!15} 
    \citeap{zali2012realarticle} & \citeyear{zali2012realarticle} & \xmark & \xmark & \cmark & DR, AC & S-ML & NIDS & ASB & DDoS & $\divideontimes$ & ND & \o \\
    \citeap{zali2012realconf} & \citeyear{zali2012realconf} & \xmark & \cmark & \xmark & AC, RD & S-ML & NIDS & SB & DDoS & $\divideontimes$ & ND & \o \\
    \rowcolor{gray!15} 
    \citeap{chung2013nice} & \citeyear{chung2013nice} & \xmark & \xmark & \cmark & AC, RD, Res & S-ML & NIDS & VB & DDoS & Cus & CC & \o \\
    \citeap{fayyad2013attack} & \citeyear{fayyad2013attack} & \xmark & \xmark & \cmark & DR, AC & S-ML & NIDS & SB & ND & Cus & ND & \o \\
    \rowcolor{gray!15} 
    \citeap{saad2013extracting} & \citeyear{saad2013extracting} & \cmark & \xmark & \xmark & AC & S-ML & NIDS & ASB & DDoS & $\divideontimes$ & ND & \o \\
    \citeap{shittu2015intrusion} & \citeyear{shittu2015intrusion} & \xmark & \xmark & \cmark & DR, AC & S-ML & NIDS & SB & ND & Cus & ND & \o \\
    \rowcolor{gray!15} 
    \citeap{RAMAKI2015206} & \citeyear{RAMAKI2015206} & \xmark & \cmark & \xmark & DR, AC & S-ML & NIDS & ASB & DDoS & $\divideontimes$ & ND & \o \\
    \citeap{shameli2015orcef} & \citeyear{shameli2015orcef} & \xmark & \xmark & \cmark & RD, Res & S-ML & NIDS & VB & \makecell{DoS,\\U2R,\\R2L} & Sim & ND & \o \\
    \rowcolor{gray!15} 
    \citeap{fredj2015realistic} & \citeyear{fredj2015realistic} & \cmark & \xmark & \xmark & AC & S-ML & NIDS & ASB & Mul & $\spadesuit$ & ND & MC \\
    \citeap{shameli2016dynamic} & \citeyear{shameli2016dynamic} & \xmark & \xmark & \cmark & RD, Res & S-ML & NIDS & SB & \makecell{DoS,\\U2R,\\R2L} & Sim & CC & \o \\
    \rowcolor{gray!15} 
    \citeap{ghasemigol2016comprehensive} & \citeyear{ghasemigol2016comprehensive} & \cmark & \xmark & \cmark & AC, Res & S-ML & NIDS & VB & Mul & Sim & CPS & \o \\
    \citeap{franccois2016bayesian} & \citeyear{franccois2016bayesian} & \xmark & \cmark & \xmark & RD & A+ML & HIDS & HB & Mul & Sim & ND & BN \\
    \rowcolor{gray!15} 
    \citeap{ahmadian2016causal} & \citeyear{ahmadian2016causal} & \cmark & \cmark & \xmark & AC, RD & S-ML & NIDS & ASB & Mul & $\divideontimes$, $\bowtie$ & ND & BN \\
    \citeap{wang2016alert} & \citeyear{wang2016alert} & \cmark & \xmark & \xmark & AC & S-ML & NIDS & ASB & DDoS & $\divideontimes$ & ND & \o \\
    \rowcolor{gray!15} 
    \citeap{hoque2016alert} & \citeyear{hoque2016alert} & \cmark & \xmark & \xmark & AC & S-ML & NIDS & SB & DDoS & $\divideontimes$ & ND & \o \\
    \citeap{8245631} & \citeyear{8245631} & \cmark & \xmark & \xmark & AC & Ag & Ag & ASB & Mul & Cus & SG & \o \\
    \rowcolor{gray!15} 
    \citeap{holgado2017real} & \citeyear{holgado2017real} & \cmark & \cmark & \xmark & VA, RD & S-ML & NIDS & SB & DDoS & $\divideontimes$ & ND & MC \\
    \citeap{8613951} & \citeyear{8613951} & \xmark & \cmark & \xmark & DR, AC & S-ML & NIDS & ASB & Mul & Sim & ICS & \o \\
    \rowcolor{gray!15} 
    \citeap{10.1145/3154273.3154311} & \citeyear{10.1145/3154273.3154311} & \xmark & \cmark & \xmark & DR, RD & S-ML & NIDS & HB & Mul & Sim & ND & \o \\
    \citeap{melo2019cloud} & \citeyear{melo2019cloud} & \xmark & \xmark & \cmark & DR, AC & S+ML & NIDS & ASB & Mul & $\divideontimes$ & CC & AIS \\
    \rowcolor{gray!15} 
    \citeap{li2019complex} & \citeyear{li2019complex} & \cmark & \xmark & \xmark & AC, VA & S-ML & NIDS & HB & Mul & Cus & EC & \o \\
    \citeap{zhang2019multi} & \citeyear{zhang2019multi} & \cmark & \xmark & \xmark & AC & S-ML & NIDS & HB & DDoS & $\divideontimes$ & SG & \o \\
    \rowcolor{gray!15} 
    \citeap{10.1145/3325061.3325062} & \citeyear{10.1145/3325061.3325062} & \cmark & \xmark & \xmark & AC & S-ML & Ag & ASB & Mul & Sim & ND & \o \\
    \citeap{melo2019novel} & \citeyear{melo2019novel} & \xmark & \xmark & \cmark & AC & S+ML & NIDS & ASB & DDoS & $\divideontimes$ & ND & AIS \\
    \rowcolor{gray!15} 
    \citeap{bendiab2020advanced} & \citeyear{bendiab2020advanced} & \xmark & \cmark & \cmark & RD, Res & A+ML & NIDS & SB & \makecell{DDoS,\\KL,\\OS} & Cus & AMI & CNN \\
    \citeap{haque2020integrating} & \citeyear{haque2020integrating} & \cmark & \xmark & \xmark & AC & S-ML & NIDS & SB & Mul & Cus & CPS & \o \\
    \rowcolor{gray!15} 
    \citeap{eom2020framework} & \citeyear{eom2020framework} & \xmark & \cmark & \cmark & RD, Res & S-ML & NIDS & VB & ND & Cus & SDN & \o \\
    \citeap{hu2020attack} & \citeyear{hu2020attack} & \xmark & \xmark & \cmark & AC, VA & S-ML & NIDS & SB & Mul & Cus & ND & \o \\
    \rowcolor{gray!15} 
    \citeap{pivarnikova2020early} & \citeyear{pivarnikova2020early} & \cmark & \xmark & \xmark & AC & S-ML & NIDS & ASB & Mul & $\heartsuit$ & ND & \o \\
    \citeap{bajtovs2020multi} & \citeyear{bajtovs2020multi} & \cmark & \xmark & \xmark & AC & S-ML & NIDS & ASB & Mul & $\gtrdot$ & ND & \o \\
    \rowcolor{gray!15} 
    \citeap{nadeem2021sage} & \citeyear{nadeem2021sage} & \cmark & \xmark & \xmark & AC & S-ML & Ag & SB & Mul & $\divideontimes$ & CPS & MC \\
    \citeap{sahu2021structural} & \citeyear{sahu2021structural} & \cmark & \xmark & \xmark & AC & Ag & Ag & HB & Mul & ND & ICS & \o \\
    \rowcolor{gray!15} 
    \citeap{sen2021towards} & \citeyear{sen2021towards} & \cmark & \xmark & \cmark & AC, Res & S-ML & NIDS & ASB & Mul & Sim & SG & \o \\
    \citeap{DBLP:journals/access/AminSNTK21} & \citeyear{DBLP:journals/access/AminSNTK21} & \cmark & \xmark & \cmark & AC,DR,Res & S-ML & Ag & SB & Mul & Sim & ND & MC \\
    \rowcolor{gray!15} 
    \citeap{mao2021mif} & \citeyear{mao2021mif} & \cmark & \xmark & \xmark & AC & A+ML & NIDS & SB & \makecell{DoS,\\U2R,\\R2L} & $\divideontimes$ & ND & CNN \\
    \citeap{rose2022ideres} & \citeyear{rose2022ideres} & \xmark & \xmark & \cmark & Res & S-ML & NIDS & HB & Mul & Sim & CPS & CNN \\
    \rowcolor{gray!15} 
    \citeap{JADIDI2022103741} & \citeyear{JADIDI2022103741} & \xmark & \xmark & \cmark & DR & A+ML & NIDS & ASB & Mul & Cus & ICS & LSTM \\
    \citeap{nadeem_alert-driven_2022} & \citeyear{nadeem_alert-driven_2022} & \cmark & \xmark & \xmark & AC & S-ML & NIDS & SB & Mul & $\triangledown$, $\unlhd$ & ND & PA \\
    \rowcolor{gray!15} 
    \citeap{wang2022end} & \citeyear{wang2022end} & \cmark & \xmark & \xmark & AC & S-ML & NIDS & ASB & Mul & $\divideontimes$ & ICS & \o \\
    \citeap{sen2022holistic} & \citeyear{sen2022holistic} & \xmark & \cmark & \xmark & DR, AC & A+ML & NIDS & SB & Mul & Cus & SG & BN \\
    \rowcolor{gray!15} 
    \citeap{sen2022using} & \citeyear{sen2022using} & \cmark & \xmark & \cmark & AC, Res & S-ML & NIDS & ASB & Mul & Sim & SG & \o \\
    \citeap{yang2022multi} & \citeyear{yang2022multi} & \cmark & \cmark & \xmark & AC & S-ML & NIDS & HB & ND & Cus & SG & \o \\
    \rowcolor{gray!15} 
    \citeap{mouwen2022robust} & \citeyear{mouwen2022robust} & \cmark & \xmark & \xmark & AC & S-ML & NIDS & SB & Mul & ND & SHS & PA \\
    \citeap{melo2022ism} & \citeyear{melo2022ism} & \cmark & \xmark & \cmark & AC, DR & S+ML & HIDS & ASB & \makecell{DoS,\\U2R,\\R2L,\\Pr} & ND & SDN & AIS \\
    \rowcolor{gray!15} 
    \citeap{shawly2023detection} & \citeyear{shawly2023detection} & \xmark & \xmark & \cmark & DR, Res & S-ML & NIDS & VB & Mul & Sim & CPS & \o \\
    \citeap{sharadqh2023hybrid} & \citeyear{sharadqh2023hybrid} & \cmark & \xmark & \xmark & AC & A+ML & NIDS & SB & Mul & Cus & ICS & CNN \\
    \rowcolor{gray!15} 
    \citeap{alrehaili2023attack} & \citeyear{alrehaili2023attack} & \cmark & \xmark & \xmark & AC, VA & S-ML & NIDS & ASB & Mul & ND & IoT & \o \\
    \citeap{jia2023artificial} & \citeyear{jia2023artificial} & \cmark & \xmark & \cmark & AC, DR & S-ML & Ag & ASB & Mul & Sim & SC & \o \\
    \rowcolor{gray!15} 
    \citeap{PresekalSRP23} & \citeyear{PresekalSRP23} & \cmark & \xmark & \cmark & AC, Res & A+ML & NIDS & HB & DDoS & Cus & SG & \makecell{GNN,\\LSTM,\\CNN} \\
    \citeap{dang2023research} & \citeyear{dang2023research} & \xmark & \xmark & \cmark & Res & Ag & NIDS & SB & ND & Cus & ND & \o \\
    \rowcolor{gray!15} 
    \citeap{wu2023network} & \citeyear{wu2023network} & \xmark & \cmark & \xmark & RD & A+ML & NIDS & SB & ND & $\heartsuit$ & ND & DT \\
    \citeap{mohammadzad2023magd} & \citeyear{mohammadzad2023magd} & \cmark & \xmark & \xmark & VA & S-ML & Ag & HB & Mul & ND & ND & \o \\
    \rowcolor{gray!15} 
    \citeap{zhang2023correlating} & \citeyear{zhang2023correlating} & \cmark & \xmark & \xmark & AC & S-ML & Ag & HB & Mul & Sim & ND & \o \\
    \citeap{alhaj2023effective} & \citeyear{alhaj2023effective} & \cmark & \xmark & \xmark & AC & S-ML & NIDS & ASB & DDoS & $\divideontimes$, $\bowtie$ & ND & \o \\
    \rowcolor{gray!15} 
    \citeap{zhao2024graph} & \citeyear{zhao2024graph} & \cmark & \xmark & \xmark & AC & S-ML & Ag & ASB & DDoS & $\divideontimes$ & CPS & GNN \\
    \citeap{bhardwaj2024attack} & \citeyear{bhardwaj2024attack} & \cmark & \cmark & \xmark & AC & A+ML & HIDS & HB & \makecell{DoS,\\DDoS,\\PS} & Sim & IoT & MLP \\
    \rowcolor{gray!15} 
    \citeap{buabualuau2024forecasting} & \citeyear{buabualuau2024forecasting} & \cmark & \xmark & \xmark & AC & S-ML & Ag & SB & Mul & $\triangledown$ & SOCs & MC \\
    \citeap{kazeminajafabadi2024optimal} & \citeyear{kazeminajafabadi2024optimal} & \xmark & \cmark & \xmark & RD & A+ML & NIDS & SB & ND & Cus & ND & BN \\
    \rowcolor{gray!15} 
    \citeap{tayouri2025coral} & \citeyear{tayouri2025coral} & \cmark & \xmark & \xmark & VA & Ag & Ag & SB & Mul & Cus & CC & \o \\
    \citeap{presekal2025anomaly} & \citeyear{presekal2025anomaly} & \cmark & \xmark & \cmark & AC, Res & A+ML & NIDS & SB & Mul & Sim & CPS & \makecell{GNN,\\LSTM} \\
    \rowcolor{gray!15} 
    \citeap{LANIGAN2025200606} & \citeyear{LANIGAN2025200606} & \xmark & \cmark & \cmark & AC, DR, Res & S-ML & NIDS & ASB & Mul & Cus & ND & DT \\
\end{supertabular}

} 


\section{Closing AG-\gls{ids} Integration Loop}\label{sec:lifecycle}
Our systematization reveals that existing \gls{ag}-\gls{ids} integration approaches primarily seek to mitigate the inherent limitations of standalone \glspl{ag} and \glspl{ids}.
These methods predominantly operate within static detection environments, where the integration is performed as a one-time configuration. 
Such a paradigm overlooks dynamic context changes and the necessity for continuous refinement of detection and response mechanisms.
Conversely, while the identified handful of hybrid approaches represent a valid attempt to address these issues, they only partially alleviate the constraints of static integration. 
This is because they focus on either the \gls{ag} or \gls{ids} side in isolation, lacking a cohesive, bidirectional perspective.
%
%
%
Consequently, they fail to facilitate the continuous adaptation of the integrated \gls{ag}-\gls{ids} mechanism itself, typically introducing hybridization to only a single component.
Real-world environments require security mechanisms capable of adapting to emerging threats, evolving vulnerabilities, and fluid system configurations.
This dynamism requires a shift toward leveraging a multitude of up-to-date \glspl{ids} and \glspl{ag} to enable robust detection and response.
Building on our systematization and to meet the above requirement, we propose a continuously integrated \gls{ag}-\gls{ids} lifecycle that transcends semi-static, one-time, or unidirectional integration paradigms.
This lifecycle orchestrates diverse \gls{ag}-\gls{ids} methodologies across multiple stages, enabling the iterative refinement of both \glspl{ag} and \glspl{ids} in tandem.
Rather than isolating specific integration strategies -- such as $IDS[AG]$, $AG|IDS$, $IDS \rightarrow AG$ or hybrid models -- our vision unifies these models within a cohesive, feedback-driven framework engineered for sustained adaptability and operational resilience.

\begin{figure}[t]
    \centering
    \includegraphics[width=0.75\linewidth]{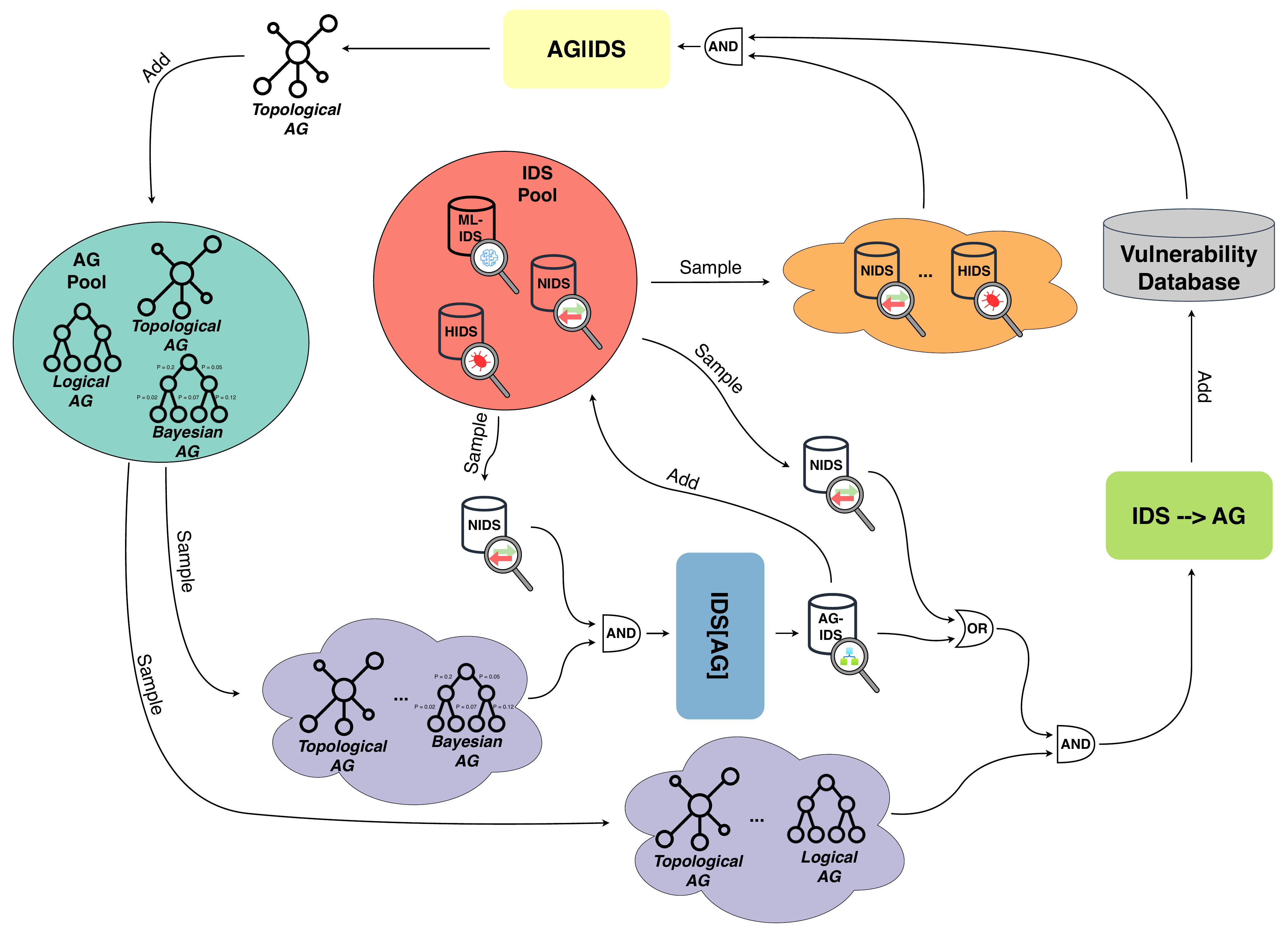}
    \caption{\gls{ag}-\gls{ids} lifecycle envisioned in this SoK. Legend: IDS $-$ Intrusion Detection System; NIDS $\-$ Network Intrusion Detection System; HIDS $\-$ Host Intrusion Detection System; AG $\-$ Attack Graph; $AG|IDS -$ IDS-based AG Generation; $IDS[AG] -$ AG-integrated IDS; $IDS \rightarrow AG$ AG-based IDS Refinement.}
    \label{fig:lifecycle}
\end{figure}

The proposed lifecycle serves as a proof-of-concept of our vision and is structured as shown in \Cref{fig:lifecycle}.
The core objective is to enable the continuous update of the \gls{ids} model and detection process (see \emph{Add} function to the IDS pool in \Cref{fig:lifecycle}), as well as \gls{ag} generation and analysis (see the \emph{Add} function to the AG pool in \Cref{fig:lifecycle}).
By establishing a complete feedback loop (represented by the information flow, logical ports, and actions in \Cref{fig:lifecycle}), the lifecycle aims to fulfill the requirements for continuous protection against novel threats and vulnerabilities.
Specifically, the lifecycle introduces feedback loops that:
\begin{inlinelist}
    \item define new detection schemes based on updated \glspl{ag} to identify emerging or modified attacks (represented by the output of the IDS[AG] module),
    \item analyze \gls{ids} outputs to uncover novel vulnerability paths (represented by the output of the $IDS \rightarrow AG$ module), and
    \item define or update AGs after uncovering new vulnerability paths (represented by the output of AG|IDS module).
\end{inlinelist}
The envisioned lifecycle performs integration iteratively, selecting \glspl{ag} and \glspl{ids} from a shared pool -- continuously adapted based on real-time alerts and vulnerabilities -- and updating a vulnerability database that records system knowledge in a context-aware manner.

At each iteration, the system selects one or more optimized \glspl{ids} (the output of the sample action) to guide the generation of custom \glspl{ag}, which is then added to the \gls{ag} pool. 
This process incorporates both known vulnerabilities from the Vulnerability Database and \gls{ids} predictions outputted from the sampled orange IDS cloud, thereby embedding $AG|IDS$ techniques into the lifecycle. 
Conversely, the system refines \gls{ids} models by integrating them with one or more \glspl{ag} outputted from the sampled violet AG clouds and returns the enhanced detectors to the \gls{ids} pool, effectively embedding IDS[AG] techniques into the lifecycle.
Similarly, the system can utilize any \gls{ag} from the pool, including those derived from \gls{ids} alerts, to validate resulting \glspl{ids} using updated \glspl{ag} via $IDS \rightarrow AG$ strategies.
To illustrate these dynamics, we provide a running example that maps the lifecycle to the logical ports, pools, and actions in \Cref{fig:lifecycle}.
Consider a production network where a previously unknown Remote Code Execution (RCE) vulnerability exists in a common logging library. 
Initially, the Vulnerability Database contains no record of this flaw, and the IDS Pool consists of standard NIDS or HIDS.
The cycle begins when an attacker exploits this zero-day flaw to establish a reverse shell. 
While the specific exploit signature is unknown, a sampled NIDS from the IDS Pool identifies anomalous outbound traffic. 
This alert is fed through the IDS[AG] module. 
By combining this real-time alert with current AG models (via an AND logical port), the system generates a new AG from the AG|IDS that includes paths updated from the vulnerability database via $IDS \rightarrow AG$ for the compromised web server as a verified stepping stone. 
This new graph is then committed to the AG Pool via the Add action.
With the threat landscape updated, the lifecycle moves to refine the detection layer. 
The system performs a Sample action on the AG Pools to retrieve the newly updated AG. 
Simultaneously, a more specialized NIDS is sampled from the IDS Pool.
These components are integrated through an AND port into the $IDS[AG]$ module. 
This results in a context-aware AG-IDS sensor specifically tuned to monitor the lateral movement paths identified in the updated attack graph.
As the attacker attempts to move from the web server to the database, this refined AG-IDS sensor detects the specific lateral movement payload. 
This detection event, potentially combined with other sensor outputs via an OR port, is fed into the $IDS \rightarrow AG$ engine. 
This process closes the loop in two ways: first, the refined detection model is returned to the IDS Pool via the Add function to ensure permanent protection; second, the specific exploit characteristics and verified attack paths are used to \emph{Add} a new entry to the Vulnerability Database and generate updated AGs. 
Through this iterative reinforcement, the lifecycle transforms a single anomalous alert into a hardened, context-aware defense posture.

In the following section, we implement an experimental proof-of-concept to show the potential of the proposed lifecycle.

\section{AG-IDS Lifecycle Proof-of-Concept}\label{sec:case_study}

To evaluate the effectiveness of the proposed \gls{ag}-\gls{ids} integration lifecycle, we define a proof-of-concept based on the CIC-IDS2017 dataset~\cite{sharafaldin2018toward}.
We selected the CIC-IDS2017 dataset because it provides publicly available, labeled data with diverse attack types. 
Its structural metadata is essential for deriving attack paths and aligning network events with AG. 
Its wide adoption in the IDS community ensures our evaluation remains comparable to existing NIDS baselines.
It includes both benign traffic and a wide range of cyberattacks in the form of packet capture files (PCAPs) and corresponding flow-based CSV files generated using the CICFlowMeter tool\footnote{\url{https://github.com/ahlashkari/CICFlowMeter}}. 
The dataset includes the generation of naturalistic background traffic through the B-Profile system~\cite{gharib2016evaluation}, which models human-like behavior in network usage across common protocols.
Collected data comprises simulations of $25$ users interacting with the network during real enterprise activities.
The dataset spans five days and comprehensive setting configurations, including heterogeneous hosts (i.e., routers, firewalls, switches, Windows, Ubuntu, and macOS operating systems), attack diversity, based on real-world threat intelligence, and an extensive feature set, with over 80 flow-based features extracted from traffic.

Our analysis focuses on the three core components of \gls{ag}-\gls{ids} integration, examined across successive iterations of the simulated lifecycle.
We here emphasize that this proof-of-concept serves as a demonstrative implementation of the proposed lifecycle, clarifying its key stages and functionalities, while a formal prototype and in-depth evaluation remain open future research directions (see \Cref{sec:opportunities}).
Finally, we make the source code of our lifecycle proof-of-concept publicly available\footnote{\url{https://github.com/Ale96Pa/AG4IDS}}.
%

\subsection{Experimental Setup}\label{sec:appendix_experimental_settings}
For the \gls{ids}, we consider a multi-class detection module which aims to predict if a given packet is benign or malicious, while predicting the attack category as well.
Due to limitations of existing rule sets in the Emerging Threats\footnote{\url{https://community.emergingthreats.net/}} (which will later be used for AG generation and IDS signatures), we limit out evaluation to FTP Patator and DoS attack classes of the CIC-IDS 2017 dataset.
Since we focus on the above-mentioned attack classes, we consider the data samples of CIC-IDS coming from the Monday, Tuesday, and Wednesday split and remove the packet traces that are neither benign nor belong to one of the attack classes that we considered.
Throughout our experiments, we consider a \gls{dt} classifier and -- if not mentioned otherwise -- we set its maximum depth to 20, the minimum number of samples required to split an internal node to two, the minimum number of samples required to be at a leaf node to one, and the optimization criterion to the Gini impurity.
We train the \gls{dt} classifier on 60\% of the dataset and test its performance on the remaining 40\% using a random split -- except for the experiments of \Cref{fig:orange_box_train_percs_exp,fig:blue_box_train_percs_exp} -- and select the 20 most relevant features according to the ANOVA F-value\footnote{\url{https://scikit-learn.org/stable/modules/generated/sklearn.feature_selection.SelectKBest.html}}. 
Concerning the \gls{ag} used, if not mentioned otherwise, we use the graph constructed from the Emerging Threats rule set and set the probability of adding a random attack path $p=0$, which correspond to leveraging a very reliable \gls{ag}.
To ensure a fair evaluation, we measure the performance of the trained \gls{dt}-based \gls{ids} using accuracy, F1-score, and \gls{fpr}.
%
%
For each experiment, we compute the performance gain by measuring the difference in accuracy/F1/\gls{fpr} between the standard \gls{ids} and its refined counterpart, and plot this delta over the barplot of each setting.
An improvement in \gls{ids} performance is given by a positive delta value for accuracy and F1-score and a negative delta for \gls{fpr}.

\subsection{IDS-based AG Generation ($AG|IDS$)}\label{ssec:exp_green_box}
\begin{wrapfigure}{r}{0.6\textwidth}
    \centering
    \includegraphics[width=0.98\linewidth]{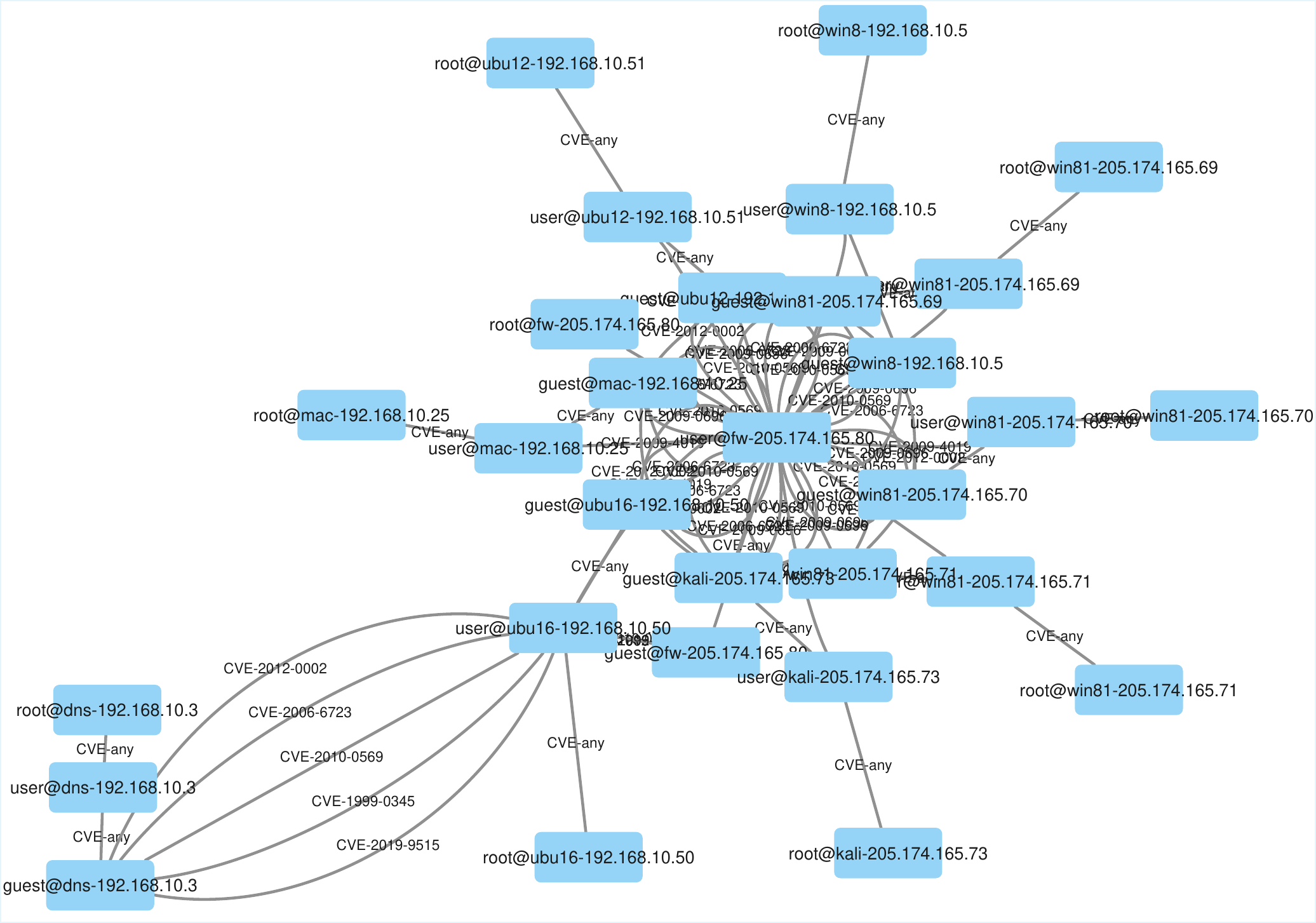}
    \caption{Example of \gls{ag} for the CIC-IDS dataset.}
    \label{fig:ag_example}
\end{wrapfigure}
CIC-IDS2017 includes information on running services and their configurations, enabling the modeling of a classical \gls{ag} generation process through the collection of known vulnerabilities from \glspl{cve}. 
We use this traditional \gls{ag} generation as a baseline and construct an \gls{ids}-based \gls{ag} alternative by incorporating signatures and rules from the Emerging Threats (ET) rule set. 
These rules explicitly identify known vulnerabilities, exploits, and attack behaviors observed in network traffic.
The \gls{ids}-based \gls{ag} maps the vulnerabilities described in the \gls{ids} rules to corresponding edges in the \gls{ag}, thereby enriching the graph with real-world observations from the \gls{ids}. 
\Cref{fig:ag_example} visualizes an example of \gls{ag} constructed for the CIC-IDS dataset. 
To ensure consistency, we consider only the alerts triggered by ET rules that correspond to attacks modeled in the CIC-IDS2017 dataset.
Finally, we simulate successive lifecycle iterations by progressively limiting the number of ET alerts used to construct the \gls{ag}. This approach mimics early-stage scenarios where the historical log of \gls{ids} alerts is sparse. 
We then compare the baseline \gls{ag} with its \gls{ids}-enhanced counterpart.

%

\begin{wraptable}{r}{0.34\textwidth}
    \centering
    \caption{Standard AG generation against $AG|IDS$.}
    \label{tab:ag_generation}
    \resizebox{0.98\linewidth}{!}{%
        \begin{tabular}{lcc}
            \hline
             & \textbf{AG generation} & \textbf{\begin{tabular}[c]{@{}c@{}}IDS-based\\ AG generation\end{tabular}} \\ \hline
            Number of paths       & 11842 & 360 \\ \hline
            Computation time & 32.23 s & 0.37 s \\ \hline
            Avgerage cyber risk  & 0.72 & 0.77 \\ \hline
        \end{tabular}
    }
\end{wraptable}
\Cref{tab:ag_generation} presents the \gls{ag} size -- i.e., the number of attack paths --, the \gls{ag} computation time, and the average cyber risk for the \glspl{ag} generated with and without the proposed lifecycle.
The results reveal two key findings:
\begin{inlinelist}
    \item classical \gls{ag} generation produces a vast number of attack paths, as it considers all possible vulnerabilities, quickly becoming computationally impractical. In contrast, $AG|IDS$ focuses only on vulnerabilities that have resulted in actual attacks, making it significantly more efficient.
    \item Only a small fraction of the thousands of vulnerabilities present in the network are relevant in the given context (just 2\% of those analyzed). Many vulnerabilities are either non-exploitable or affect local services that do not impact the broader network. Consequently, $AG|IDS$ enables a more targeted \gls{ag} generation process by prioritizing relevant vulnerabilities over exhaustive enumeration.
\end{inlinelist}

\begin{experimentaltakeaways}
\phantomsection
{{\bf \faFlask\ \textit{Experimental Finding \#1:}\label{finding:1}} The iterative use of $AG|IDS$ in our lifecycle enables faster generation of tailored \glspl{ag}, improving scalability.}
\end{experimentaltakeaways}

%
\begin{wrapfigure}{r}{0.5\textwidth}
    \centering
    \includegraphics[width=0.98\linewidth]{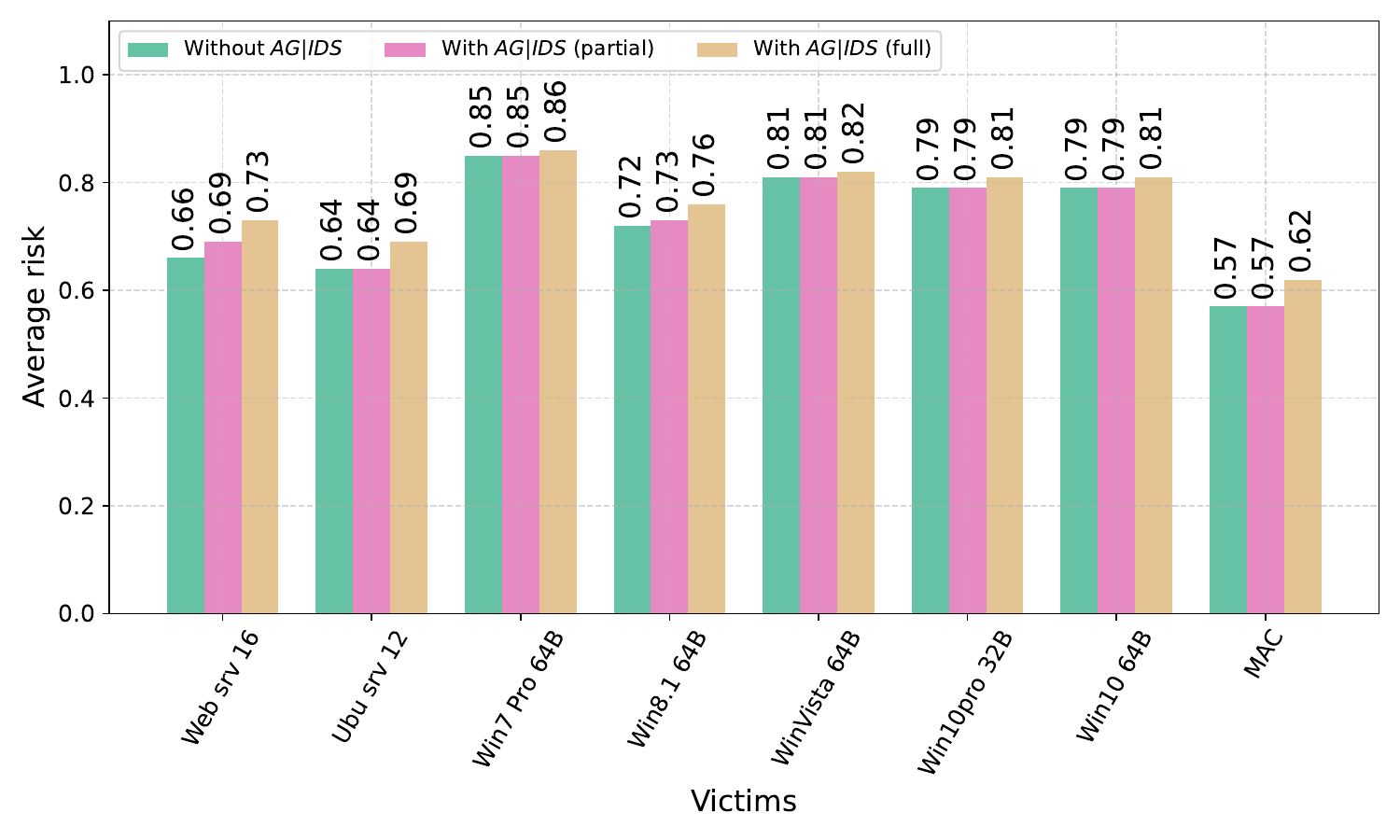}
    \caption{Comparison between standard AG generation and $AG|IDS$ for risk computation on the CIC-IDS victims.}
    \label{fig:risk_comparison}
\end{wrapfigure}
To complement the analysis presented in \Cref{tab:ag_generation}, \Cref{fig:risk_comparison} presents the average cyber risk associated with different victim hosts in the dataset.
We compare the classical \gls{ag} generation with the proposed lifecycle at two stages: an early iteration ($AG|IDS$ partial) and the complete execution ($AG|IDS$ full).
A consistent trend is the increase in average risk when \gls{ids} data is integrated into the \gls{ag}, especially when the full alert set is used.
Although the numerical differences are modest, the increase is systematic, indicating improved situational awareness.
%
%
This enables more precise, continuous, and context-aware risk assessments, supporting better-informed security decisions and demonstrating the operational benefits of the proposed lifecycle.
Finally, the progressive refinement of \gls{ids} alerts -- and the discovery of new vulnerabilities -- yields increasingly detailed risk measurements.
These findings support the lifecycle’s positive impact on continuous risk evaluation and defense strategy formulation.

\begin{experimentaltakeaways}
\phantomsection
{{\bf \faFlask\ \textit{Experimental Finding \#2:}\label{finding:2}} The iterative refinement of \gls{ag} generation across lifecycle iterations supports the creation of \glspl{ag} with more grounded risk measures.}
\end{experimentaltakeaways}

\subsection{AG-integrated IDS (IDS[AG])}\label{ssec:exp_blue_box}
To evaluate the effectiveness of the $IDS[AG]$ component within the proposed lifecycle, we compare a standard \gls{dt}-based \gls{ids} with its $IDS[AG]$-enhanced counterpart.
Without loss of generality, we consider a multi-class \gls{ids} module that classifies each packet as benign or malicious and also identifies the attack category. 
Due to limitations in the Emerging Threats rule set, we focus on the FTP Patator and DDoS attack classes from the CIC-IDS2017 dataset. 
The $IDS[AG]$ integration is enabled by converting the presence of an attack path in the \gls{ag} -- continuously generated as described in \Cref{ssec:exp_green_box} -- into a binary feature added to the \gls{dt} training set.
Specifically, for each packet observed during training, if its source and destination correspond to nodes connected by an attack path in the \gls{ag}, the feature is set to one; otherwise, it is set to zero.
This allows the \gls{ids} model to incorporate knowledge of whether an attack path exists along the packet’s route. 
To reflect the dynamic nature of the lifecycle, we also assess how the quality of the \gls{ag} and the availability of data affect detection performance.
%

\begin{figure}[tb]
\centering

\begin{minipage}[t]{0.49\textwidth}
    \centering
    \includegraphics[width=\textwidth]{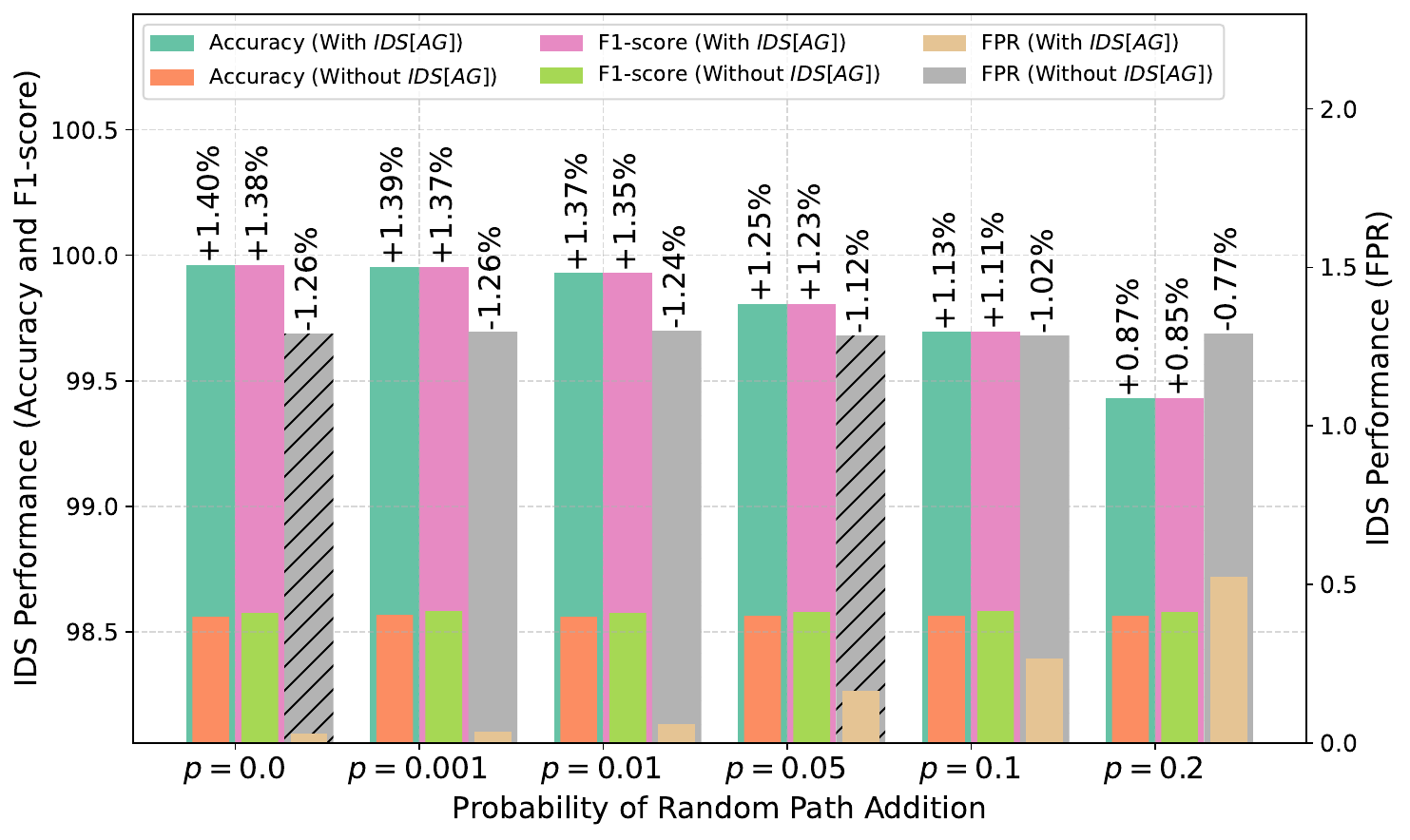}
    \caption{IDS detection effectiveness with and without $IDS[AG]$ against the probability of randomly adding unreliable attack paths in the \gls{ag}. $IDS[AG]$ outperforms the standard \gls{ids}.}
    \label{fig:blue_box_path_prob_exp}
\end{minipage}%
\hfill
\begin{minipage}[t]{0.49\textwidth}
    \centering
    \includegraphics[width=\textwidth]{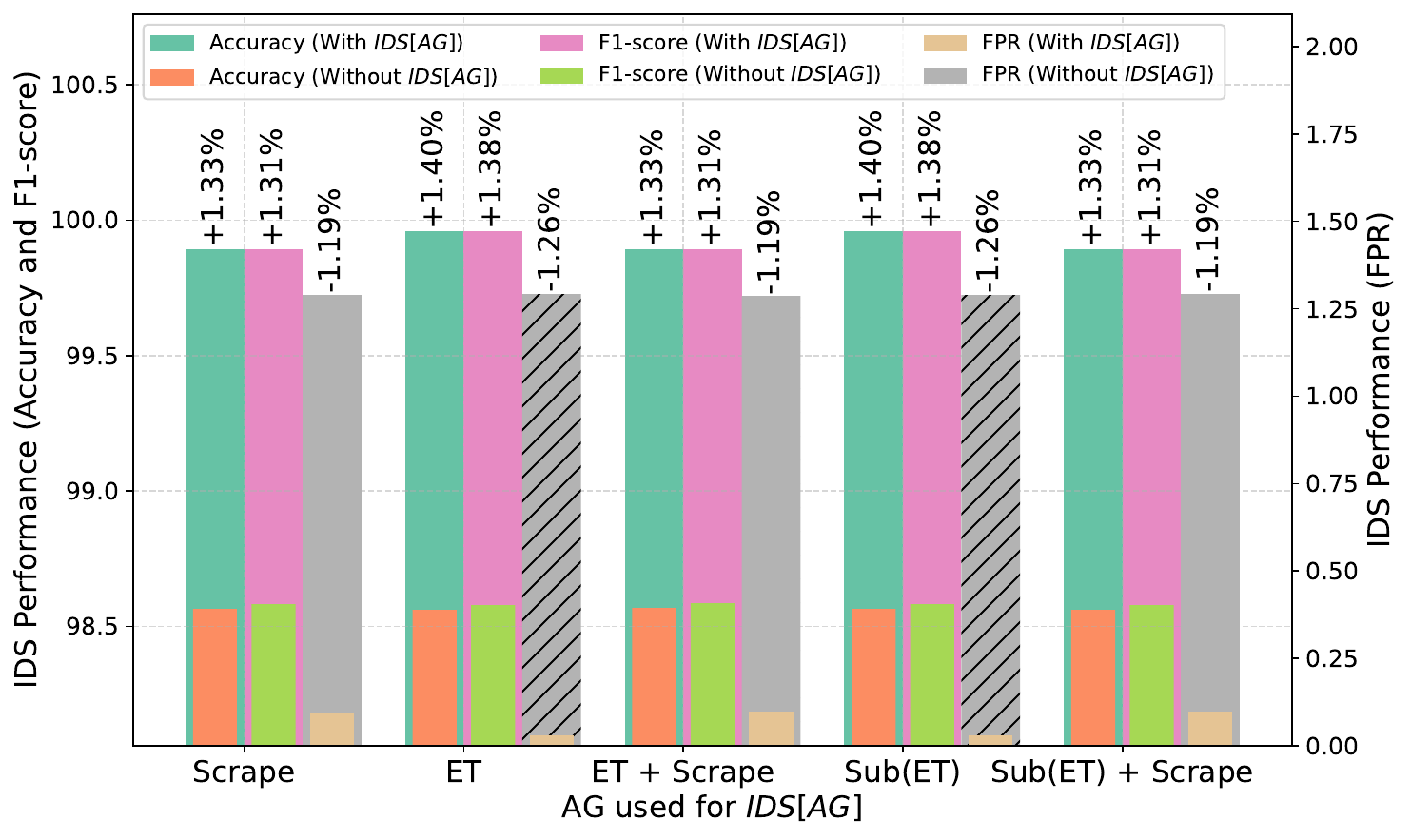}
    \caption{IDS detection effectiveness with and without $IDS[AG]$ against different \glspl{ag}. More tailored \glspl{ag} (e.g., ET and Sub(ET)) achieve higher performance gains.}
    \label{fig:blue_box_ags_exp}
\end{minipage}

\end{figure}

\Cref{fig:blue_box_path_prob_exp} shows the $IDS[AG]$ performance as the \gls{ag} quality is degraded by randomly adding non-existent attack paths with probability $p$.
A value of $p=0$ corresponds to a perfectly reliable \gls{ag} containing only real attack paths, while $p=0.2$ represents a highly connected \gls{ag} where most paths do not reflect actual threats (e.g., because some vulnerabilities cannot be exploited).
Across all $p$ values, $IDS[AG]$ consistently outperforms the baseline, improving detection accuracy and F1-score by over $1\%$, and reducing the \gls{fpr} by a similar margin. 
As expected, lower $p$ -- i.e., more reliable \glspl{ag} -- yields greater performance gains.
These results confirm that the iterative \gls{ag} refinement central to the proposed lifecycle enhances \gls{ids} detection. 
In early iterations (higher $p$), the \gls{ag} may be incomplete, yielding modest improvements.
However, as the \gls{ag} is refined in later iterations (lower $p$), the detection gains become more substantial.
\Cref{fig:blue_box_ags_exp} shows similar trends when varying the \gls{ag} used in $IDS[AG]$, considering:
\begin{inlinelist}
    \item the \gls{ag} derived from vulnerabilities in the dataset's service descriptions (\emph{Scrape}),
    \item the \gls{ag} built from Emerging Threats signatures (\emph{ET}), and
    \item three combinations of \emph{Scrape} and \emph{ET} or its subset (\emph{Sub(ET)}).
\end{inlinelist}
The \emph{ET} \gls{ag} is the optimal \gls{ag} to be used for this task -- as it is constructed directly from effective \gls{ids} signatures -- and results in the highest performance boost, further confirming that iterative \gls{ag} refinement improves attack detection.

\begin{experimentaltakeaways}
\phantomsection
{{\bf \faFlask\ \textit{Experimental Finding \#3:}\label{finding:3}} The iterative refinement of \glspl{ag} across lifecycle iterations leads to progressive improvements in \gls{ids} performance.}
\end{experimentaltakeaways}

\begin{figure}[tb]
\centering

\begin{minipage}[t]{0.49\textwidth}
    \centering
    \includegraphics[width=\textwidth]{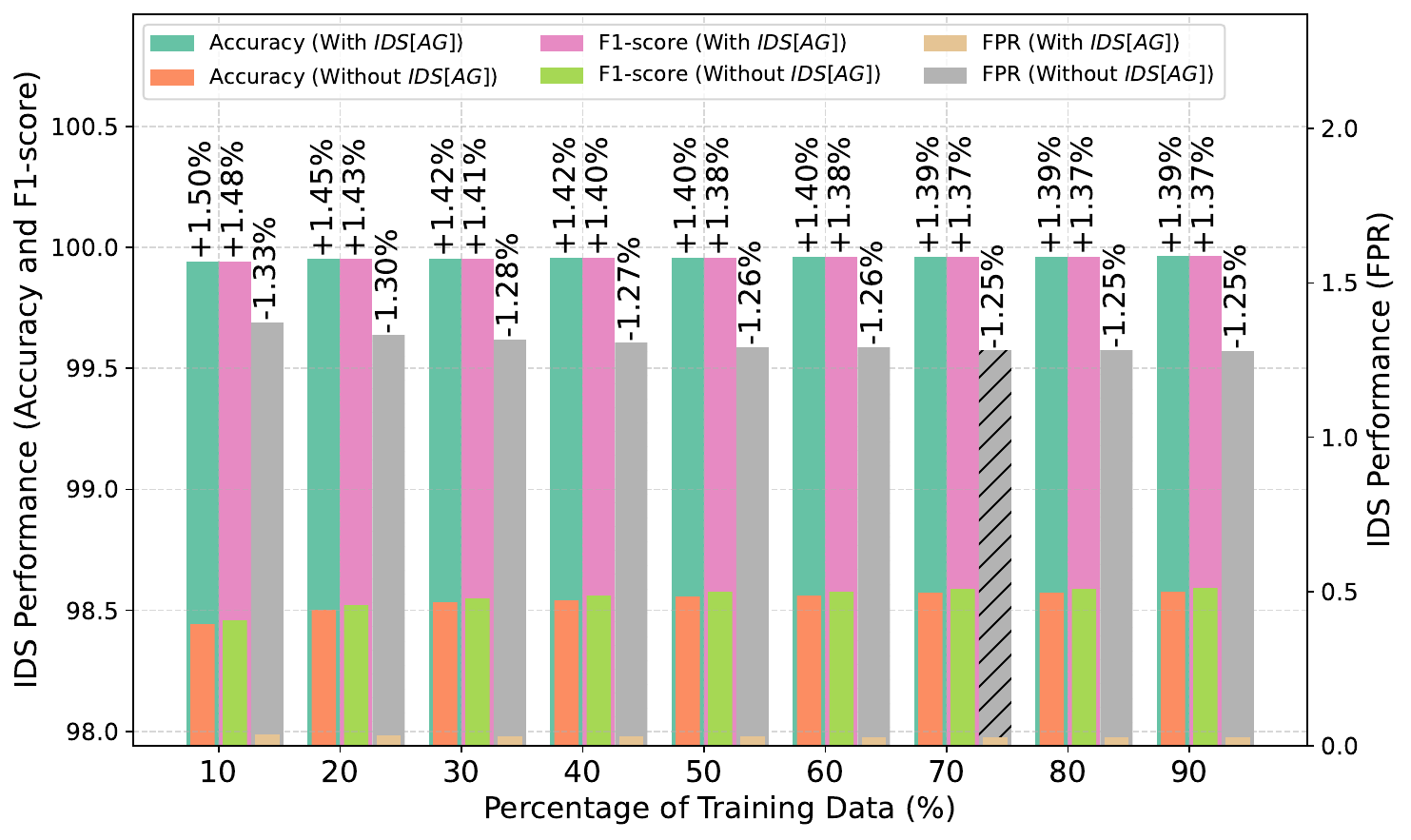}
    \caption{IDS detection effectiveness with and without $IDS[AG]$ against percentage of training data. $IDS[AG]$ achieves higher performance gains when fewer data are available.}
    \label{fig:blue_box_train_percs_exp}
\end{minipage}%
\hfill
\begin{minipage}[t]{0.49\textwidth}
    \centering
    \includegraphics[width=\textwidth]{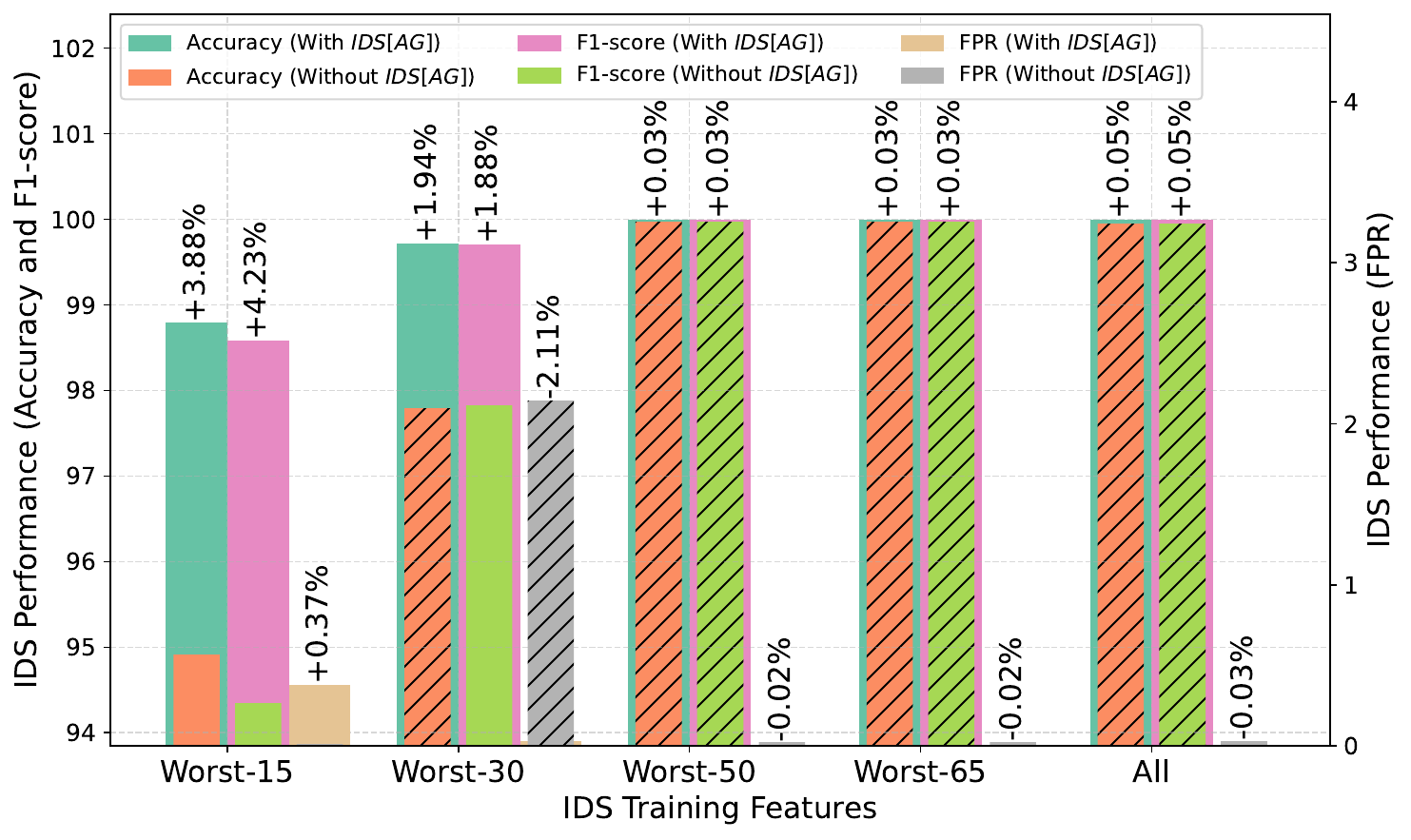}
    \caption{IDS effectiveness with and without $IDS[AG]$ against different sets of the worst-$K$ features. $IDS[AG]$ achieves higher performance gains when few features are available.}
    \label{fig:blue_box_worst_features_exp}
\end{minipage}

\end{figure}

\Cref{fig:blue_box_train_percs_exp} illustrates the impact of data availability on \gls{ids} detection performance when using the $IDS[AG]$ integration.
We vary the percentage of data used to train the \gls{ids}, both with and without $IDS[AG]$. 
Across all settings, $IDS[AG]$ consistently outperforms the baseline.
The performance gain is especially notable when less training data is available---a highly desirable property, given that data collection and labeling are major bottlenecks in network security.
This advantage is particularly relevant in our lifecycle, where early iterations typically involve limited datasets. 
As more data becomes available in later stages, the performance boost from $IDS[AG]$ persists, demonstrating its robustness throughout the lifecycle.
Overall, the results show that the proposed lifecycle enables rapid deployment of an effective and reliable \gls{ids}, even with limited data and partially reliable \glspl{ag}. 
This supports prompt security responses while maintaining consistent performance improvements as the lifecycle progresses.
\begin{experimentaltakeaways}
\phantomsection
{{\bf \faFlask\ \textit{Experimental Finding \#4:}\label{finding:4}} The lifecycle implementation allows prompt \gls{ids} optimization even with limited data.}
\end{experimentaltakeaways}
%

%

\begin{figure}[tb]
\centering

\begin{minipage}[t]{0.49\textwidth}
    \centering
    \includegraphics[width=\textwidth]{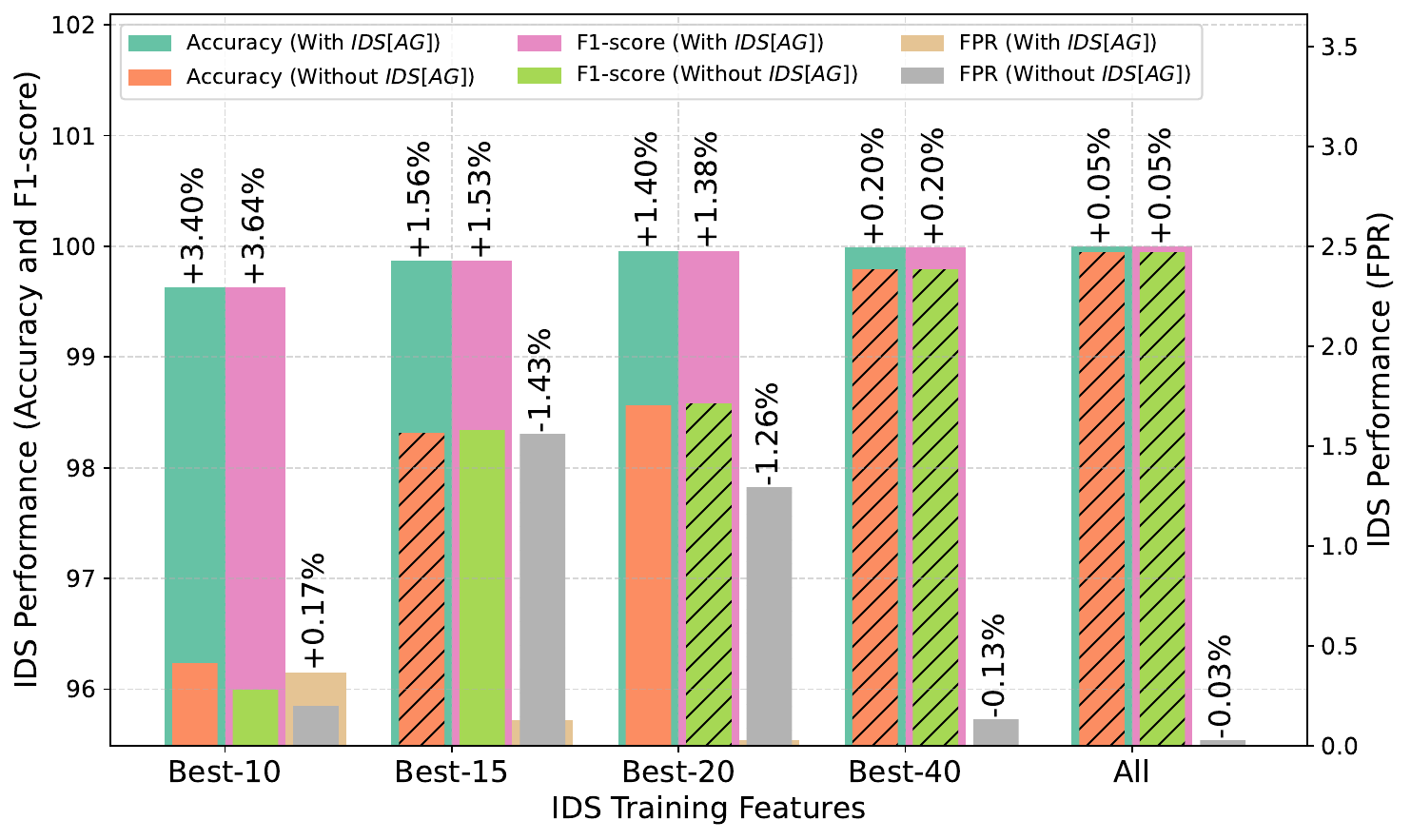}
    \caption{IDS effectiveness with and without $IDS[AG]$ against different sets of the best-$K$ features. $IDS[AG]$ achieves higher performance gains when few features are available.}
    \label{fig:blue_box_top_features_exp}
\end{minipage}%
\hfill
\begin{minipage}[t]{0.49\textwidth}
    \centering
    \includegraphics[width=\textwidth]{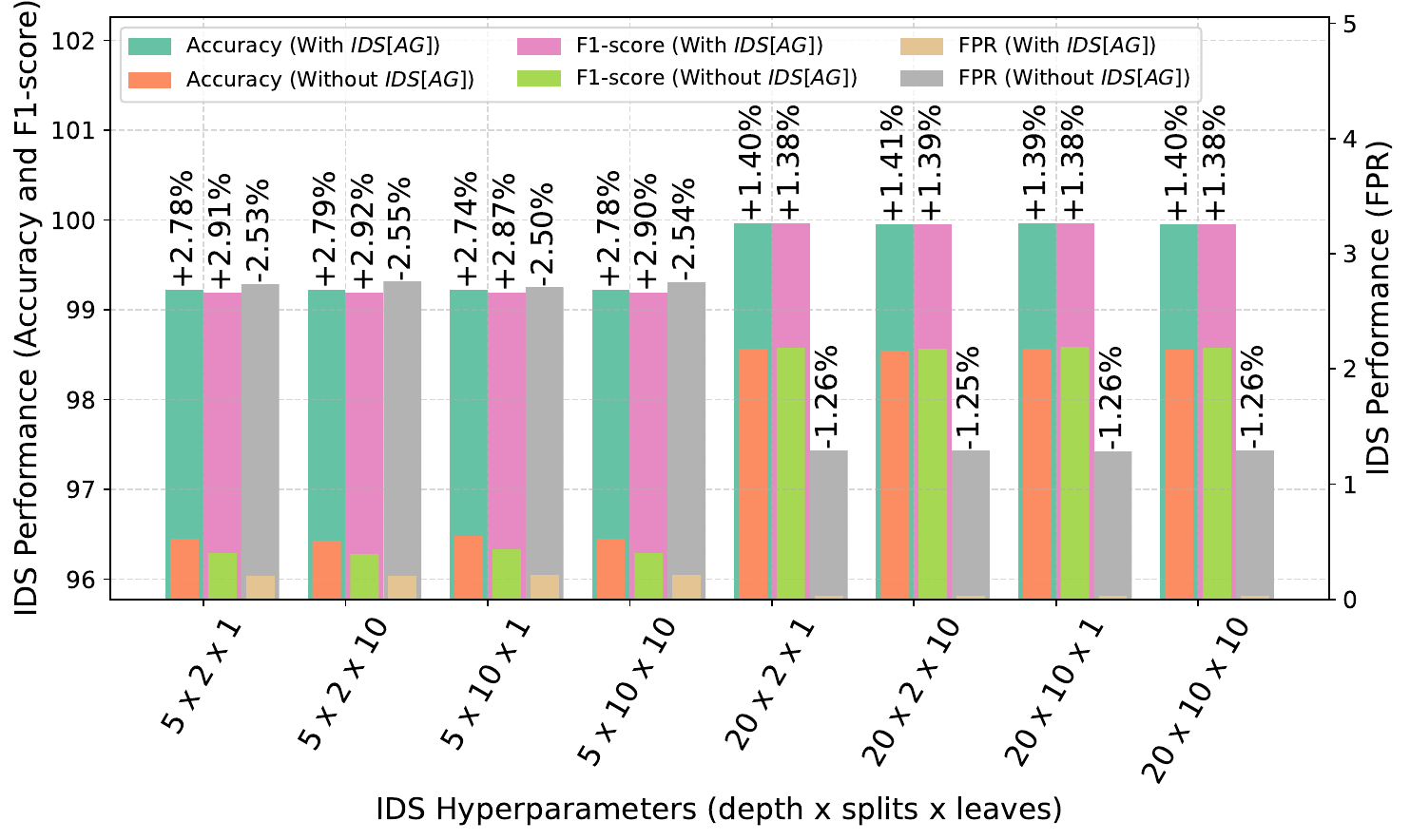}
    \caption{IDS detection effectiveness with and without $IDS[AG]$ against different hyperparameter values for the \gls{dt}-based \gls{ids}. $IDS[AG]$ achieves higher performance gains when the \gls{dt} settings are not optimal (e.g., shallower tree).}
    \label{fig:blue_box_ids_params_exp}
\end{minipage}

\end{figure}

To analyze the impact of the chosen features on the performance gain achieved by the $IDS[AG]$ integration mechanism, we consider training several \glspl{dt}, altering the set of features selected.
We consider selecting either the $K$ best or worst features according to the ANOVA F-value and varying $K$ between 10 and 80 (the CIC-IDS dataset contains 80 features in total).
\Cref{fig:blue_box_worst_features_exp} presents the results obtained when training with a varying range of the $K$ worst features.
Similarly, \Cref{fig:blue_box_top_features_exp} presents the results obtained when training with the $K$ best features.
The obtained results highlight how $IDS[AG]$ achieves higher performance gains when few features are available, especially if these features are unreliable.
Even when relying only on the very best features available in the dataset, the standardly trained \gls{ids} underperforms against its $IDS[AG]$ counterpart, proving the importance of adding the \gls{ag}'s information into the \gls{ids} decision.
These results are expected, as the addition of a single feature that defines if an attack path exists is relevant to boost the classification performance mostly if the information available from the other features is limited.
Meanwhile, whenever the information available from a large set of features is sufficient to make the \gls{dt} model fit well and reach a high level of performance, the addition of a single -- yet relevant -- feature does not impact the model performance much.
However, it is also relevant to notice that although the performance gains achieved when a large set of features is available, it is still larger than zero, meaning that it is still worth embedding the \gls{ag}'s information into the \gls{ids} decision process to ensure a more reliable detection.
Finally, it is relevant to notice that whenever the set of available features is limited, the performance gains achieved by $IDS[AG]$ is very large -- up to 4\% F1-score --, which represents a very desirable property, as it allows to define a light-weight detection scheme, relying on few features that are easy to extract from the data during early iterations of the proposed lifecycle.
%

Lastly, we here analyze the impact that the \gls{dt} hyperparameter values have on the performance gains achieved using $IDS[AG]$.
We vary the three most relevant parameters of the \gls{dt} model that implements the \gls{ids}, namely its maximum depth, the minimum number of samples required to split an internal node, and the minimum number of samples required to be at a leaf node.
\Cref{fig:blue_box_ids_params_exp} presents the results obtained.
As expected, $IDS[AG]$ achieves higher performance gains whenever the \gls{dt} settings are not optimal.
For example, a shallower tree results in a less accurate classification by the model, where the \gls{ag}-based knowledge injection process can help a lot in improving the \gls{ids}'s decision-making process.
Meanwhile, whenever the \gls{dt} hyperparameters allow for the construction of a very reliable \gls{ids}, the \gls{ag}-based knowledge injection process can only partially improve the detection performance.
Nevertheless, $IDS[AG]$ always brings positive performance gains, achieving more than 1\% accuracy and F1 boost even when the best \gls{dt} parameters are selected -- e.g., depth set to 20, splits to 2, and leaves to 1 -- and defining a detection module that achieves almost 100\% detection accuracy.
Therefore, these results prove that the usage of $IDS[AG]$ refinement is always desirable, backing our lifecycle definition.
%

\subsection{AG-based IDS Refinement (IDS $\rightarrow$ AG)}\label{ssec:exp_orange_box}
We evaluate the effectiveness of the IDS $\rightarrow$ AG component in the proposed lifecycle, using the same experimental settings as in \Cref{ssec:exp_blue_box}.
%
We extend the \gls{dt}-based \gls{ids} module with an IDS $\rightarrow$ AG refinement layer that leverages prior knowledge from the \gls{ag} constructed in \Cref{ssec:exp_green_box}.
Specifically, the \gls{dt} is trained on the CIC-IDS2017 dataset following standard \gls{ml}-based \gls{ids} procedures. 
During inference, if the model predicts an attack, we check whether the packet’s source and destination are part of an existing attack path in the \gls{ag}. 
If such a path exists, the prediction is retained; otherwise, it is flipped to benign.
This refinement assumes that if no viable attack path exists in the \gls{ag}, the \gls{ids} prediction is likely a false positive, as no threat could feasibly exploit that flow. 
Conversely, we do not flip benign predictions to attacks, since the presence of an attack path only indicates a potential strategy, not an active threat.
Therefore, the \gls{ids}$\rightarrow$\gls{ag} strategy is designed to correct false positives---i.e., benign samples misclassified as attack-based on structural threat knowledge encoded in the \gls{ag}.
%


\begin{figure}[tb]
\centering

\begin{minipage}[t]{0.49\textwidth}
    \centering
    \includegraphics[width=\textwidth]{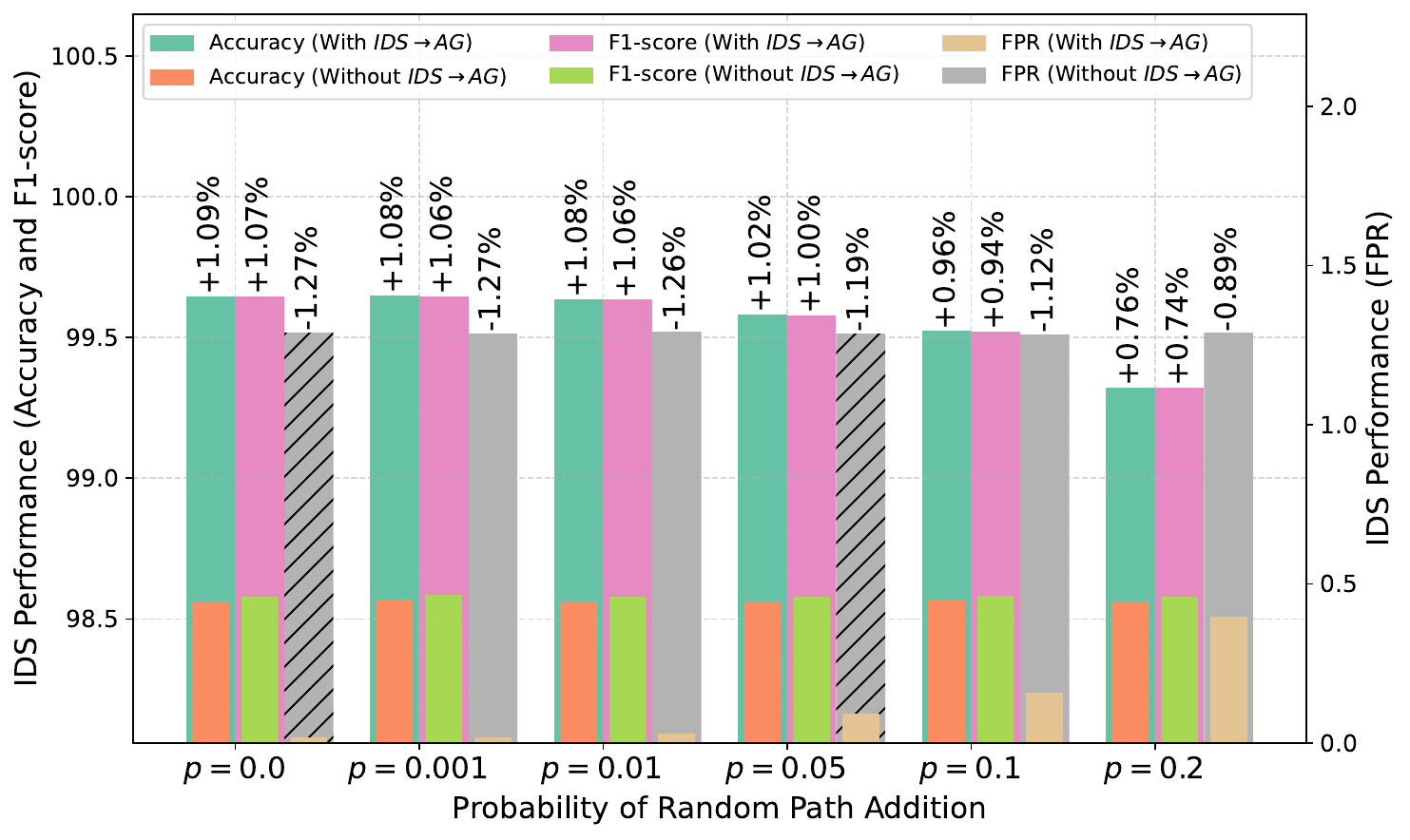}
    \caption{IDS effectiveness with and without IDS $\rightarrow$ AG against the probability of randomly adding unreliable attack paths in the \gls{ag}. IDS $\rightarrow$ AG outperforms the standard \gls{ids}.}
    \label{fig:orange_box_path_prob_exp}
\end{minipage}%
\hfill
\begin{minipage}[t]{0.49\textwidth}
    \centering
    \includegraphics[width=\textwidth]{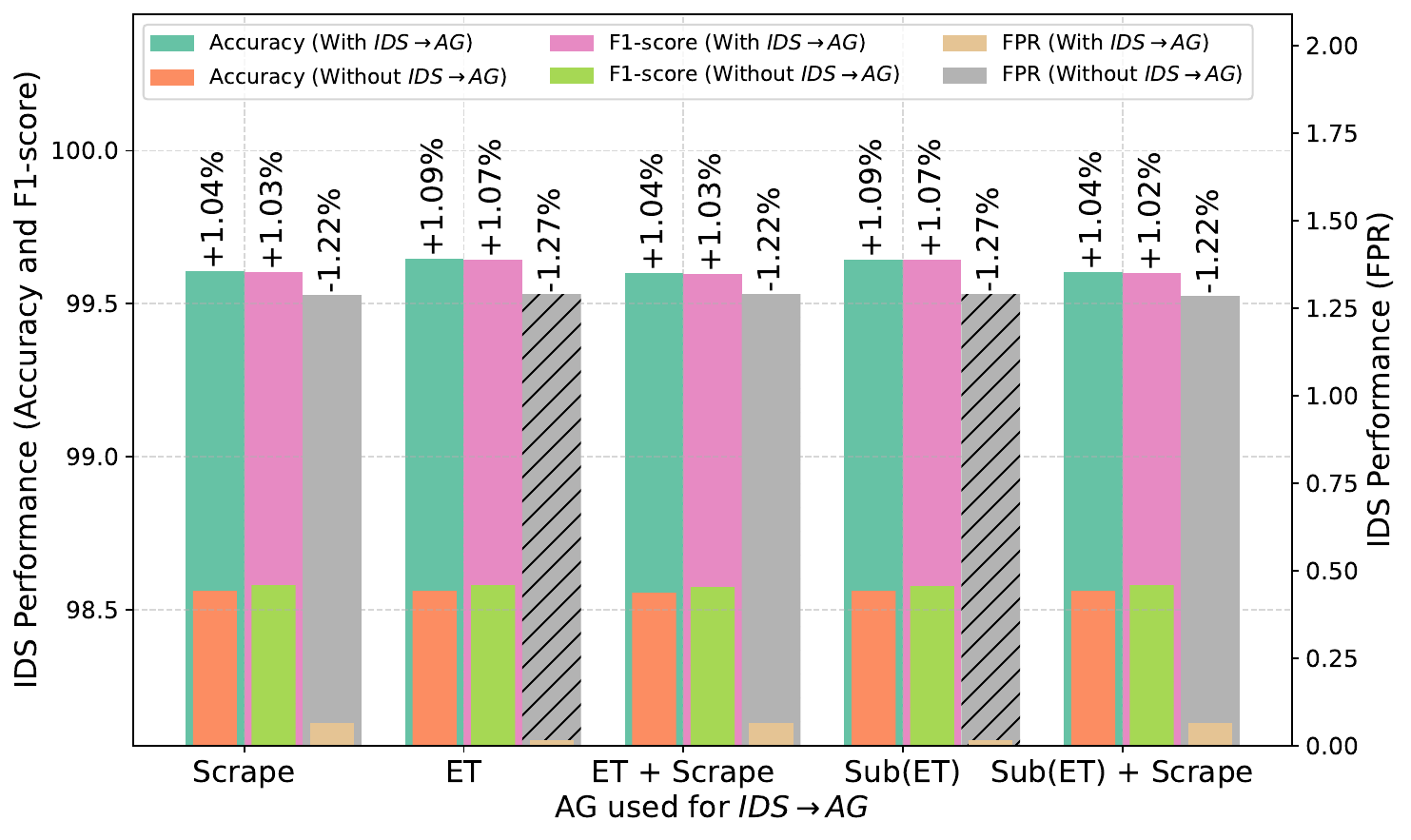}
    \caption{IDS effectiveness with and without IDS $\rightarrow$ AG against different \glspl{ag}. More tailored \glspl{ag} (e.g., ET and Sub(ET)) achieve higher performance gains.}
    \label{fig:orange_box_ags_exp}
\end{minipage}

\end{figure}

We conduct experiments similar to those in \Cref{ssec:exp_blue_box}, analyzing the impact of \gls{ag} quality on the \gls{ids}$\rightarrow$\gls{ag} component by randomly adding non-existent attack paths and using different \glspl{ag}.
\Cref{fig:orange_box_path_prob_exp,fig:orange_box_ags_exp} present the results of these experiments.
The findings support our intuition regarding the benefits of \gls{ag}-based refinement: the \gls{ids}$\rightarrow$\gls{ag} module consistently outperforms the baseline across all settings.
Moreover, more reliable (i.e., lower $p$) and tailored \glspl{ag} -- such as \emph{ET} or \emph{Sub(ET)} -- lead to greater performance gains.
These results confirm that the iterative \gls{ag} refinement process central to the proposed lifecycle yields progressively higher detection performance, as \glspl{ag} become increasingly accurate over time.

\begin{experimentaltakeaways}
\phantomsection
{{\bf \faFlask\ \textit{Experimental Finding \#5:}\label{finding:5}} Both $IDS[AG]$ and IDS $\rightarrow$ AG enhance the \gls{ids} attack detection performance, with the \gls{ag} iterative refinement leading to incremental improvements.}
\end{experimentaltakeaways}
%

\begin{figure}[tb]
\centering

\begin{minipage}[t]{0.49\textwidth}
    \centering
    \includegraphics[width=\textwidth]{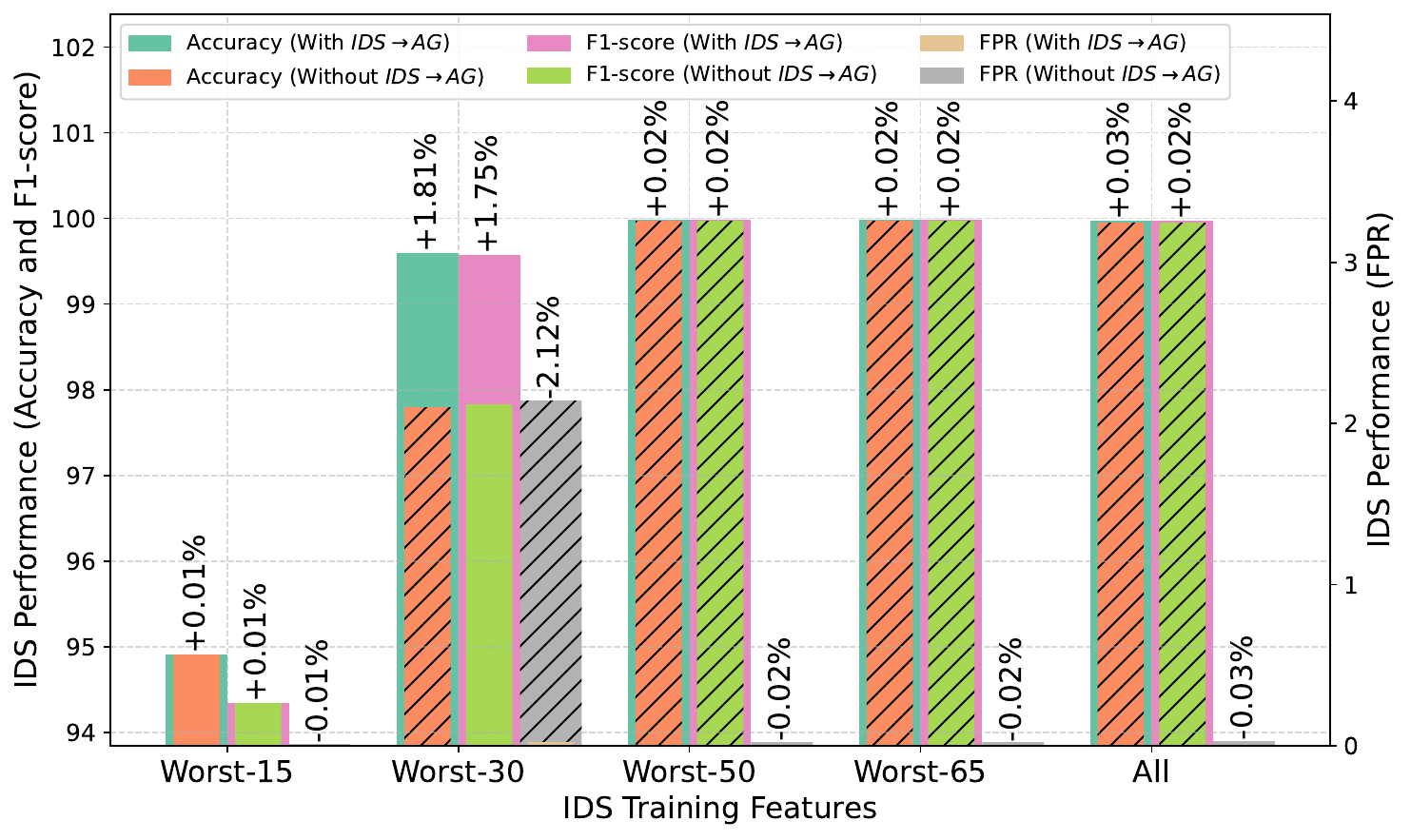}
    \caption{IDS effectiveness with and without IDS $\rightarrow$ AG against different sets of the worst-$K$ features.}
    \label{fig:orange_box_worst_features_exp}
\end{minipage}%
\hfill
\begin{minipage}[t]{0.49\textwidth}
    \centering
    \includegraphics[width=\textwidth]{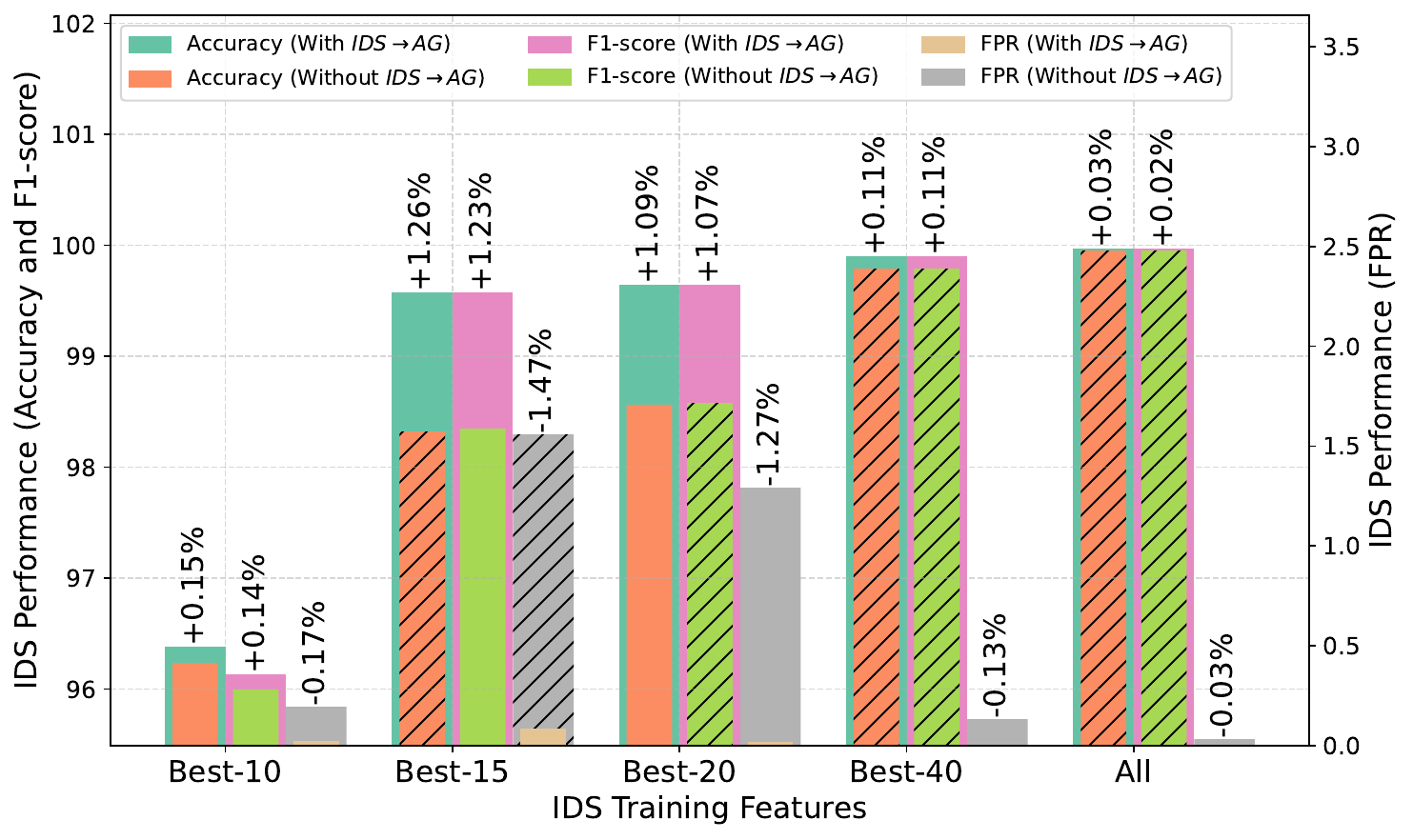}
    \caption{IDS effectiveness with and without IDS $\rightarrow$ AG against different sets of the best-$K$ features.}
    \label{fig:orange_box_top_features_exp}
\end{minipage}

\end{figure}

%
An additional analysis concerns the impact of the chosen features on the performance gain achieved using IDS $\rightarrow$ AG by training several \glspl{dt} over various sets of selected features.
We consider selecting either the $K$ best or worst features according to the ANOVA F-value and varying $K$ between 10 and 80.
\Cref{fig:orange_box_worst_features_exp} (\Cref{fig:orange_box_top_features_exp}) presents the results obtained when training with a varying range of the $K$ worst (best) features.
The obtained results show how IDS $\rightarrow$ AG outperforms a standard \gls{ids} counterpart across all settings, validating our lifecycle potential.
Interestingly, using only very few features result in a very marginal gain -- different from the findings of \Cref{fig:blue_box_worst_features_exp,fig:blue_box_top_features_exp} --, since the IDS $\rightarrow$ AG can only refine the \gls{ids} predictions and not completely alter its decision-making process.
Therefore, for IDS $\rightarrow$ AG to achieve the optimal performance gain, the underlying \gls{ids} must achieve a reasonably high detection performance, with attacks being detected in several cases.
Due to the unbalanced nature of the dataset and task, whenever only very few unreliable features are used, the \gls{dt} model learns to predict benign instances across almost all packets seen -- as proven by the lower accuracy and almost zero \gls{fpr} --, thus making the refinement process less relevant---as it allows only to refine falsely predicted attack patterns. 
%


\begin{figure}[tb]
\centering

\begin{minipage}[t]{0.49\textwidth}
    \centering
    \includegraphics[width=\textwidth]{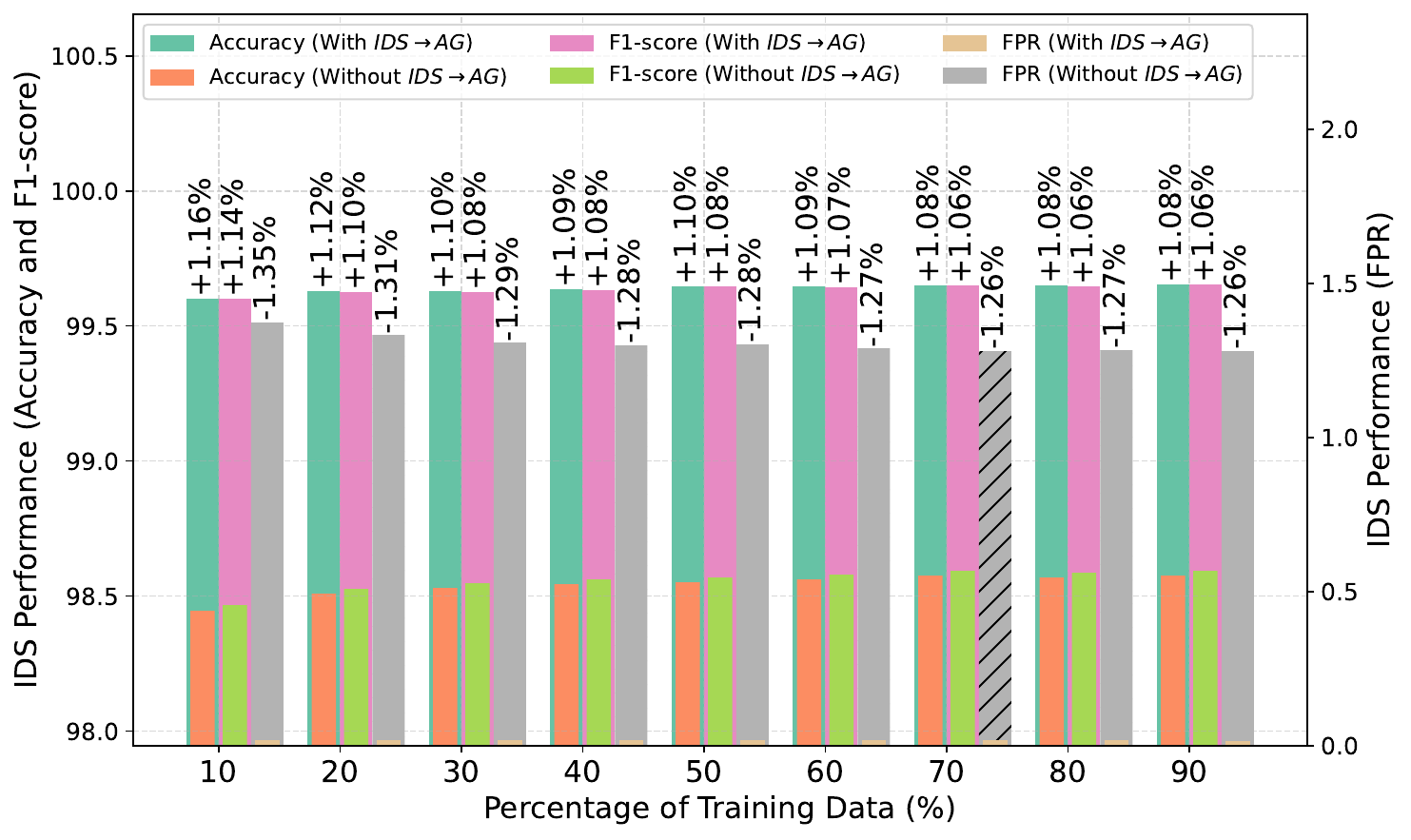}
    \caption{IDS detection effectiveness with and without IDS $\rightarrow$ AG against percentage of training data. IDS $\rightarrow$ AG achieves higher performance gains when fewer data are available.}
    \label{fig:orange_box_train_percs_exp}
\end{minipage}%
\hfill
\begin{minipage}[t]{0.49\textwidth}
    \centering
    \includegraphics[width=\textwidth]{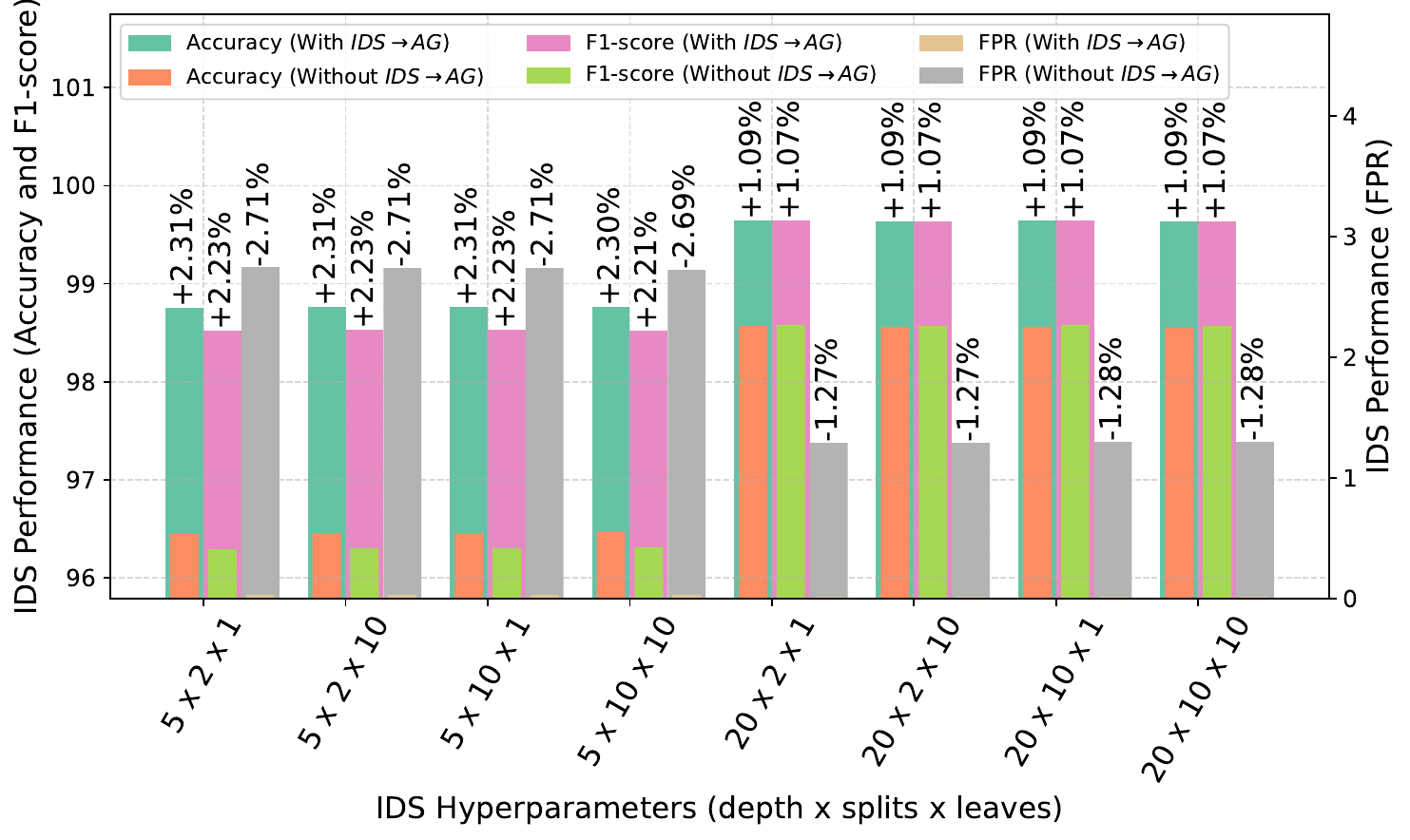}
    \caption{IDS detection effectiveness with and without IDS $\rightarrow$ AG against different hyperparameter values for the \gls{dt}-based \gls{ids}. IDS $\rightarrow$ AG outperforms the standard counterpart across all settings. IDS $\rightarrow$ AG achieves higher performance gains when the \gls{dt} settings are not optimal (e.g., shallower tree).}
    \label{fig:orange_box_ids_params_exp}
\end{minipage}

\end{figure}

We also analyze the effect that data availability has on the \gls{ids} detection performance when using the IDS $\rightarrow$ AG integration.
Similarly to what was done for $IDS[AG]$, we vary the percentage of training data, testing the obtained model on the remaining test data, and measuring the performance delta.
The results are in \Cref{fig:orange_box_train_percs_exp} and confirm the findings of \Cref{ssec:exp_orange_box}, highlighting how IDS $\rightarrow$ AG results in higher performance gains when fewer data are available.
Therefore, leveraging IDS $\rightarrow$ AG enables reducing the amount of data to be collected, proving the reliability of our lifecycle.
%


Finally, we here analyze the impact that the \gls{dt} hyperparameter values have on the performance gains achieved using IDS $\rightarrow$ AG.
We vary the depth, splits, and leaves parameters of the \gls{dt} model that implements the \gls{ids} and present the corresponding results in \Cref{fig:orange_box_ids_params_exp}.
The results are similar to the ones obtained for $IDS[AG]$, showing that the IDS $\rightarrow$ AG refinement process achieves higher performance gains whenever the \gls{dt} settings are not optimal---e.g., a shallower tree resulting in a less accurate classification.
Once again, these results prove that the usage of IDS $\rightarrow$ AG refinement is desirable across \gls{ids} configurations, backing our lifecycle definition.
%

\section{Opportunities}\label{sec:opportunities}
Our systematization highlights promising areas for advancement: achieving dataset standardization, addressing model heterogeneity, and enhancing integration strategies throughout the system lifecycle.
In this perspective, the proposed systematic \gls{ag}-\gls{ids} lifecycle provides a clear roadmap of opportunities to fully integrate these technologies from a comprehensive, end-to-end perspective.
Building on these insights, we here outline more in detail the promising research directions to advance adaptive and scalable \gls{ag}-\gls{ids} systems.

\mypara{\gls{ag}-\gls{ids} system deployment}
The proof-of-concept in \Cref{sec:case_study} showcases the effectiveness of the proposed lifecycle.
However, its thorough development and deployment still represent an open research opportunity for the future.
Indeed, to deploy the proposed lifecycle, it is required to define flexible frameworks capable of dealing with the management of both \glspl{ag} and \glspl{ids} over various network scenarios, independently of their types.
Moreover, such a framework should enable multiple levels of integration, while also being flexible to allow defining more sophisticated combination approaches.
Existing Security Information and Event Management (SIEM) or cyber threat intelligence platforms, such as Microsoft XDR~\footnote{\url{https://tinyurl.com/MicrosoftXDR}}, MITRE ATT\&CK Mapping~\footnote{\url{https://attack.mitre.org}}, or GraphWeaver~\footnote{\url{https://graphweaver.com}}, can be a starting base for eventually building an \gls{ag}-\gls{ids} system. 

\mypara{Multi-source integration}
%
Our systematization reveals that current integration approaches typically rely on single \gls{ag} and \gls{ids} sources, limiting contextual coverage and detection diversity.
Additionally, real-world environments -- for instance, security operations centers -- monitor and respond to different sources.  
In this context, future research should explore and support integration across multiple \glspl{ag} and \glspl{ids} following a similar roadmap to existing solutions, e.g., Elastic Stack~\footnote{\url{https://www.elastic.co}} or Security Onion~\footnote{\url{https://securityonionsolutions.com}}.

%
%


\mypara{Datasets, Benchmarking and Reproducibility}
Advancing \gls{ag} and \gls{ids} integration requires addressing key foundational opportunities, starting with the scarcity of comprehensive datasets. 
Designing \glspl{ag} needs vulnerability inventories, while optimizing \glspl{ids} requires system snapshots (e.g., network traffic). 
Robust datasets combining both are scarce.
Thus, collecting such data and developing methods to map network traffic to reachability inventories are critical.
Furthermore, the lack of open-source implementations hinders reproducibility. 
This necessitates creating a standardized benchmarking platform for fair comparison, which will incentivize further research in the field.



\mypara{Machine Learning}
The systematization highlights three compelling opportunities for advancing the use of \gls{ml} in \gls{ag}-\gls{ids} integration systems:
\begin{inlinelist}
    \item\label{item:ml} integrating advanced \gls{ml} models -- e.g., \glspl{gnn} \cite{AgiolloEurosp2023}, \glspl{cnn} \cite{VinayakumarIccai17} and \glspl{lstm} \cite{LaghrissiJbd21} -- which have demonstrated encouraging performance for detection and response tasks; 
    
    \item\label{item:retrain} implementing continuous retraining to proactively address concept drift and challenges posed by evolving traffic; 
    
    \item\label{item:deploy} increasing attention to the constraints and requirements for real-world deployment; and,

    \item\label{item:agentic} leveraging emerging tools like LangChain~\footnote{\url{https://www.langchain.com}} for developing modular agentic \gls{ag} and \gls{ids} integration.
\end{inlinelist}
Regarding \ref{item:ml}, moving beyond simplistic \gls{ml} solutions allows models to exploit critical contextual information (e.g., network topology and temporal dynamics), significantly boosting accuracy.
With respect to \ref{item:retrain}, continuous learning must replace the static train-and-deploy paradigm to maintain model performance against dynamic traffic and data distribution shifts.
Concerning \ref{item:deploy}, models must be developed with real-world constraints in mind to ensure practical deployment.
Finally, \ref{item:agentic} offers a novel path for automating complex investigation steps through custom agentic systems.

\mypara{Enhanced \gls{cti}}
In the context of threat intelligence, the \gls{ag}-\gls{ids} integration is worth investigating for the automation of Tactics, Techniques and Procedures (TTPs) mapping~\cite{al2024mitre}. 
%
%
Specifically, the raw alerts generated by the \glspl{ids} are correlated and mapped to MITRE ATT\&CK semantics on TTPs by leveraging the logical structure of \glspl{ag}.
In another note, the \gls{ag}-\gls{ids} integration system can serve as a validation engine.
In particular, the identified attack path or attack scenario is confirmed by a sequence of \gls{ids} alerts, which can be provided as feedback for threat intelligence platforms.

\mypara{Neuro-Symbolic integration for \gls{ag}-\gls{ids}}
Neuro-symbolic (NeSy) AI combines symbolic reasoning, which handles structured knowledge and logical constraints, with neural learning, which excels at pattern recognition and generalization from data \cite{CiattoCsur2024,marra2024statistical}. 
In the \gls{ag}–\gls{ids} integration context, NeSy can enable systems that reason over \glspl{ag} while learning from \gls{ids} alerts and network traffic, creating adaptive and interpretable detection pipelines. 
This dual capability is particularly valuable for correlating alerts with feasible attack paths and anticipating complex intrusion strategies.
A concrete example is dynamic alert prioritization: a NeSy-based system could use \gls{ag} constraints (e.g., prerequisite vulnerabilities and reachable states) to rank \gls{ids} alerts by their likelihood of forming a valid attack path, while neural components learn patterns of alert sequences that historically led to successful compromises. 
This would reduce false positives and help SOC operators focus on the most critical threats.
\section{Conclusions}
\label{sec:conclusions}
\glspl{ag} and \glspl{ids} are fundamental to network security, underpinning both detection and response. 
While each has evolved independently and in partial integration, research on their combined use remains fragmented, limiting a unified understanding of their full potential.
This paper provides a comprehensive systematization of \gls{ag}-\gls{ids} integration, categorizing existing approaches into three main scopes: \gls{ids}-based \gls{ag} generation, \gls{ag}-integrated \gls{ids}, and \gls{ag}-based \gls{ids} refinement.
Our analysis reveals that no current solution addresses all three dimensions cohesively, leaving a gap for holistic integration.
To bridge this gap, we propose a dynamic \gls{ag}-\gls{ids} lifecycle that enables continuous adaptation and iterative refinement of detection and response mechanisms. 
We further provide an experimental proof-of-concept of this envisioned lifecycle, empirically demonstrating its advantages in improving scalability, detection accuracy, and contextual risk assessment.
This vision offers a pathway to overcome persistent challenges such as scalability, high false positive rates, and limited situational awareness. 
Moreover, it establishes a foundation for future research, highlighting opportunities in benchmarking, dataset development, and real-world deployment.

\section*{Acknowledgments}
This research was supported by the Netherlands’ National Growth Fund through the Dutch 6G flagship project “Future Network Services”.

\bibliographystyle{abbrvnat}
\bibliography{bibliography}
\end{document}